%% file: main.tex
\documentclass[twoside,journey]{IEEEtran}

\PassOptionsToPackage{table}{xcolor}
\usepackage[utf8]{inputenc}
\usepackage[T1]{fontenc}
\usepackage{graphicx,amsmath,amssymb,amsfonts}
\usepackage[noadjust]{cite}
\usepackage{tikz} 
\usepackage[numbers,sort&compress]{natbib}
\usepackage[edges]{forest}
\usepackage{makecell}

\usepackage[nointegrals]{wasysym} 
\usepackage{array}

\usepackage{float}
\usepackage{multicol}

\usepackage{setspace}
\usepackage{subfig}
\usepackage{url}
\usepackage{stfloats}

\usepackage{amsthm,pifont}
\usepackage{flushend}

\usepackage{cases,subeqnarray}
\usepackage{bm,multirow,bigstrut}
\usepackage{amsmath, amsthm, amssymb}
\usepackage{textcomp}
\usepackage{latexsym,bm}
\usepackage{booktabs}
\usepackage{pdfpages}

\usepackage{mathtools}
\usepackage{dsfont}
\usepackage{extarrows}
\usepackage{epsfig}
\usepackage{epstopdf}
\usepackage[noend]{algpseudocode}
\usepackage{algorithmicx,algorithm}
\usepackage{amsmath}
\usepackage{xr}
\usepackage{threeparttable}

\usepackage[pagebackref=false,breaklinks=true,colorlinks,citecolor=blue,linkcolor=blue,bookmarks=true]{hyperref}
\usepackage[nameinlink]{cleveref}

\def\RevisionPreambleLight{1}
\input{preamble}
\let\RevisionPreambleLight\undefined

\usepackage{fontawesome5} 
\usepackage{adjustbox} 
\usetikzlibrary{shapes.geometric, arrows, positioning,fit,calc}
\definecolor{lime}{HTML}{A6CE39}

\definecolor{level0}{HTML}{F8CECC}
\definecolor{level1}{HTML}{FFE6CC}
\definecolor{level2}{HTML}{E1D5E7}
\definecolor{level3}{HTML}{F5F5F5}
\definecolor{level4}{HTML}{F5F5F5}
\definecolor{hidden-draw}{RGB}{150,150,150}
\definecolor{lightkeycolor}{RGB}{255,240,240}

\usetikzlibrary{arrows.meta}
\forestset{
    .style={
        for tree={
            font=\sffamily,
            edge={thick},
        }
    }
}

\DeclareRobustCommand{\orcidicon}{%
\begin{tikzpicture}
\draw[lime, fill=lime] (0,0) 
circle [radius=0.16] 
node[white] {{\fontfamily{qag}\selectfont \tiny ID}};\draw[white, fill=white] (-0.0625,0.095) 
circle [radius=0.007];\end{tikzpicture}
\hspace{-2mm}}
\foreach \x in {A, ..., Z}{%
\expandafter\xdef\csname orcid\x\endcsname{\noexpand\href{https://orcid.org/\csname orcidauthor\x\endcsname}{\noexpand\orcidicon}}
}

\hypersetup{
    colorlinks=true,
    linkcolor=black,
    filecolor=black,
    urlcolor=black,
}

\crefname{section}{Sec.}{Secs.}
\crefname{subsection}{Sec.}{Secs.}
\crefname{subsubsection}{Sec.}{Secs.}
\crefname{figure}{Fig.}{Figs.}
\crefname{table}{Table}{Tables}
\crefname{equation}{Eq.}{Eqs.}
\crefname{algorithm}{Alg.}{Algs.}
\crefname{theorem}{Theorem}{Theorems}
\crefname{lemma}{Lemma}{Lemmas}
\crefname{proposition}{Proposition}{Propositions}
\crefname{corollary}{Corollary}{Corollaries}
\crefname{remark}{Remark}{Remarks}
\crefname{definition}{Definition}{Definitions}
\crefformat{equation}{#2Eq.~#1#3}
\crefmultiformat{equation}{#2Eqs.~#1#3}{ and~#2#1#3}{, #2#1#3}{ and~#2#1#3}
\Crefformat{equation}{#2Eq.~#1#3}
\Crefmultiformat{equation}{#2Eqs.~#1#3}{ and~#2#1#3}{, #2#1#3}{ and~#2#1#3}

\providecommand{\redcross}{\textcolor{red!70!black}{\ding{55}}}

\IEEEoverridecommandlockouts
\begin{document}

\title{A Survey on Cloud-Edge-Terminal\\Collaborative Intelligence in AIoT Networks}

\author{
 Jiaqi Wu\orcidA{},
 Jing Liu\orcidB{},
 Yang Liu\orcidC{},~\IEEEmembership{Member,~IEEE},
Lixu Wang\orcidI{},
 Zehua Wang\orcidD{},~\IEEEmembership{Member,~IEEE},
 Wei Chen\orcidF{},~\IEEEmembership{Member,~IEEE},
 Zijian Tian\orcidE{},
 Richard Yu\orcidG{},~\IEEEmembership{Fellow,~IEEE},
 Victor C. M. Leung\orcidH{},~\IEEEmembership{Life Fellow,~IEEE}
\IEEEcompsocitemizethanks{
This work is supported in part by the Guangdong Pearl River Talents Recruitment Program under Grant No. 2019ZT08X603, the Guangdong Pearl Rivers Talent Plan under Grant No. 2019JC01X235, the Natural Sciences and Engineering Research Council (NSERC) of Canada under Grants Nos. RGPIN-2021-02970 and DGECR-202100187, and the Mitacs Project under Grant No. IT44479, and the Jiangsu Provincial Distinguished Professor Program under Grant No. 140923070. (Corresponding author: Jing Liu).}
 \IEEEcompsocitemizethanks{
\IEEEcompsocthanksitem Jiaqi Wu is with the School of Artificial Intelligence, China University of Mining and Technology (Beijing), Beijing 100083, China and with the Department of Automation, Tsinghua University, Beijing 100084, China (e-mail: wujiaqi@mail.tsinghua.edu.cn).
\IEEEcompsocthanksitem Jing Liu is with the Department of Electrical and Computer Engineering, The University of British Columbia, Vancouver, BC V6T 1Z4, Canada, and with the Division of Natural and Applied Sciences, Duke Kunshan University, Kunshan 215316, China, and also with the College of Future Information Technology, Fudan University, Shanghai 200433, China (e-mail: jing.liu@ieee.org).
\IEEEcompsocthanksitem Yang Liu is with the College of Electronic and Information Engineering, Tongji University, Shanghai 201804, China (e-mail: yang\_liu@ieee.org)
\IEEEcompsocthanksitem Lixu Wang is with the College of Computing and Data Science, Nanyang Technological University, Singapore 639798 (e-mail: wanglixu4334@gmail.com).
\IEEEcompsocthanksitem Zehua Wang is with the Department of Electrical and Computer Engineering, The University of British Columbia, Vancouver, BC V6T 1Z4, also with the School of Computer Science and Technology, China University of Mining and Technology, Xuzhou 221116, Jiangsu province, China. (e-mail: zwang@ece.ubc.ca).
\IEEEcompsocthanksitem Zijian Tian is with the School of Artificial Intelligence, China University of Mining and Technology (Beijing), Beijing 100083, China.(e-mail: tianzj0726@126.com).
\IEEEcompsocthanksitem Wei Chen is with the School of Computer Science and Technology, China University of Mining and Technology, Xuzhou 221116, Jiangsu province, China. (e-mail: chenwdavior@163.com).
\IEEEcompsocthanksitem F. Richard Yu is with the Department of Systems and Computer Engineering, Carleton University, Ottawa, ON K1S 5B6, Canada (e-mail: richard.yu@carleton.ca).
\IEEEcompsocthanksitem Victor C. M. Leung is with the Artificial Intelligence Research Institute, SMBU, Shenzhen 518172, China, the College of Computer Science and Software Engineering, Shenzhen University, Shenzhen 528060, China, and the Department of Electrical and Computer Engineering, The University of British Columbia, Vancouver, BC V6T 1Z4, Canada (e-mail: vleung@ece.ubc.ca).
}
}

\maketitle

\markboth{IEEE COMMUNICATIONS SURVEYS \& TUTORIALS}%
{Wu \MakeLowercase{\textit{et al}}: A Survey on Cloud-Edge-Terminal Collaborative Intelligence in AIoT Networks}

\begin{abstract}
The proliferation of Internet of things (IoT) devices and AI-driven services have established cloud-edge-terminal collaborative intelligence (CETCI) as a critical paradigm in the artificial intelligence of things (AIoT). Existing surveys often address this domain from a fragmented perspective, focusing on two-tier architectures or specific technologies, while failing to holistically analyze the synergistic interplay across the complete three-tier continuum. Many reviews lack deep analysis of intelligent mechanisms enabling true end-to-end collaboration, particularly with emerging technologies like large language models (LLMs) and generative AI. To bridge this gap, we provide comprehensive analysis of CETCI with core focus on intelligent resource management and full-lifecycle collaborative learning. We uniquely dissect the dual nature of collaboration: AI as enabler for infrastructure optimization and distributed infrastructure as platform for collaborative AI model evolution. We systematically analyze architectural components spanning cloud, edge, and terminal layers, examining core technologies including network virtualization, container orchestration, and software-defined networking, while presenting multi-perspective categorizations investigating task offloading, resource allocation, and optimization techniques across heterogeneous infrastructures. Furthermore, we review advances in federated learning, distributed deep learning, edge-cloud model evolution, and reinforcement learning-based approaches. Finally, we discuss challenges (e.g., scalability, heterogeneity, interoperability) and future trends (e.g., 6G+, agents, quantum computing, digital twin), examining how distributed computing and communication technologies integration can address research challenges and promote opportunities, serving as guide for researchers developing robust, efficient, and secure collaborative AIoT systems.
\end{abstract}

\section{Introduction}\label{sec1}
\subsection{Background}\label{sec1.1}
\IEEEPARstart{T}{he} evolution of computing paradigms has witnessed a significant shift from traditional cloud computing to more distributed architectures in response to the explosive growth of Internet of things (IoT) devices and the increasing complexity of communication network infrastructures~\cite{li2025edgeenhanced, wang2024endedgecloud}. Traditional cloud computing, characterized by centralized data processing and storage, faces substantial challenges with the increasing number of devices generating massive amounts of heterogeneous data at the network edge, creating unprecedented demands on communication bandwidth and network resource management \cite{zhou2024enhancing}. To address these limitations, edge computing has emerged as a novel approach that brings computational resources closer to data sources, delivering lower service latency and higher quality of service (QoS) to end devices while optimizing network traffic patterns and reducing communication overhead \cite{raj2021reliable}. Furthermore, the integration of artificial intelligence with IoT has accelerated architectural evolution, giving rise to artificial intelligence of things (AIoT), a paradigm that combines AI technologies with IoT infrastructure to enable intelligent operations and decision making \cite{zhu2023pushing}. Consequently, AIoT empowers IoT systems to perceive complex environments, process collected data intelligently, and make smart decisions in a timely manner, making the overall system faster, smarter, greener, and safer \cite{zhang2021empowering}.

It is important to clarify that AIoT is not merely a concatenation of AI and IoT technologies but represents a deep convergence where AI empowers IoT with intelligent decision-making capabilities (AI for IoT), while IoT provides the massive data infrastructure necessary to train and evolve AI models (IoT for AI) \cite{zhang2021empowering}. Through such bidirectional synergy, traditional connected devices are transformed into intelligent agents capable of autonomous perception, learning, and action.
The recognition that neither cloud computing nor edge computing alone can effectively address all the requirements of modern intelligent systems has led to the emergence of cloud-edge-terminal collaboration architectures. Meanwhile, hybrid paradigms, including edge-cloud computing and fog computing, leverage the complementary strengths of each layer in the computational hierarchy \cite{vo2022edge}. Specifically, the cloud layer provides powerful computing capabilities and global insights, whereas the edge layer offers reduced latency and localized processing, and the terminal layer (IoT devices) handles data collection and preliminary analysis \cite{yang2024differentially}. As a result, collaborative architectures have demonstrated superior performance in various domains, including industrial applications where the industrial Internet of things (IIoT) is revolutionizing traditional processes \cite{chalapathi2021industrial}. Furthermore, the collaboration facilitates enhanced security and privacy protection by enabling processing and analyzing data at the edge while sharing only privacy-preserving model parameters rather than raw data \cite{kim2024privacy}. As computing capabilities of IoT hardware continue to advance, the vision of an edge-cloud continuum is emerging, where software components can seamlessly move between different levels of the computational hierarchy to optimize system performance \cite{khalyeyev2023characterization}.

\begin{figure}[t!]
    \centering
  \includegraphics[width=0.48\textwidth]{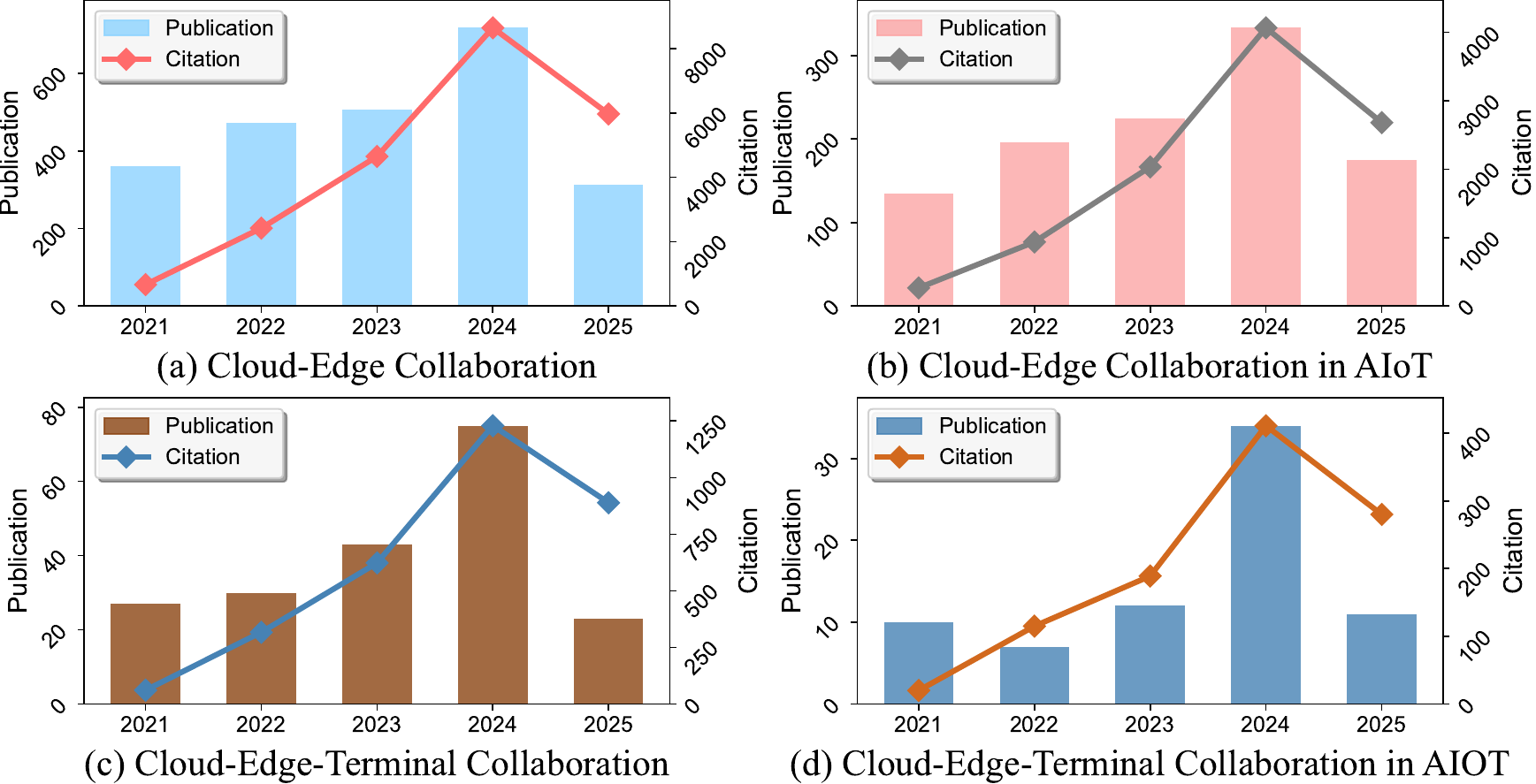}
\caption{%
\protect Research trends in computing paradigms and AIoT from 2021 to 2025, measured by publications and citations.}

    \label{fig:1}
    \vspace{-10px}
  \end{figure}

\subsection{Motivation}\label{sec1.2}

Despite the significant potential of cloud-edge-terminal collaborative intelligence (CETCI) for revolutionizing AIoT systems, several fundamental challenges must be addressed to realize seamless collaborative operation across distributed environments. CETCI represents a specific paradigm of collaborative intelligence systems for AIoT (CISAIoT), which encompasses the orchestrated integration of cloud computing capabilities with intelligent edge and terminal devices to enable adaptive, context-aware AIoT services. As evidenced in \cref{fig:1}, which tracks the growth in publications and citations from 2021 to 2025, with a marked surge in cloud-edge-terminal collaboration research and its AIoT network applications (statistics collected as of 2025/08/21). Terminal devices, including IoT sensors and actuators, face inherent computational limitations that restrict their capacity for complex AI task processing \cite{yang2024differentially}. Additionally, the resource-constrained nature of these devices, characterized by limited processing power, memory, and energy, creates significant barriers to deploying sophisticated machine learning models locally \cite{chi2024trusted}. Consequently, efficient task offloading and resource allocation strategies become necessary. 
The significance of the growth shown in \cref{fig:1} therefore lies not only in publication volume but in the convergence of several pressures that increasingly require a three-tier design. AIoT deployments now produce far more multimodal data than terminals can process locally, yet sending everything to the cloud violates latency budgets, strains bandwidth, and increases privacy exposure. At the same time, modern AI workloads such as foundation-model inference and continual adaptation require distributed memory, energy-aware execution, and cross-site coordination that neither pure edge nor pure cloud architectures handle well in isolation. CETCI becomes the natural response because it aligns where data are generated, where time-critical decisions must be made, and where global model training or orchestration remains most efficient.

Moreover, network bandwidth limitations present a critical bottleneck, as the substantial data volume generated by AIoT devices can overwhelm network infrastructure, especially in real-time data streaming scenarios \cite{raj2021reliable}. Therefore, optimizing communication protocols and data transmission is crucial for effective collaboration. Furthermore, the real-time processing demands of many AIoT applications, such as autonomous vehicle navigation and real-time health monitoring \cite{ahmed2023power}, necessitate distributed intelligence at the edge to mitigate the prohibitive latency of cloud processing. In addition, scalability challenges in communication networks arise as AIoT deployments expand to support millions of interconnected devices, requiring adaptive network architectures that can dynamically scale communication resources, optimize network topology, and maintain consistent performance across varying loads and network conditions~\cite{hossain2024quantumedge}. Privacy and security concerns are also paramount in AIoT systems \cite{alwarafy2021survey}, given the sensitive nature of collected data, which includes personal health information or manufacturing process data \cite{zhou2024enhancing}. Robust security mechanisms are thus essential to prevent unauthorized access and data breaches, while simultaneously preserving user privacy during collaborative learning and data sharing.

The rapid migration of large language models (LLMs) and generative AI tasks to the network edge introduces novel challenges that existing architectures struggle to address, particularly as diffusion-based anomaly detection surveys show the expanding role of generative models in operational monitoring and security analytics \cite{liu2025anomaly, liu2025survey}. Traditional resource management paradigms are often ill-equipped to handle the massive computational and memory requirements of these modern AI workloads. Consequently, there is an urgent need for a comprehensive survey that not only reviews established cloud-edge collaboration but also specifically addresses the integration of these advanced AI paradigms within the constraints of next-generation AIoT networks.
Unlike traditional IoT paradigms that primarily focus on connectivity and data collection, this survey defines ``Collaboration'' as a sophisticated synergy of \textit{Resource Collaboration} (optimizing computation, storage, and networking as discussed in \cref{sec3}) and \textit{Intelligence Collaboration} (co-evolving AI models and knowledge across the cloud-edge-terminal continuum as detailed in \cref{sec4}). Accordingly, the distinction is crucial for understanding how CETCI transcends simple data transmission to enable distributed, autonomous intelligence.

\subsection{Related Surveys}\label{sec1.3}
\input{t1_survey_1.tex}
As illustrated in \cref{tab:1}, existing surveys have explored various facets of cloud computing, edge computing, and AIoT \cite{hong2019resource}, each contributing valuable insights within their specific research domains. Surveys on edge computing have established strong foundations in resource management and application deployment \cite{duc2019machine}, while cloud computing surveys have provided comprehensive coverage of centralized resource provisioning and service management \cite{souza2024maintenance}. Similarly, existing AIoT surveys have made significant contributions by addressing AI integration into IoT systems, encompassing data acquisition, model training, and application development \cite{ahmad2023deep}. For example, Tuli et al. \cite{tuli2023ai} investigate AI-augmented edge and fog computing with particular emphasis on resource management challenges, and Zhang and Tao \cite{zhang2021empowering} provide a comprehensive AIoT overview, discussing its progress and challenges. Recent works like Duan et al. \cite{duan2023distributed} have advanced the understanding of distributed AI systems, with their primary focus on distributed computing aspects. Moreover, application-specific surveys such as Kumar and Agrawal \cite{kumar2023analysis} for industrial IoT and Qu et al. \cite{qu2024mobile} for mobile edge intelligence with LLMs offer valuable specialized domain insights within their respective application areas. Additionally, emerging surveys on generative AI in edge-cloud systems highlight the growing interest in advanced AI deployment across distributed architectures \cite{xu2024unleashing}, with their emphasis on generative AI technologies.

To better ground this recent literature, CETCI is connected to foundational studies from 2018--2021 that shaped the field's historical trajectory. Early surveys established the end-edge-cloud orchestration viewpoint \cite{ren2019survey}, cataloged the core edge-computing system stack \cite{liu2019survey}, and framed AIoT management as an adaptive, scalable learning problem rather than a static networking problem \cite{chen2019learning}. In parallel, early work on edge offloading and collaborative intelligence clarified the latency-energy trade-offs that still govern present CETCI designs \cite{wang2019edge, lei2020deep}. Newer 2022--2025 works extend this earlier foundation, particularly in how they expand from two-tier edge-cloud optimization toward three-tier cloud-edge-terminal collaboration with stronger privacy, autonomy, and model-lifecycle requirements.

However, a critical gap remains in the existing literature. As summarized in \cref{tab:1}, many recent surveys on edge ai or cloud-edge computing \cite{wang2024endedgecloud, liu2025edgecloud} primarily focus on two-tier interactions and often overlook the terminal layer's emerging role as an active participant in computation and intelligence. Consequently, such works tend to treat AI models as workloads to be optimized rather than as enablers of system-wide collaboration. Furthermore, most reviews address either resource management \cite{yang2025edge} or model training \cite{qu2025mobile} in isolation. Our survey distinguishes itself by providing a holistic analysis of the three-tier CETCI architecture, focusing on cross-layer resource orchestration and the full lifecycle of collaborative intelligence, thereby filling a crucial gap by examining how terminal-layer context awareness can inform and optimize edge-cloud operations.
This distinction becomes clearer when compared with the recent Agentic AI tutorial by Jiang \textit{et al.} \cite{jiang2026agentic}, which is included in \cref{tab:1}. That work is valuable for explaining how large AI models evolve toward planner-tool-memory agent loops in future communication systems, but its primary lens is agent architecture for intelligent communications rather than AIoT-wide cloud-edge-terminal collaboration. Our survey therefore complements it by highlighting heterogeneous terminal participation, cross-layer resource management, and how collaborative learning, security, and application constraints interact when AI agents must coexist with sensing, control, and privacy-critical edge operations.

Beyond conventional edge computing surveys, recent advancements in mobile edge intelligence specifically for LLMs have been comprehensively documented by Qu et al. \cite{qu2025mobile}, who provide a contemporary analysis of edge LLM caching, training, and inference while highlighting cost-effectiveness and privacy preservation through on-device deployment. Additionally, Van Huynh et al. \cite{vanhuynh2022edge} present an innovative integrated communication, computing, and storage model for digital twin-enabled metaverse applications by leveraging mobile edge computing (MEC) and ultra-reliable low-latency communications, thereby demonstrating the practical implementation of edge intelligence for demanding real-time interactive services. Furthermore, Wang et al. \cite{wang2024endedgecloud} contribute significantly by analyzing collaborative elements within end-edge-cloud systems specifically for deep learning, systematically investigating key enabling technologies including model compression, partition, and knowledge transfer across the computational continuum. 

Meanwhile, Yang et al. \cite{yang2025edge} offer an advanced exploration of reinforcement learning methodologies for mobile edge computing networks, detailing how RL strategies enhance offloading, caching, and communication within MEC networks while addressing critical issues related to resource optimization, security, and privacy. More recently, Liu et al. \cite{liu2025edgecloud} provide a comprehensive examination of edge-cloud collaborative computing with a focus on distributed intelligence and model optimization, systematically analyzing model compression, adaptation, and neural architecture search alongside AI-driven resource management strategies that balance performance, energy efficiency, and latency requirements. While some research has explored edge-cloud collaboration \cite{ren2019survey}, these works have primarily concentrated on specialized applications such as federated learning for QoS enhancement \cite{zhou2024enhancing} or blockchain for secure decision-making \cite{chi2024trusted}, each making valuable contributions within their specific domains. Consequently, as evident from \cref{tab:1}, our survey complements existing literature by providing comprehensive analysis of CETCI in AIoT networks, systematically addressing key dimensions including task offloading, resource allocation, model compression, knowledge transfer, security and privacy concerns, and diverse application domains, thereby extending the current understanding of collaborative intelligence systems.

\subsection{Survey Scope and Contributions}\label{sec1.4}
\begin{figure*}[t!]
  \centering
\includegraphics[width=0.95\textwidth]{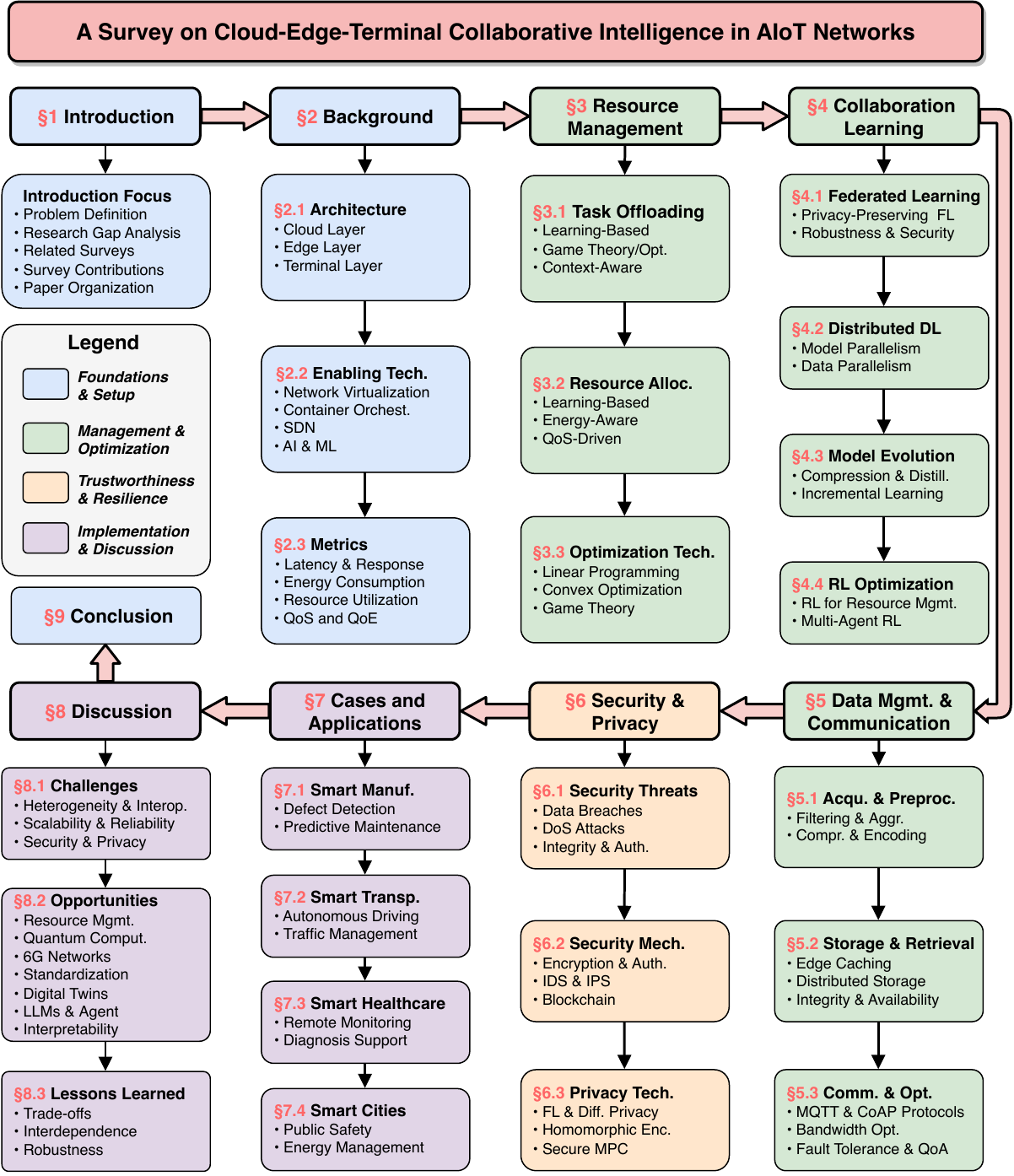}
\caption{Illustrative structure of cloud-edge-terminal collaborative intelligence (CETCI) in AIoT networks, highlighting the end-to-end survey scope that connects architectural layers, enabling technologies, intelligent management, and collaborative learning within a single cross-layer framework.}

  \label{fig:2}
\vspace{-10px}
\end{figure*}

\begin{figure*}[t!]
  \centering
\includegraphics[width=0.95\textwidth]{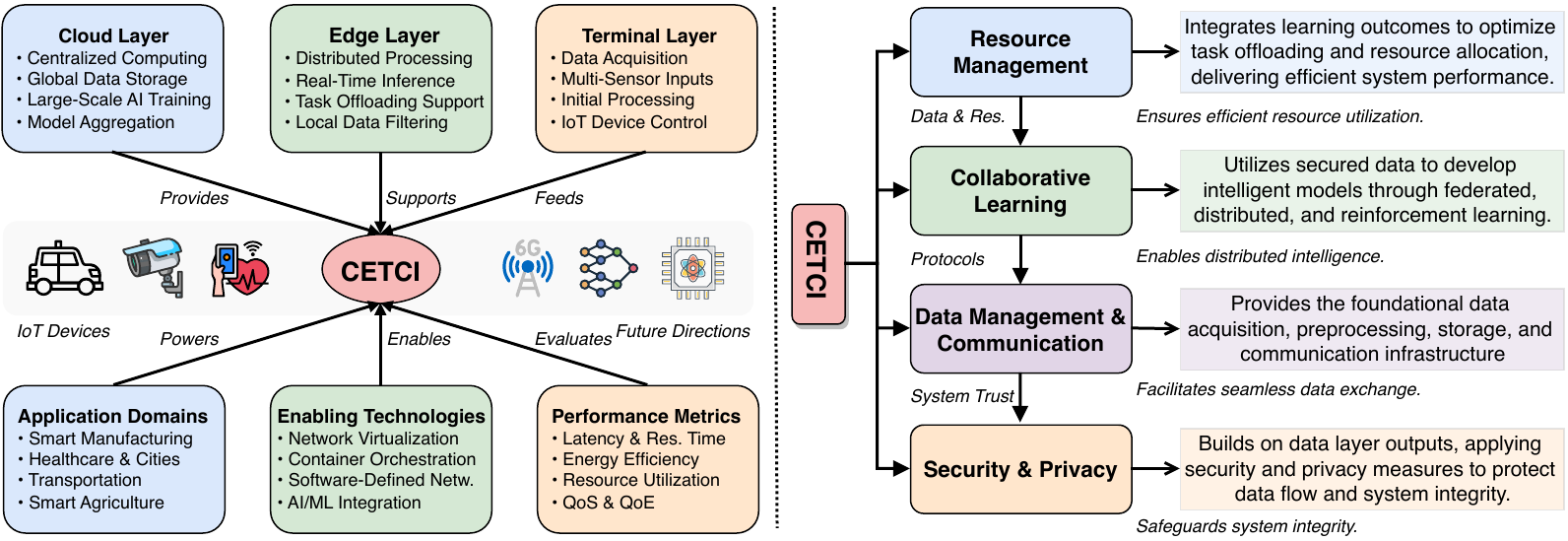}
\caption{Topology diagram of CETCI research scope and supporting components, showing how deployability is shaped by the coupling among resource management, collaborative learning, security and privacy protection, and data management.}
  \label{fig:3}
\vspace{-10px}
\end{figure*}

In this survey, we extend the conventional scope of cloud-edge-terminal computing from isolated layer analysis to practical collaborative intelligence for real-world AIoT applications, which we term CETCI. Specifically, we define ``Collaborative Intelligence'' as a dual-faceted paradigm: (1) \textit{Infrastructure-level Collaboration (AI for Collaboration)}, where AI algorithms coordinate distributed components for efficient resource management and task offloading, as discussed in \cref{sec3}; and (2) \textit{Model-level Collaboration (Collaboration for AI)}, where AI models themselves become the collaborative workload through distributed training and inference mechanisms such as federated learning, as detailed in \cref{sec4}. Our main contribution is to explain when these two dimensions should be co-designed rather than optimized independently.
We introduce and define key terminologies such as ``Intelligent Management'' and ``Collaborative Learning'' not merely as buzzwords, but as essential architectural components that bridge the gap between physical infrastructure and application requirements. Moreover, we consistently apply these terms throughout the survey to categorize technologies that enable autonomous decision-making and distributed knowledge evolution, respectively.

According to research objectives and involved domains, CETCI is delineated into the hierarchical framework shown in \cref{fig:2}, encompassing: 1) \textit{Architectural Paradigms}, consisting of cloud, edge, and terminal layers with their respective computing capabilities, communication interfaces, and resource characteristics, responsible for distributed task execution, data processing, and intelligent decision-making across the computing continuum; 2) \textit{Enabling Technologies}, targeting network virtualization, container orchestration, software-defined networking, and AI/ML integration platforms for large-scale AIoT deployments, linking the physical infrastructure and intelligent algorithms, supporting configurable deployment of collaborative tasks across heterogeneous environments; 3) \textit{Intelligent Management}, focusing on resource optimization strategies driven by mathematical optimization, machine learning, and distributed algorithms, especially encompassing task offloading, resource allocation, and performance optimization; and 4) \textit{Collaborative Learning}, encompassing federated learning, distributed deep learning, edge-cloud model evolution, and reinforcement learning paradigms that enable privacy-preserving knowledge sharing and distributed intelligence across the computing hierarchy. Compared with cloud-edge computing or edge-cloud continuum studies that mostly emphasize infrastructure federation, and compared with AIoT or generic collaborative intelligence studies that often describe application ecosystems without linking them to a joint optimization logic, CETCI explicitly treats cross-layer resource orchestration and distributed model evolution as two coupled design axes that must be analyzed together in deployment-grade AIoT systems~\cite{qu2024mobile, du2024enhancing, asghari2024server}.

Furthermore, as illustrated in \cref{fig:3}, our survey adopts a topology-driven perspective that organizes CETCI into a route map rather than a second paper outline. The core architecture shows how cloud, edge, and terminal layers interact with application domains, enabling technologies, and performance metrics, while the supporting components highlight where resource management, collaborative learning, security and privacy protection, and data management directly shape deployability. In turn, the perspective helps identify which subsystem should be strengthened first when latency, privacy, or scalability becomes the dominant bottleneck in a target AIoT deployment~\cite{rahman2023blockchainbased, nezami2021decentralized}. The resulting boundary is therefore stronger than a naming exercise: CETCI is needed when an AIoT service cannot be judged solely by where computation runs, because deployment feasibility also depends on how model synchronization, communication budgets, privacy controls, and closed-loop actuation interact across the three layers.

Thanks to the development of distributed AI and edge computing technologies \cite{duan2023distributed, ren2019survey, maia2024survey}, CISAIoT networks have made significant progress in recent years, evolving from traditional centralized cloud computing to sophisticated collaborative paradigms that leverage the unique advantages of each computational layer. Nevertheless, advancements have liberated AIoT applications from the constraints of single-layer processing limitations and addressed critical challenges in latency, bandwidth, privacy, and scalability~\cite{vo2022edge}. Compared to conventional isolated approaches~\cite{becker2021edgepier, cardoso2023softwaredefined, zhang2024stackelberggamebased}, collaborative architectures can optimize resource utilization, enhance system resilience, and enable real-time intelligent processing across diverse AIoT scenarios. Therefore, researchers are currently focusing on the design of collaborative frameworks, optimization strategies, and intelligent coordination mechanisms that can seamlessly integrate multi-layer capabilities while addressing the inherent challenges of distributed heterogeneous environments \cite{hong2019resource}. The contributions of this survey are:

\begin{itemize}
  \item We provide a comprehensive survey that systematically reviews cloud-edge-terminal collaborative intelligence systems for AIoT networks (CISAIoT). It covers the full spectrum of recent research across all three architectural layers and provides a holistic view of CISAIoT development without restricting to specific technologies.
  \item We present a unified framework to systematically categorize existing works from architectural, technological, and application-oriented perspectives. Our survey encompasses architectural paradigms, enabling technologies, intelligent management, collaborative learning, and security mechanisms, and it evaluates them through shared dimensions including assumptions, computational and communication complexity, convergence behavior, scalability, energy-latency trade-offs, and deployment maturity.
  \item We comprehensively review collaborative learning frameworks, specifically dissecting the interplay between federated learning, split learning, and knowledge distillation across the cloud-edge-terminal continuum. We analyze how these mechanisms enable privacy-preserving intelligence and efficient model evolution in resource-constrained environments, and we clarify which mechanisms are better suited for bandwidth-limited, privacy-sensitive, or compute-imbalanced deployments.
  \item We identify critical open challenges and outline promising future research directions, including emerging trends in 6G integration, quantum computing, digital twins, and large language models, offering valuable insights and guidance for advancing the state-of-the-art in collaborative AIoT systems.
\end{itemize}

\subsection{Paper Organization}\label{sec1.5}

The remainder of this survey is organized as follows. \cref{sec2} introduces the CETCI architecture and enabling technologies, \cref{sec3} reviews intelligent resource management, and \cref{sec4} examines collaborative learning paradigms. \cref{sec5} and \cref{sec6} then cover security, privacy, data management, and communication issues, \cref{sec7} summarizes representative application domains, and \cref{sec8}--\cref{sec9} conclude with open challenges, future directions, and the overall takeaways of the survey.

\input{t1_acronyms.tex}

\section{Architecture and Key Concepts}\label{sec2}

The cloud-edge-terminal architecture serves as the physical substrate that supports the dual facets of collaborative intelligence. It not only provides the heterogeneous computing and networking resources required for \textit{Infrastructure-level Collaboration} but also facilitates the distributed data flow and model synchronization essential for \textit{Model-level Collaboration}. Accordingly, the following sections briefly review key architectural components and enabling technologies, focusing specifically on their roles in supporting intelligent resource management and collaborative learning, rather than providing generic definitions. A conventional architecture survey could stop at layer descriptions, whereas CETCI requires analyzing how architectural choices propagate into algorithm selection, synchronization cost, privacy exposure, and service robustness across the same end-to-end workflow.

\subsection{Cloud-Edge-Terminal Architecture}\label{sec2.1}

\subsubsection{Cloud Layer}\label{sec2.1.1}
Centralized computing infrastructure defines the cloud layer's primary function, offering scalable resources that form the core of distributed CETCI systems.  Consequently, this massive computational capacity supports complex model training, analytics, and data warehousing for interconnected AIoT applications \cite{levin2019aiops}.  Typically managed by third-party providers, cloud platforms offer on-demand access to computing power, memory, and storage, thus enabling AIoT applications to scale dynamically \cite{yosuf2021cloud}.  Furthermore, the centralized architecture facilitates efficient resource management and allocation across diverse tasks and applications.  The cloud's capacity for large-scale data processing is essential for AIoT applications handling massive datasets generated by numerous interconnected devices.  In addition, the cloud readily provides the substantial processing power and memory often required for complex computations, such as training sophisticated machine learning models.  Furthermore, its role extends to long-term data storage, archiving historical data from AIoT devices for future analysis, model training, and system optimization \cite{blythman2022libraries}.  Consequently, the centralized repository ensures data integrity, availability, and accessibility across the AIoT system, facilitating data-driven insights and predictive models \cite{zhou2024enhancing}.

\subsubsection{Edge Layer}\label{sec2.1.2}

Intermediate processing nodes strategically positioned between cloud and terminals characterize the edge layer's architecture.  Rather than relying solely on distant cloud resources, the distributed nodes enable localized computation, significantly reducing data transmission requirements and improving responsiveness \cite{liang2020ai}.  The edge layer's proximity to data sources enables computations closer to where data originates, thereby reducing the volume of data transmitted to the cloud \cite{baucas2020using}. Additionally, the localization minimizes latency and enhances real-time processing capabilities.  Formally, if $L_c$ and $L_e$ denote the latency for processing data in the cloud and at the edge, respectively, and $T_t$ and $t_t$ represent the data transmission times between the terminal device and the cloud and edge, respectively, then the total latencies are $L_{total\_c} = L_c + T_t$ and $L_{total\_e} = L_e + t_t$. Because edge nodes are closer to terminal devices, $t_t$ is significantly smaller than $T_t$.  In many applications, specialized edge hardware allows $L_e$ to be comparable to, or even smaller than, $L_c$ \cite{kennedy2021avec}, thus generally making $L_{total\_e}$ smaller than $L_{total\_c}$ and improving real-time performance.  Additionally, by offloading computationally intensive tasks from both terminal devices and the cloud, the edge layer balances the network workload \cite{ren2022caching}, optimizing energy consumption and resource utilization, particularly crucial for applications with stringent real-time requirements, such as autonomous driving and industrial automation \cite{becker2020aiops, zhu2023pushing}.  Pre-processing and filtering data at the edge further reduces the transmitted data volume, conserving network bandwidth and reducing communication costs, which is essential for large-scale AIoT deployments \cite{raj2021reliable, zhou2024enhancing}.

\subsubsection{Terminal layer}\label{sec2.1.3}
Physical world interaction defines the terminal layer's essential role through diverse IoT devices and sensors.  Beyond simple data collection, the devices transform environmental phenomena into actionable digital information while enabling bidirectional communication with their surroundings \cite{chang2023exploration}.  As a result, the data acquisition transforms physical phenomena into digital signals, encompassing various modalities like temperature, pressure, humidity, location, images, and audio, depending on the application \cite{qiu2024improving}.  Subsequent to acquisition, initial data processing within the terminal layer often involves tasks such as data cleaning, filtering, aggregation, and feature extraction to reduce data volume, enhance data quality, and prepare relevant information for later stages \cite{mihai2018wsn}.  For example, a smart thermostat might collect temperature readings and calculate an average over a specific period, thereby reducing communication overhead and computational burden on the edge and cloud layers \cite{nawandar2021standalone}. Beyond data collection, the terminal layer enables bidirectional interaction with the physical environment through IoT-controlled actuators that can affect the physical world based on received instructions or processed data, enabling real-time control and automation \cite{lei2020deep}. For instance, smart irrigation system automatically adjusts the irrigation schedule based on collected soil moisture data, creating a feedback loop for dynamic adaptation and optimization \cite{raj2021reliable}.

\subsection{Enabling Technologies}\label{sec2.2}

Effective collaboration across distributed architectures relies on fundamental technologies that enable dynamic resource management and intelligent orchestration. Network virtualization and software-defined networking (SDN) provide the necessary abstraction and programmability to manage heterogeneous infrastructure. Network virtualization creates logical networks independent of physical hardware, enabling dynamic resource provisioning and network slicing tailored to specific AIoT application needs \cite{maia2024survey, khalafi2024network}. Network slicing capability is crucial for isolating critical traffic and ensuring QoS in multi-tenant edge environments. SDN complements this by separating the control plane from the data plane, allowing for centralized, programmable traffic management \cite{nunez2023brief}. In CETCI systems, SDN controllers dynamically optimize data flows between cloud, edge, and terminal layers, facilitating efficient task offloading and load balancing based on real-time network conditions \cite{alnajjar2020network}.

Container orchestration platforms, such as Kubernetes and its lightweight variants (e.g., K3s), automate the deployment, scaling, and management of containerized applications across the distributed continuum \cite{bogo2020componentaware}. Kubernetes and similar platforms are essential for managing the lifecycle of AI models and microservices, ensuring consistent behavior across diverse hardware \cite{wang2021integration}. By leveraging declarative management models, container orchestration systems optimize cluster scheduling and resource allocation, dynamically placing containers on the most suitable nodes based on resource availability and latency constraints \cite{rodriguez2020container}. Automated orchestration is vital for maintaining system responsiveness and scalability in large-scale AIoT deployments. For a more detailed technical overview of these enabling technologies, please refer to \cref{sec:supp_tech.1} of the \textit{supplementary material}.

\subsubsection{Artificial Intelligence and Machine Learning}\label{sec2.2.4}

Intelligent decision-making capabilities emerge from AI/ML algorithms deployed across computational layers.  Adaptive learning systems enable autonomous operations while supporting predictive analytics and real-time optimization across CISAIoT frameworks \cite{zhang2021empowering}.  A fundamental concept in ML is the model, represented as a function $f_\theta(\mathbf{x})$, where $\mathbf{x}$ denotes the input data and $\theta$ represents the model's parameters. Learning aims to find optimal parameters $\theta^*$ that minimize a loss function $L(\theta)$, measuring the discrepancy between model predictions and ground truth \cite{toussaint2020machine}.  Specifically, $\theta^* = \arg\min_{\theta} L(\theta) = \arg\min_{\theta} \frac{1}{N} \sum_{i=1}^{N} l(f_\theta(\mathbf{x}_i), y_i)$, where $N$ is the number of training samples, $\mathbf{x}_i$ is the $i$-th input sample, $y_i$ is the corresponding ground truth label, and $l(\cdot, \cdot)$ is a specific loss function. Importantly, the learning process allows systems to adapt and improve their performance over time.  Depending on application requirements and resource constraints, AI/ML models can be deployed at different layers within the cloud-edge-terminal collaborative framework \cite{raj2021reliable}. For example, complex deep learning models, given their computational demands, might be trained in the cloud, while simpler models can be deployed at the edge or terminal layers for real-time inference.

In the context of CETCI, we explicitly distinguish between two forms of collaboration: \textit{Model collaboration}, where AI models themselves are the workload (e.g., split learning, federated learning), and \textit{Component collaboration via AI}, where AI serves as the optimizer for resource management (e.g., using multi-agent reinforcement learning (MARL) for task offloading). Building upon this distinction, our survey primarily focuses on the latter, investigating how AI/ML algorithms empower cloud, edge, and terminal components to collaboratively optimize resource scheduling and system performance \cite{liu2026collaborative, liu2026enhancing, liu2026projecting}. Within this framing, emerging LLM and agentic workflows should not be treated as isolated future topics: they expand model collaboration through split inference and parameter-efficient adaptation, while also expanding component collaboration through tool-use planning, memory synchronization, and policy-level fallback control across heterogeneous layers \cite{xu2024when, jiang2026agentic, shen2024agilequant}.
Collaborative learning paradigms, such as federated learning, enable distributed model training across multiple devices without sharing raw data, thus preserving privacy and reducing communication overhead \cite{yang2024differentially}.  Reinforcement learning (RL) provides a framework for optimizing decision-making in AIoT systems, where agents learn to interact with an environment to maximize cumulative reward \cite{lei2020deep}.  Specifically, the goal is to find the optimal policy $\pi^* = \arg\max_{\pi} \mathbb{E} \left[ \sum_{t=0}^{\infty} \gamma^t r_t | \pi \right]$, where $r_t$ is the reward at time $t$ and $\gamma$ is a discount factor.
Section II now makes the hardware substrate more explicit as well: practical CETCI deployments increasingly rely on heterogeneous edge accelerators, including NPUs, TPUs, embedded GPUs, and domain-specific inference engines, to execute AI pipelines under strict latency and energy budgets \cite{jiang2020mnn, kennedy2021avec, hao2021enabling}. As a result, the accelerators improve local inference throughput and make partial on-device training or adaptation feasible, but they also introduce new placement constraints because memory capacity, operator support, and accelerator virtualization differ substantially across devices. In privacy-sensitive AIoT settings, stronger local acceleration can reduce raw-data uplink requirements by allowing feature extraction, filtering, or lightweight model adaptation to remain near the data source, whereas cloud coordination is still needed for global aggregation, large-model training, or cross-site orchestration.
\input{t_metrics.tex}

\subsection{Performance Metrics}\label{sec2.3}
\begin{figure}[t!]
    \centering
  \includegraphics[width=0.48\textwidth]{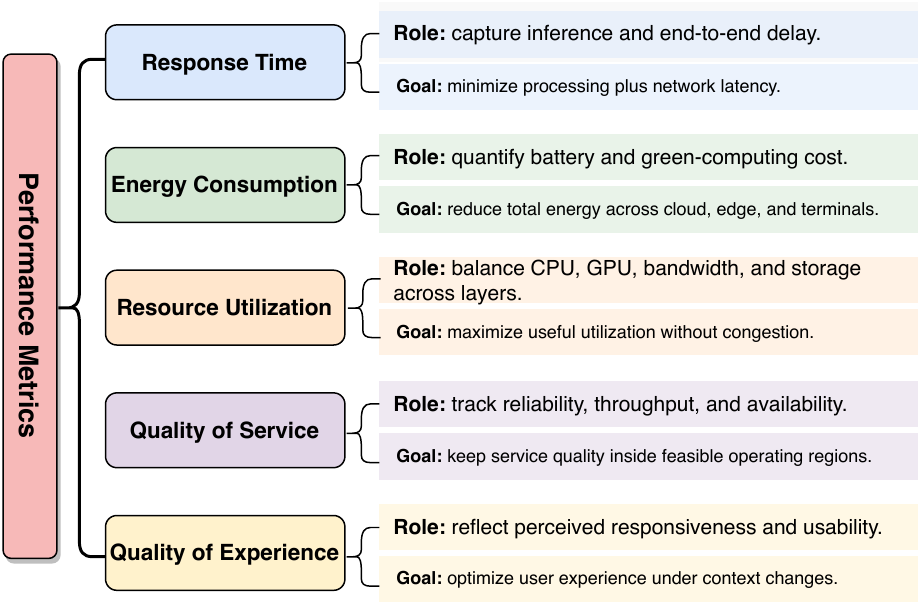}
\caption{%
\protect Overview of performance metrics for evaluation.}
  \label{fig:performance_metrics}
  \vspace{-10px}
\end{figure}
Response time and energy consumption are paramount for delay-sensitive and battery-powered AIoT devices. CETCI optimizes these by strategically offloading tasks to edge or cloud layers, balancing the trade-off between transmission latency and processing speed \cite{raj2021reliable}. Resource utilization and QoS ensure efficient infrastructure usage and reliable service delivery, respectively. Intelligent resource allocation algorithms maximize utilization while adhering to QoS constraints defined in service level agreements (SLAs) \cite{zhou2024enhancing}. Finally, quality of experience captures the user's subjective satisfaction, which is enhanced by context-aware adaptation and low-latency processing at the edge \cite{zhang2021empowering}.
Comprehensive performance evaluation of cloud-edge-terminal collaborative intelligence requires systematic assessment across multiple dimensions to ensure effective system operation in AIoT environments. As illustrated in \cref{fig:performance_metrics}, we present a performance metrics evaluation framework for CETCI systems. Rather than repeating generic definitions, we summarize the metrics that most directly affect AIoT deployment decisions, such as inference latency, training energy efficiency, and QoS stability~\cite{zhou2024enhancing, zhang2021empowering}. Throughout the remainder of this survey, the main comparative lens therefore tracks six recurring dimensions: problem assumptions, computational and communication complexity, convergence behavior, scalability, energy-latency trade-off, and deployment maturity. For detailed mathematical formulations and general definitions, see \cref{sec:supp_tech.2} of the \mbox{\textit{supplementary material}}.

\section{Intelligent Resource Management Paradigms}\label{sec3}

\input{fig3_sec3.tex}
Intelligent resource management in CETCI moves beyond optimizing isolated layers (e.g., edge-only or cloud-only) to adopting a holistic, end-to-end perspective. By considering the entire data lifecycle as a closed-loop process, from data generation at the terminal, through multi-stage processing and offloading across the edge and cloud, to the final delivery of intelligent feedback to the terminal, effective management requires synergistic coordination of task offloading, resource allocation, and optimization techniques across all three layers. The core challenge lies in jointly optimizing network, computation, and storage resources along the full data path to meet end-to-end performance constraints, such as latency and energy efficiency, in highly dynamic and heterogeneous AIoT environments.
We focus on \textit{Infrastructure-level Collaboration (AI for Collaboration)}, exploring how AI-driven strategies enable intelligent coordination among distributed components. Unlike traditional resource management, which often relies on static rules, CETCI leverages learning and optimization to dynamically schedule tasks and allocate resources, addressing the inherent trade-offs between local (terminal/edge) and global (cloud) objectives while ensuring collaborative decision-making for system-wide efficiency. The discussion in this section emphasizes when a method should be preferred, not merely which algorithm family exists, by comparing the same families through assumptions, complexity, convergence, scalability, communication burden, and deployment readiness.

\subsection{Task Offloading Strategies}\label{sec3.1}
Strategic allocation of computational tasks across distributed resources forms the cornerstone of effective resource management in collaborative intelligence systems. As summarized in \cref{tab:2}, recent task offloading strategies leverage a diverse set of techniques, from game theory to deep reinforcement learning, each presenting distinct trade-offs. For instance, while learning-based approaches offer high adaptability to dynamic environments~\cite{binh2024reinforcement, sharma2024deep}, they often face challenges related to scalability and training overhead. Conversely, traditional optimization methods provide theoretical guarantees~\cite{duan2024binary, zhang2024stackelberggamebased} but may struggle with the complexity of real-world AIoT systems. The comparative analysis in \cref{tab:2} systematically evaluates these approaches based on their core algorithms, optimization objectives, communication and computational burden, convergence conditions, and inherent limitations.
\input{t2_task_3.1.tex}
\subsubsection{Learning-Based Offloading Strategies}\label{sec3.1.1}
Adaptive decision-making algorithms enable autonomous task allocation through machine learning techniques.  RL frameworks train intelligent agents to optimize offloading choices by continuously learning from environmental feedback and performance metrics \cite{ren2022caching}.  Specifically, RL algorithms, such as deep Q-networks (DQN) \cite{ji2024task}, train agents to make offloading decisions by interacting with the environment and receiving rewards based on metrics like latency, energy consumption, and task completion time \cite{binh2024reinforcement}.
Beyond single-agent approaches, MARL has emerged as a key enabler for collaborative decision-making in decentralized AIoT networks. In these frameworks, each edge node or terminal acts as an independent agent, learning to coordinate with others through shared state representations or cooperative reward structures \cite{she2024efficient}. For instance, by designing a global reward function that reflects system-wide performance (e.g., total latency or energy efficiency), agents are incentivized to make offloading decisions that benefit the collective system rather than just their local objectives. By employing distributed cooperative learning processes, components can dynamically adapt to varying network conditions and workload distributions without relying on a central controller.
Researchers have also explored more sophisticated RL techniques, including deep meta Q-learning \cite{sharma2024deep} for faster adaptation to new scenarios and multi-agent deep deterministic policy gradient (MADDPG) \cite{she2024efficient} for coordinated decision-making in multi-user environments.  In addition, dependency-aware mechanisms have been incorporated into the learning process to handle tasks with dependencies and optimize overall workflow \cite{chen2024dynamica}.

\subsubsection{Game Theory and Optimization-Based Strategies}\label{sec3.1.2}

Strategic interaction modeling provides mathematically rigorous frameworks for offloading optimization.  Game-theoretic approaches enable competitive and cooperative resource allocation while addressing conflicting objectives among multiple participants \cite{zhang2024stackelberggamebased}.  Similarly, Duan \textit{et al.}~\cite{duan2024binary} leverage game theory to optimize offloading strategies for cloud robots in a cloud-edge collaboration system, minimizing both system completion time and energy consumption through a Nash equilibrium solution.  In addition to game-theoretic approaches, optimization techniques, often integrated with these models, employ algorithms such as evolutionary algorithms and swarm intelligence.  Specifically, Bandyopadhyay \textit{et al.}~\cite{bandyopadhyay2024delaysensitive} propose DARC-DE, a Delay-Aware Resource-Constrained Offloading algorithm based on discretized differential evolution, to maximize bandwidth utilization and minimize delay in edge-cloud systems.  Chen \textit{et al.}~\cite{chen2022resource} introduce a hybrid simulated annealing-binary particle swarm optimization (SA-BPSO) algorithm to minimize total user overhead, considering latency, energy, and computing costs.  Likewise, Jiao \textit{et al.}~\cite{jiao2024sraeabco} present E-ABCO, an elite-artificial bee colony offloading method, further demonstrating the application of swarm intelligence for optimizing offloading strategies.

\subsubsection{Context-Aware Offloading Strategies}\label{sec3.1.3}
Situational intelligence drives adaptive offloading through environmental and application-specific awareness.  Contextual factors, including device mobility, task dependencies, and security requirements influence optimal placement decisions \cite{feng2024dependencyaware}.  Additionally, device mobility necessitates dynamic adaptation of offloading locations for efficient task execution \cite{li2024blockchainbased}. Security requirements influence offloading decisions by prioritizing secure environments for sensitive tasks.  Similarly, energy constraints, especially for resource-limited devices, drive strategies that balance computational load and minimize energy consumption across the architecture, such as those in \cite{bai2024delayaware} and \cite{li2024twotimescale}.
Furthermore, a critical aspect of context awareness in CETCI is recognizing the hierarchical structure of the architecture, which enables vertical collaborative offloading. In this paradigm, a task can be dynamically partitioned and processed in multiple hops, such as initial filtering at the terminal, feature extraction at the edge, and final model inference in the cloud. Implementing these multi-hop processing strategies requires sophisticated decision-making to manage end-to-end latency and data dependencies across the terminal-edge-cloud path.

\textit{Summary and Design Implications}: While traditional optimization-based offloading strategies (e.g., game theory) offer theoretical performance guarantees, they often suffer from high computational complexity and slow convergence in highly dynamic AIoT environments. In contrast, learning-based approaches (e.g., DRL) demonstrate superior adaptability~\cite{sharma2024deep, ji2024task} but incur significant training overhead and may lack performance guarantees. For practitioners, the main decision rule is to prefer optimization-driven offloading when workload patterns are stable, state observability is reliable, and SLA compliance dominates, and to prefer learning-driven offloading when mobility, burstiness, or environmental uncertainty changes the decision boundary online. Heuristic or swarm-based schemes remain attractive when millisecond-level reaction is required but exact optimality is secondary, whereas MARL becomes more compelling when coordination across multiple offloading agents matters~\cite{chen2024dynamica} more than single-node optimality. A key design implication for future systems, such as those in 6G networks, is the adoption of hybrid architectures that combine meta-learning with traditional optimization to enable rapid adaptation with limited retraining overhead while retaining a degree of analytical tractability.

\subsection{Resource Allocation Mechanisms}\label{sec3.2}
\input{t3_res_alloc_3.2.tex}

Optimal distribution of computing, networking, and storage resources across heterogeneous environments stands as a fundamental requirement for successful collaborative intelligence. Accordingly, \cref{tab:3} provides a comparative summary of recent resource allocation techniques, highlighting the key algorithms employed, their primary optimization objectives (including latency, energy consumption, and QoS), alongside a qualitative analysis of their respective trade-offs. Notably, a significant trend emerges toward adopting DRL to handle the dynamic nature of AIoT environments; however, learning-based methods often incur higher training complexity compared to heuristic or traditional optimization methods.
Effective resource allocation in CETCI extends beyond isolated optimization within a single layer and necessitates joint optimization across the entire end-to-end path. Specifically, such optimization involves orchestrating a diverse set of resources, including the computational power at terminal, edge, and cloud nodes; the available storage at each layer; and the network bandwidth of the communication links connecting them (i.e., terminal-to-edge and edge-to-cloud). A key challenge is to develop allocation strategies that can dynamically balance these resources to satisfy application-specific QoS requirements while adapting to fluctuating network conditions and workloads.
Learning-based strategies utilize machine learning to dynamically adjust resource allocation based on real-time system states and historical patterns. Meanwhile, energy-aware methods focus on minimizing energy usage while maintaining performance, addressing the constraints of edge nodes and terminal devices. Furthermore, QoS-driven approaches prioritize metrics like latency, reliability, and throughput, ensuring performance guarantees for delay-sensitive applications.

\subsubsection{Learning-Based Resource Allocation}\label{sec3.2.1}
Intelligent resource distribution emerges through machine learning algorithms that adapt to dynamic system conditions.  Adaptive allocation frameworks leverage historical patterns and real-time monitoring to optimize resource utilization across heterogeneous computing environments \cite{fang2023large}.  Combining optimal transport and federated actor-critic learning, Gan et al.~\cite{gan2024optimal} introduced a dual-driven resource optimization mechanism (OTFAC) to optimize bandwidth and computation resource allocation, achieving significant reductions in delay and energy consumption through offline and online learning.  Deep reinforcement learning (DRL) has also become a prominent technique. Specifically, Huang et al.~\cite{huang2024joint} developed a decision-assisted hybrid action space DRL algorithm for joint optimization of offloading and resource allocation in space-air-ground integrated networks, effectively managing resources across a complex hybrid cloud and multi-access edge computing scenario.  Addressing security, Zhang et al.~\cite{zhang2024securityaware} proposed a security-aware DRL-based resource allocation scheme in a cloud-edge-terminal cooperative vehicular network, enhancing security during virtual network embedding with a dynamic trust evaluation mechanism.  In addition, hybrid approaches like the simulated annealing-binary particle SA-BPSO proposed by Chen et al.~\cite{chen2022resource} offer further optimization avenues in cloud-edge collaborative systems, minimizing total user overhead while considering latency, energy, and computing costs.

\subsubsection{Energy-Aware Resource Allocation}\label{sec3.2.2}
Sustainability considerations drive power-efficient resource distribution across energy-constrained devices.  Battery life optimization requires balanced allocation strategies that maintain performance while minimizing power consumption through intelligent workload management \cite{li2024energyaware}.  Similarly, an energy-efficient scheme for low-latency multi-type service provisioning in a local-edge-cloud environment is proposed in \cite{chen2024fast}, leveraging the alopex-based differential evolution (Alopex-DE) algorithm to optimize resource allocation for minimal latency and energy consumption.  In \cite{hua2024energyefficient}, energy efficiency in heterogeneous edge-cloud computing is addressed by jointly optimizing power control, transmission scheduling, and offloading decisions to minimize mobile device energy consumption, specifically considering user mobility.  Addressing a different network architecture, \cite{nguyen2024integrated} tackles energy consumption in a space-air-ground integrated network (SAGIN) through joint optimization of computation offloading, UAV trajectory control, and resource allocation.  Finally, \cite{wang2024cooperative} introduces a digital twin (DT) assisted end-edge-cloud collaborative computing scheme for 6G Industrial IoT, where the DT monitors the physical environment and informs resource allocation strategies to minimize system cost under latency and energy constraints.

\subsubsection{QoS-Driven Resource Allocation}\label{sec3.2.3}
Performance guarantee enforcement requires resource allocation mechanisms that prioritize service quality metrics.  Latency-sensitive applications demand specialized allocation strategies that ensure consistent performance under varying operational conditions \cite{alieldin2021hidden}. Minimizing latency is paramount in these applications, while maintaining reliable communication and sufficient throughput is essential for applications with continuous data streams, such as remote patient monitoring in smart healthcare.  For instance, prioritizing low-latency workloads for edge execution while offloading less time-sensitive tasks to the cloud can significantly improve overall QoS in edge-cloud collaborative systems \cite{zhai2024edgecloud}.  Similarly, adaptive strategies that dynamically adjust resource assignments based on real-time QoS demands, coupled with efficient scheduling algorithms \cite{asghar2022survey}, can further enhance system responsiveness and efficiency \cite{zhou2024enhancing}.

\textit{Critical Analysis and Main Findings}: A critical limitation of many existing resource allocation models is the implicit assumption of perfect and instantaneous channel or system state information, which is often impractical in real-world, dynamic edge networks due to signaling overhead and delays. Consequently, this gap between theory and practice can lead to suboptimal performance. Our main finding is that while joint optimization of communication and computation resources is a clear trend for maximizing end-to-end performance, its real-time implementation is challenging because such problems are often NP-hard~\cite{nguyen2024integrated, bandyopadhyay2024delaysensitive}. Learning-based allocation is therefore preferable when state transitions are hard to model but enough interaction data can be collected, while optimization-based allocation is preferable when constraints are explicit and certifiable feasibility is essential. A key design implication is the need for lightweight, distributed algorithms that can find near-optimal solutions with incomplete or delayed state information, potentially by leveraging predictive models to forecast future resource availability and demand.

\subsection{Optimization Techniques}\label{sec3.3}
Mathematical and learning-based optimization approaches address the complex challenges of task offloading and resource allocation in distributed computing environments. Linear programming provides tractable solutions for resource allocation problems under linearity assumptions by transforming complex distribution challenges into mathematically solvable frameworks with guaranteed optimal solutions \cite{bandyopadhyay2024delaysensitive}. Convex optimization extends to nonlinear problems with reliable convergence guarantees through gradient-based methods and interior-point algorithms, enabling globally optimal solutions for complex resource distribution challenges \cite{nguyen2024integrated, feng2024dependencyaware}. Game theory modeling provides equilibrium-based optimization frameworks through competitive and cooperative interaction modeling, enabling strategic decision-making among rational participants through concepts like Nash equilibrium and Stackelberg leadership \cite{yu2024game, zhang2024stackelberggamebased}. Across heterogeneous AIoT environments, however, these analytical families still depend on simplified coupling assumptions, tractable state abstractions, or equilibrium observability that may weaken under mobility, non-stationarity, and cross-domain policy constraints. For comprehensive coverage of optimization techniques including detailed algorithm analysis, application scenarios, and performance evaluation results with specific implementations and case studies, see \cref{sec:supp_3} of the \textit{supplementary material}.

\subsubsection*{Summary and Insights}
The evolution of resource management in CETCI has shifted from static, rule-based heuristics to dynamic, AI-driven optimization. While traditional mathematical optimization (e.g., convex optimization and game theory) provides theoretical guarantees~\cite{yu2024game, zhang2024stackelberggamebased} and stability, it often struggles with the high dimensionality and non-stationarity of modern AIoT networks. In contrast, learning-based approaches, particularly DRL and MARL, offer superior adaptability~\cite{sharma2024deep} and can learn complex policies directly from experience without explicit system modeling, but they face challenges in convergence, sample efficiency, and interpretability. The practical lesson is that \textit{Infrastructure-level Collaboration} should be chosen according to operational uncertainty: deterministic environments reward analytical models, whereas volatile environments benefit from adaptive learning with lightweight online correction. Across the reviewed literature, no single family dominates all six evaluation dimensions, so method selection should be treated as a deployment decision that trades convergence guarantees and interpretability against scalability and online adaptability. Future work should therefore prioritize hybrid control-and-learning pipelines that can degrade gracefully when measurements are delayed or imperfect.

\section{Intelligent Collaboration Learning Paradigms}\label{sec4}

We examine \textit{Model-level Collaboration (Collaboration for AI)}, focusing on how distributed entities collaboratively train, evolve, and execute AI models. Collaborative intelligence (CI) represents the synergistic fusion of resources and knowledge across the cloud-edge-terminal hierarchy. While foundational technologies such as federated learning (FL) and distributed deep learning (DDL) are well-established collaborative paradigms in literature \cite{hong2019resource, liu2025edgecloud}, our analysis shifts from their algorithmic principles to their role as enablers of advanced collaboration mechanisms, specifically analyzing how they facilitate \textit{Cross-layer Collaboration} (e.g., cloud-to-edge) and \textit{Peer-to-Peer Collaboration} (e.g., edge-to-edge), distinguishing these sophisticated interactions from simple parallel computation. The taxonomy is organized by what is exchanged across layers: model updates in FL, activations or partitions in split and model-parallel learning, distilled or incrementally updated model states in evolution methods, and policy/value information in RL-driven coordination.
\begin{figure}[t!]
    \centering
  \includegraphics[width=0.48\textwidth]{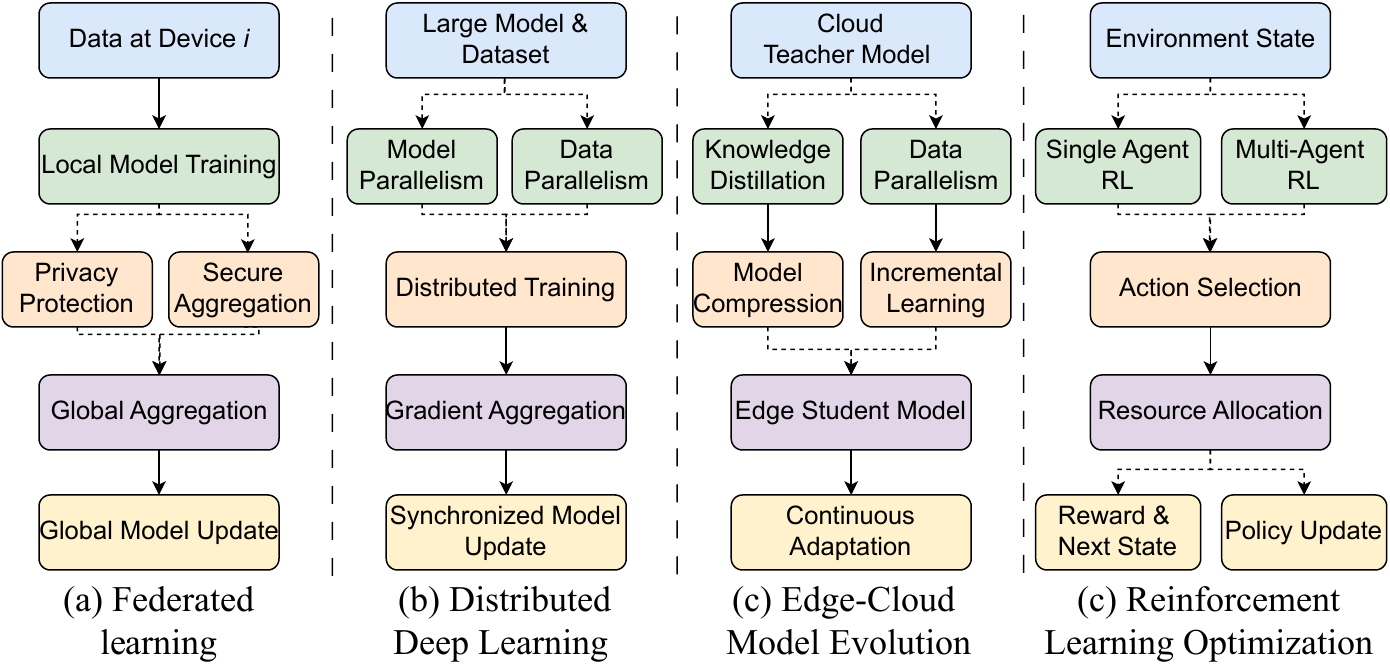}
\caption{Different types of collaboration learning paradigms.}
    \label{fig:learning_paradigms}
    \vspace{-10px}
  \end{figure}
As illustrated in \cref{fig:learning_paradigms}, CETCI employs four primary intelligent collaboration learning paradigms: federated learning with privacy-preserving aggregation, distributed deep learning through model and data parallelism, edge-cloud model evolution via knowledge distillation and compression, and reinforcement learning-based optimization. Their complementarity is important: FL prioritizes data sovereignty, split and model-parallel DDL prioritize fitting large models into heterogeneous hardware, edge-cloud evolution prioritizes continual adaptation under update budgets, and RL/MARL prioritizes sequential decision quality under uncertain environments. The taxonomy and its detailed components are further summarized in \cref{fig:4}.

\subsection{Federated Learning}\label{sec4.1}
FL serves as a quintessential example of a ``Data-Private, Model-Collaborative'' paradigm, directly addressing the data silo challenge inherent in the distributed architecture described in \cref{sec2}. By enabling collaborative model training without exchanging raw, sensitive data, FL functions as a critical mechanism for knowledge collaboration. Instead of centralizing data, terminals and edge nodes train local models on their respective datasets and contribute only model updates (e.g., gradients) to a central aggregator. By enabling privacy-preserving distributed learning, FL not only protects sensitive information but also transforms isolated data islands into a unified and intelligent system. FL accommodates diverse data scenarios, including horizontal (shared feature space), vertical (shared sample space), and transfer learning to leverage pre-existing knowledge in heterogeneous environments.

\subsubsection{Privacy-Preserving Federated Learning}\label{sec4.1.1}
Data protection mechanisms ensure secure collaborative learning without compromising individual privacy.  Differential privacy and secure aggregation techniques enable model training while safeguarding sensitive information across distributed participants \cite{stevens2022efficient}.  DP introduces calibrated noise into shared model updates, masking individual contributions while maintaining accuracy, particularly beneficial when the central server's trustworthiness is uncertain \cite{liu2021communicationefficient}.  Variants like zero-concentrated DP (zCDP) offer strong privacy guarantees \cite{hu2020concentrated}, and methods like the multi-stage adjustable private algorithm (MAPA) dynamically adjust the privacy-utility trade-off, improving accuracy and convergence in asynchronous FL \cite{li2019asynchronous}.  Alternatively, secure aggregation employs cryptographic techniques, such as multi-party computation (MPC) \cite{gupta2021hierarchical} and learning with errors \cite{chatel2024verifiable}, to aggregate updates without revealing individual contributions, thereby eliminating the need for a trusted third party and enhancing scalability in large AIoT networks.  Differentially private federated tensor completion, combined with secure aggregation techniques like objective perturbation and parallel tensor decomposition, provides a robust solution for protecting sensitive data in collaborative AIoT data prediction \cite{zhang2024efficient}.

\subsubsection{Robustness and Security of Federated Learning}\label{sec4.1.2}
Adversarial attack resistance becomes critical in decentralized learning environments.  Model poisoning threats and communication vulnerabilities require robust defense mechanisms to maintain system integrity and reliability \cite{ferrag2023poisoning}.  Additionally, unreliable AIoT communication networks exacerbate these issues. Communication failures, such as dropped packets, disrupt the FL process, leading to incomplete aggregation and reduced accuracy.  While asynchronous FL offers flexibility and efficiency \cite{liu2024quasyncfl}, it introduces further security risks by allowing attackers to exploit timing discrepancies for injecting poisoned models. Consequently, robust aggregation techniques \cite{zhao2024flexiblefl}, anomaly detection methods \cite{ma2022privacypreserving}, and secure communication protocols \cite{liu2021communicationefficient} are essential defense mechanisms for ensuring secure and reliable FL in distributed heterogeneous AIoT networks.
\input{fig4_sec4.tex}

\subsection{Distributed Deep Learning}\label{sec4.2}
While DDL traditionally refers to scaling training processes within data centers, its application in CETCI is fundamentally different. Here, DDL is repurposed as a collaborative mechanism to overcome the resource limitations of individual layers. Our analysis distinguishes this paradigm from conventional distributed training by focusing on collaborative inference and training across the cloud-edge-terminal continuum. Key strategies include \textit{Model Parallelism}, which facilitates ``split computing'' or ``co-inference'' by partitioning a single large model across heterogeneous devices (e.g., initial layers on an edge node and final layers in the cloud). Complementarily, \textit{Data Parallelism} enables distributed training on geographically dispersed data, accelerating model convergence by leveraging the collective computational power of the network.

\subsubsection{Model Parallelism}\label{sec4.2.1}

Large-scale model partitioning enables distributed deployment across heterogeneous computing resources.  Computational load distribution through model segmentation allows efficient utilization of available processing capacity while overcoming individual device limitations \cite{li2023alpaserve}.  For instance, AlpaServe \cite{li2023alpaserve} leverages model parallelism not only for scaling individual large models but also for statistically multiplexing devices serving various models, consequently improving resource utilization and reducing latency.  AutoDiCE \cite{guo2022autodice} further exemplifies this approach by automating the process of splitting and distributing CNN models across heterogeneous edge devices for efficient inference \cite{joshi2023enabling}.  Consequently, such distributed deployment empowers all-in-edge deep learning, enabling both training and inference on resource-constrained edge servers \cite{joshi2023enabling}.

\subsubsection{Data Parallelism}\label{sec4.2.2}
Training acceleration emerges through distributed data processing across multiple computing nodes.  Parallel gradient computation enables faster model convergence while leveraging the collective computational power of distributed devices \cite{liu2020hiertrain}.  Furthermore, this aggregation process often involves averaging the gradients or employing more sophisticated techniques like secure aggregation \cite{yang2024differentially}. Consequently, the training process is accelerated by leveraging the combined computational power of the distributed devices.  Within the CETCI architecture, data parallelism efficiently utilizes resources across different tiers, allowing resource-constrained terminal devices to handle smaller data chunks while edge servers aggregate local updates, thereby reducing the communication burden on the cloud \cite{qi2023arena}.  Meanwhile, the hierarchical approach enhances scalability and reduces latency, which is particularly beneficial for real-time AIoT applications.  Techniques like quantization can be combined with data parallelism to further improve communication efficiency, especially crucial in bandwidth-limited AIoT networks \cite{liu2024quasyncfl}.
Unlike FL, which shares parameter updates from full local models, split learning partitions networks across CETCI tiers by exchanging intermediate activations \cite{gao2020endend, banitalebidehkordi2021autosplit, lin2024split}. Consequently, partitioning becomes highly effective when terminal devices lack the capacity to host complete architectures, although the process inherently amplifies synchronization sensitivity and activation-transfer overhead. Regarding deployment choices, FL provides superior performance under conditions of strict data sovereignty and intermittent participation. Conversely, split learning excels within tightly coupled networks constrained by strict memory limits. As a general rule, exchanging activations remains highly sensitive to link quality, whereas sharing model updates scales better despite remaining vulnerable to data drift.

\subsection{Edge-Cloud Model Evolution}\label{sec4.3}
\input{t5_ec_mod_evo_4.3.tex}

Edge-cloud model evolution represents an advanced and dynamic form of collaborative intelligence, focusing on the entire lifecycle of AI models within the CETCI ecosystem. Continuous model adaptation across distributed computing environments is a fundamental requirement for dynamic systems. As summarized in \cref{tab:5}, various techniques enable the efficient transfer and adaptation of intelligence from powerful cloud models to resource-constrained edge devices. Our comparative analysis evaluates these methods based on their core algorithms, primary objectives such as accuracy and latency, and key trade-offs. For instance, while knowledge distillation is effective for transferring ``dark knowledge,'' it can be sensitive to the choice of teacher-student architecture. Similarly, compression techniques must carefully balance model size against potential accuracy degradation.

\subsubsection{Model Compression and Distillation}\label{sec4.3.1}

Knowledge transfer mechanisms enable deployment of sophisticated models on resource-constrained devices.  Compression techniques and teacher-student paradigms facilitate efficient model adaptation while preserving essential intelligence capabilities \cite{wang2024clouddevice}.  Knowledge distillation, in turn, transfers learned information from a larger teacher model in the cloud to a smaller student model at the edge \cite{tang2024anomaly}.  Specifically, the student model is trained to mimic the teacher model's output distribution, effectively transferring the embedded "dark knowledge" and improving generalization performance \cite{chen2023robust}.  Approaches like growth-adaptive distillation~\cite{yang2025growthadaptive} dynamically adjust the student model's architecture during this process, while others utilize methods such as adapter-based knowledge distillation (AKD) for efficient transfer \cite{yue2021inexactadmm}.
Edge LLM deployments rely on complementary compression mechanisms. Specifically, quantization and sparsity shrink memory alongside compute costs, low-rank adapters ensure bandwidth-efficient fine-tuning, and distillation transfers cloud-scale capabilities to smaller edge models \cite{shen2024agilequant, yi2025edgemoe, wang2024clouddevice}. Within CETCI environments, deployment selection depends heavily on the primary bottleneck. For example, quantization directly relieves latency, adapters significantly reduce update overhead, and distillation enables efficient cloud-to-edge intelligence transfer. Ultimately, integrating the aforementioned compression methods forms the critical systems substrate for agentic CETCI, ensuring that split inference and tool usage remain entirely within feasible edge latency and energy budgets.

\subsubsection{Incremental Learning and Model Updates}\label{sec4.3.2}
Continuous model adaptation enables learning from evolving data streams without complete retraining.  Edge-based incremental learning addresses resource constraints while preserving existing knowledge during adaptation to new environmental conditions \cite{yang2024continual}. Model updates can be implemented periodically or triggered by events like data distribution shifts or concept drift~\cite{li2024digital}. For example, the CADCL-DTM~\cite{li2024conditionadaptive} model uses incremental learning to adapt to dynamic changes in smart manufacturing, and the MEDIA algorithm~\cite{zhao2024media} employs it for efficient collaborative cloud-edge computing.  In intelligent transportation, cloud-edge frameworks leverage visual prompts and knowledge distillation for continual adaptation~\cite{lian2024cloudedge}, while EdgeC3~\cite{lin2024online} optimizes model aggregation and offloading for continuous data streams. Similarly, ADMM-FedMeta~\cite{yue2021inexactadmm} facilitates continual edge learning through knowledge transfer, and continual learning enhances digital twin synchronization~\cite{joshi2024integration}.

\subsection{Reinforcement Learning-based Optimization}\label{sec4.4}
\input{t6_rl_4.4.tex}

Advanced decision-making in complex environments increasingly relies on RL paradigms due to their ability to optimize operations without explicit supervision. To this end, \cref{tab:6} offers a comparative analysis of RL and MARL approaches in CETCI. The table contrasts different methods based on their core algorithms (e.g., policy-gradient, federated RL), their optimization objectives, and their inherent strengths and weaknesses. As shown, while single-agent RL is effective for localized optimization, MARL is increasingly adopted for complex, decentralized decision-making, though it introduces challenges in coordination and non-stationarity.

\subsubsection{Reinforcement Learning for Resource Management}\label{sec4.4.1}

Adaptive policy learning enables autonomous resource management through environmental interaction.  RL agents continuously optimize allocation strategies by learning from feedback signals and dynamic system conditions \cite{li2024load}.  Consequently, RL algorithms can learn to anticipate these changes and adaptively adjust resource allocation to maximize efficiency and minimize latency.
Researchers have explored applying RL to dynamic resource management in related contexts. For example, DRL has been employed to optimize task offloading decisions in edge-cloud systems~\cite{sharma2024deep}, where DRL agents dynamically offload tasks based on resource availability, network conditions, and task deadlines.  Similarly, multi-agent RL has been used for resource scheduling in cloud-edge computing power networks~\cite{guo2024madrlom}, enabling devices to learn optimal offloading strategies while coordinating to avoid congestion~\cite{binh2024reinforcement}.  In addition, RL can optimize energy consumption by dynamically adjusting computational load distribution across different layers, as demonstrated in studies on energy-efficient edge-cloud collaborative sensing systems~\cite{afachao2024efficient} and large-scale industrial IoT environments~\cite{zhang2024lsia3cs}.

\subsubsection{Multi-Agent Reinforcement Learning}\label{sec4.4.2}

Coordinated decision-making emerges through distributed agents that learn collaborative strategies.  MARL frameworks enable autonomous devices to optimize individual objectives while contributing to system-wide performance improvements \cite{li2024load}.  Through interactions with the environment and other agents, MARL enables devices to learn optimal policies, dynamically allocating resources like computing power and bandwidth in resource-constrained scenarios \cite{zhu2022federated}.  Consequently, MARL facilitates efficient task offloading, allowing devices to autonomously determine whether to process tasks locally or offload them to edge or cloud servers based on factors such as network conditions and task complexity \cite{suzuki2022multiagent}.  Examples of MARL's effectiveness in AIoT include controlling multiple UAVs for maximized QoS in autonomous mobile access applications using a centralized training and distributed execution approach \cite{park2022coordinated}.  In addition, studies have explored MARL for distribution network reconfiguration within a cloud-edge collaboration framework, leveraging algorithms like discrete multi-agent soft actor-critic (MASAC) to address non-stationary environments \cite{gao2024cloudedge}.  Similarly, Coop-MADRL has been employed for cooperative computing offloading and route optimization in multi-cloud-edge networks, minimizing network utilization and task latency \cite{suzuki2023multiagent}, while other research demonstrates MARL's utility in distributed transmission within collaborative cloud-edge systems for enhanced network delay performance and QoS satisfaction \cite{xu2021multiagent}.

\subsubsection*{Summary and Insights}
Collaborative learning paradigms have fundamentally transformed how intelligence is acquired and deployed in AIoT. Federated learning and distributed deep learning enable the use of massive, decentralized datasets while respecting privacy and bandwidth constraints. However, these \textit{Model-level Collaborations} impose significant burdens on the underlying infrastructure, requiring frequent, high-volume communication and synchronized computation. The main takeaway is that FL is usually preferable when data sovereignty is the primary constraint, split or model-parallel learning is more attractive when end devices cannot host full models, edge-cloud model evolution is most useful when personalization must be sustained after deployment, and RL-driven coordination is best interpreted as a policy layer that optimizes collaborative behavior rather than a replacement for supervised model training~\cite{liu2026semantic}. Future breakthroughs therefore lie in co-designing these learning paradigms with communication-efficient mechanisms, such as sparse updates, quantization, semantic alignment, and topology-aware synchronization \cite{liu2026diffusionguided}.

\section{Security and Privacy Protection}\label{sec5}

Securing the CETCI paradigm presents unique challenges that transcend traditional cybersecurity models. Unlike centralized systems, the distributed, heterogeneous, and often resource-constrained nature of CETCI architectures introduces a multifaceted threat landscape, as illustrated in \cref{fig:taxonomy_sec5}. Specific threat models emerge at different layers: the terminal-edge link is particularly vulnerable to attacks targeting physical device integrity and location privacy, whereas the edge-cloud link faces sophisticated threats like coordinated denial-of-service (DoS) attacks and data poisoning in collaborative learning. Consequently, protecting this ecosystem requires a defense-in-depth strategy that moves beyond generic solutions. Recent advancements have focused on edge-centric defense mechanisms, including lightweight encryption tailored for resource-scarce terminals, blockchain-based trust management to ensure data integrity across decentralized nodes, and the use of trusted execution environments to create secure enclaves for sensitive computations at the edge.

\input{fig5_sec5.tex}
\subsection{Security Threats and Vulnerabilities}\label{sec5.1}

Distributed collaborative architectures introduce expanded attack surfaces that must be addressed to ensure robust operation. Sensitive data across multiple layers risks unauthorized access and interception, potentially leading to breaches and eavesdropping. Moreover, denial-of-service attacks can overwhelm resources, disrupting services and disabling critical devices. Consequently, ensuring data integrity and authentication demands robust mechanisms to verify authenticity and prevent tampering within collaborative ecosystems. A comprehensive summary of security threats and vulnerabilities across diverse scenarios including vehicular networks and multimedia IoT, with detailed analysis of privacy, accuracy, and latency considerations, is provided in Table 1 of the \textit{supplementary material}.

\subsubsection{Data Breaches and Eavesdropping}\label{sec5.1.1}
Information security vulnerabilities expose sensitive data to unauthorized access and interception attacks.  Distributed system architectures create multiple attack vectors where malicious actors can exploit weak points to compromise data integrity \cite{rahman2023blockchainbased}.  For example, eavesdroppers can intercept sensitive data transmitted between terminal devices and the edge or cloud \cite{fang2024secure}. Resource-constrained terminal devices, often lacking robust security, are susceptible to breaches \cite{chi2024trusted}, potentially allowing unauthorized data access and manipulation, consequently disrupting AIoT applications.  Similarly, edge nodes, aggregating data from multiple devices, are also vulnerable.  A compromised edge node amplifies the impact of a breach, exposing a larger dataset \cite{wang2022privacypreserving}.  In addition, the increasing use of AI models introduces further risks, as attackers can exploit model vulnerabilities to infer sensitive information, even without direct data access \cite{yang2024differentially}.

\subsubsection{Denial-of-Service Attacks}\label{sec5.1.2}
Resource exhaustion attacks threaten system availability by overwhelming computational and network capabilities.  DoS vulnerabilities become amplified in CETCI systems where interdependent layers create cascading failure risks \cite{li2022fleam}.  Specifically, DoS attacks within AIoT can target any architectural layer, from resource-constrained terminal devices to powerful cloud servers.  For example, attacks on the terminal layer might flood IoT sensors, exhausting their processing power and disrupting data acquisition \cite{wang2018thingpot}.  Similarly, targeting the edge layer can hinder local processing, increasing latency and affecting real-time decision-making \cite{chen2024intelligenta}.  Attacks on the cloud layer, in contrast, can disrupt centralized data storage and analysis, impacting overall system performance \cite{raj2021reliable}. Moreover, the distributed nature of distributed systems introduces unique vulnerabilities, as the interconnected layers create multiple points of failure exploitable by attackers \cite{zhou2024enhancing}.  The heterogeneity of devices and networks further complicates attack detection and mitigation. Consequently, DoS attacks can significantly degrade key AIoT performance metrics, including latency, resource utilization, and QoS, potentially impacting critical infrastructure and services like smart healthcare and manufacturing \cite{chi2024trusted}.  Robust security mechanisms are therefore crucial for mitigating these attacks and ensuring reliable system operation \cite{yang2024differentially}.

\subsubsection{Data Integrity and Authentication}\label{sec5.1.3}
Trust establishment requires comprehensive verification mechanisms to ensure data authenticity and prevent tampering.  Multi-layered authentication protocols safeguard against unauthorized modifications while maintaining system reliability across distributed environments \cite{dui2024maintenance}.  Similarly, authentication verifies the identities of participating devices and users~\cite{yao2024efficient}, mitigating risks associated with unauthorized access.  Achieving these security goals requires a multi-faceted approach, encompassing cryptographic techniques for data protection and secure authentication protocols~\cite{xiao2024domainspecific}.  Robust system-wide breach detection and mitigation mechanisms are also essential~\cite{yang2024differentially}. For instance, blockchain can enhance data integrity and provenance~\cite{chi2024trusted}, while federated learning can improve data privacy during collaborative model training~\cite{zhou2024enhancing}.

\subsection{Security Mechanisms and Solutions}\label{sec5.2}
Protecting CETCI paradigms requires a multi-layered security approach. Foundational mechanisms include encryption and authentication, which secure data transmission and storage across the continuum \cite{wang2018secure, zhou2024enhancing}. Lightweight symmetric encryption is often employed for resource-constrained devices, while attribute-based encryption (ABE) offers fine-grained access control for sensitive AIoT data \cite{xiao2024domainspecific, fang2024secure}. Complementing these are proactive threat monitoring systems, such as intrusion detection and prevention systems (IDPS), which identify and neutralize malicious activities. Deploying IDPS at different architectural layers provides comprehensive coverage; edge-based IDPS enables faster responses to localized threats, while cloud-based systems leverage centralized resources to identify complex, large-scale attack patterns \cite{chi2024trusted}. In addition to detection, IDPS incorporates prevention mechanisms, including blocking malicious traffic and isolating compromised devices, to actively mitigate threats. Furthermore, blockchain-based decentralized trust mechanisms eliminate single points of failure and ensure immutable transaction records for secure access control \cite{rahman2023blockchainbased, gadekallu2022blockchain}. \cref{tab:7} provides an overview of methods addressing security threats and vulnerabilities, highlighting their deployment scenarios and performance metrics. For a detailed analysis of specific mechanisms, see \cref{sec:supp_1.1} of the \textit{supplementary material}.
\input{t7_security_5.1.tex}

\subsection{Privacy-Preserving Techniques}\label{sec5.3}
Collaborative privacy protection relies on foundational techniques including federated learning for exchanging updates instead of raw data \cite{li2023edgecloud, liu2024quasyncfl} alongside differential privacy for injecting calibrated noise \cite{rodriguezbarroso2020federated, yang2024differentially, wang2022privacypreserving, zhang2024efficient}. When rigorous confidentiality is demanded, homomorphic encryption allows computation directly on encrypted data \cite{kim2024privacy, gupta2021hierarchical}, whereas secure multi-party computation protects inputs across multiple participants \cite{sengupta2022sprite, chi2024trusted}. Crucially, enforcing stronger security guarantees inevitably increases communication volume, cryptographic overhead, and synchronization latency, which directly constrains the offloading and learning strategies discussed previously in \cref{sec3,sec4,sec6.3}. For comprehensive details regarding privacy optimization and hierarchical FL frameworks, readers are directed to \cref{sec:supp_1.2} of the \textit{supplementary material}.
A deployment-oriented interpretation is therefore necessary: DP is attractive when bounded privacy leakage is required with moderate extra communication~\cite{liu2026diffusionguided}, HE and SMPC are more suitable when confidentiality dominates and additional latency is acceptable, and secure aggregation offers a middle ground~\cite{liu2026semantic} for large collaborative-learning populations, particularly when personalizing foundational models~\cite{liu2026pfadapter}. Across these options, the decisive trade-off is cross-layer rather than cryptographic alone, because privacy gains must be balanced against learning accuracy, energy draw at the edge, retransmission overhead, and deadline compliance for the target AIoT service.

\section{Data Management and Communication}\label{sec6}

Effective data management and communication form the bedrock of the CETCI paradigm. Although the distributed architecture detailed in \cref{sec2} is powerful, it introduces significant data-centric challenges. The massive influx of data from terminal devices creates high concurrency, while the diversity of sensors and applications results in profound data heterogeneity in terms of formats, protocols, and quality. Furthermore, the communication links connecting terminals, edge nodes, and the cloud are often resource-constrained, facing limited bandwidth and variable latency. As a result, these challenges directly impact the efficacy of the intelligent resource management strategies discussed in \cref{sec3}. Optimal task offloading and resource allocation decisions require efficient, low-latency data transmission, because communication bottlenecks can otherwise negate the benefits of distributed processing. Therefore, to address these critical issues, we provide in this section a comprehensive analysis of the mechanisms designed to overcome such challenges, as illustrated in \cref{fig:taxonomy_sec6}. We explore techniques for data acquisition and preprocessing, such as filtering and compression, that mitigate bandwidth load. We then examine distributed storage and retrieval strategies, including edge caching, that enhance data availability. Finally, we discuss communication protocols and optimization techniques that ensure reliable and efficient data exchange across the heterogeneous CETCI ecosystem.

\input{fig6_sec6.tex}

\subsection{Data Acquisition and Preprocessing}\label{sec6.1}

Effective data management in resource-constrained environments requires streamlining data processing to optimize performance. As detailed in \cref{tab:8}, various data preprocessing and management techniques are employed to optimize data flow, ranging from federated learning for privacy-preserving aggregation to advanced compression for bandwidth reduction, each presenting unique trade-offs. For instance, while lossless compression guarantees data fidelity, it offers lower compression ratios compared to deep learning-based methods, which may introduce acceptable quality loss for certain AIoT applications. Accordingly, the table provides a comparative analysis of their techniques, application scenarios, and key limitations.
\input{t8_data_6.1.tex}

\subsubsection{Data Filtering and Aggregation}\label{sec6.1.1}
Preprocessing operations at distributed network edges optimize transmission efficiency while preserving essential information~\cite{zhu2023pushing}. Volume reduction through intelligent filtering minimizes communication overhead and latency~\cite{raj2021reliable}.  Several filtering techniques include threshold-based filtering, discarding data outside a predefined range, and statistical filtering, employing measures like mean and standard deviation to remove outliers~\cite{becker2020aiops}.  Additionally, Kalman filtering estimates system state and filters sensor noise.  For aggregation, methods such as averaging, summing, and counting over specific time windows or spatial regions apply~\cite{liu2024quasyncfl}, while more sophisticated techniques like wavelet transforms extract relevant features while reducing dimensionality.  In resource-constrained AIoT environments, the processes enable real-time processing and decision-making~\cite{chi2024trusted}, exemplified by anomaly detection and preventative maintenance in smart manufacturing~\cite{yang2024differentially}, as well as traffic flow optimization in smart transportation~\cite{zhou2024enhancing}.

\subsubsection{Data Compression and Encoding}\label{sec6.1.2}
Lossy and lossless compression techniques are fundamental for minimizing transmission overhead in resource-constrained environments. Lossy methods achieve higher compression ratios at the expense of data fidelity, and studies have explored their effectiveness in machine learning training sets, demonstrating significant compression with minimal quality loss \cite{underwood2024understanding}.  Deep learning-based super-resolution models can further enhance the quality of lossy compressed images at the receiver side, balancing size reduction and visual degradation in MIoT scenarios \cite{noura2023deep}. In contrast, lossless compression preserves data integrity but achieves lower compression ratios. For example, edge source coding techniques utilize user preferences to optimize lossless compression performance at the network edge~\cite{lu2020user}. Preprocessing techniques, particularly for floating-point time-series data, can enhance compression capabilities \cite{taurone2023change}. Specifically, methods based on data statistics and deviation have shown superior compression ratios compared to deep learning-based methods for certain time series datasets \cite{agrawal2022lossless}.  Similarly, ROOT I/O compression improvements in HEP analysis balance compression efficiency with computational costs \cite{shadura2020root}. Adaptive techniques, such as selective edge compression, dynamically adjust compression parameters based on data characteristics or network conditions, improving network transfer performance and achieving significant data savings \cite{melissaris2020optimizing}.  Alternatively, hybrid approaches combining machine learning with standard compression methods have also been explored for scientific data, aiming to preserve derived quantities with minimal error \cite{banerjee2022scalable}.

\subsection{Data Storage and Retrieval}\label{sec6.2}
Efficient data management in distributed collaborative systems encompasses three critical aspects: edge caching for proximity-based storage optimization, distributed data storage across multi-tier infrastructures, and comprehensive data integrity and availability mechanisms. Edge caching strategies enhance access latency through intelligent content placement and predictive caching algorithms, while distributed storage architectures enable efficient data distribution across cloud and edge nodes with strategic placement based on access patterns \cite{luckow2021pilotedge, raj2021reliable}. Data integrity and availability protection ensures information authenticity and continuous accessibility through robust mechanisms including blockchain applications, authentication protocols, and fault-tolerant storage designs \cite{chi2024trusted, dui2024maintenance}. For detailed coverage of edge caching techniques, distributed storage architectures, and data integrity mechanisms including specific implementation strategies and optimization approaches, see \cref{sec:supp_2.1} of the \textit{supplementary material}.

\subsection{Communication Protocols and Optimization}\label{sec6.3}
Efficient data exchange in distributed collaborative systems requires three fundamental components: resource-aware communication protocols, bandwidth optimization techniques, and fault tolerance mechanisms. Lightweight protocols such as MQTT and CoAP enable efficient data transmission in resource-constrained environments, while network coding enhances data integrity and reduces redundancy in AIoT communication \cite{paolo2023security, gundogan2020iot}. Bandwidth optimization strategies encompass traffic scheduling for critical data flow prioritization, rate adaptation for dynamic network conditions, and link aggregation for enhanced capacity and resilience \cite{li2024cloudedgedevice, nezami2021decentralized}. Comprehensive fault tolerance mechanisms ensure resilient communication through error detection and recovery, redundancy management, connection management strategies, and quality assurance practices that guarantee reliable service delivery \cite{nikolic2021selfhealing, javed2018cefiot}. Crucially, the communication properties also serve as the bottleneck variables behind several later comparisons in \cref{sec3,sec4,sec5}: an algorithm with strong learning accuracy or privacy guarantees may still be deployment-unfriendly if synchronization traffic, retransmission sensitivity, or fault-recovery overhead dominate the end-to-end workflow. For detailed analysis of communication protocol implementations, bandwidth optimization algorithms, and fault tolerance mechanisms including specific deployment strategies and performance considerations, see \cref{sec:supp_2.2} of the \textit{supplementary material}.

\section{Applications}\label{sec7}

The practical impact of CETCI is best understood through its application in various vertical AIoT domains. To systematically examine this impact, we adopt in this section a domain-specific classification to analyze how collaborative intelligence addresses unique challenges in five key areas: smart manufacturing, smart transportation, smart healthcare, smart cities, and smart agriculture. Each domain leverages a distinct ecosystem of IoT devices and has specific requirements regarding latency, bandwidth, and security, thereby showcasing the adaptability of the CETCI paradigm. Across these domains, the same architectural template is reused with different emphases: latency sensitivity and actuation risk favor edge-first loops, privacy criticality favors local processing plus protected aggregation, limited bandwidth favors aggressive filtering or compression, and high model-update frequency favors cloud-assisted evolution with lightweight edge adaptation. \cref{fig:taxonomy_sec7} provides a high-level overview of these application areas.

\input{fig7_sec7.tex}

\subsection{Smart Manufacturing}\label{sec7.1}
In smart manufacturing, CETCI transforms traditional production lines into intelligent and autonomous systems by processing data directly from IIoT devices, thereby enabling real-time process optimization and enhancing product quality.

\subsubsection{Real-Time Defect Detection}\label{sec7.1.1}
Production line monitoring demands immediate anomaly identification to minimize manufacturing waste \cite{bujari2023layered}. Terminal sensors and IoT devices capture real-time data, including images, videos, and quality-related sensor readings \cite{becker2020aiops}. Edge devices, proximate to the production process, perform initial data processing and filtering to reduce the volume of data transmitted to the cloud \cite{wu2022pecco}, enabling faster processing and reduced latency crucial for real-time defect detection \cite{zhou2024enhancing}.  Complex AI models trained on historical data in the cloud are deployed on edge devices for real-time inference, such as deep learning models analyzing images and videos for product defects \cite{zhang2024cloud}. The cloud layer provides centralized storage for large datasets and facilitates model training and updates, which allows for continuous improvement of defect detection and adaptation to new defect types \cite{hao2019smartedgecocaco}.  Additionally, the cloud offers insights into production quality and identifies trends in defect occurrence \cite{raj2021reliable}.

\subsubsection{Predictive Maintenance}\label{sec7.1.2}
Machine learning analytics enable proactive equipment failure prediction and optimized maintenance scheduling \cite{rao2021eco}. Distributed computational workloads enhance this process through real-time data processing capabilities. Specifically, sensors at the terminal layer collect equipment performance data, which is subsequently preprocessed and analyzed at the edge layer.  Complex models and historical data residing in the cloud then contribute to more accurate predictions \cite{xu2024fusion}. Equally important, the hierarchical approach reduces latency, enabling timely interventions and minimizing downtime.  For example, \cite{zhang2024cloud} presents a quality-related hierarchical fault detection framework using minimal gated units and federated bi-directional knowledge distillation to improve accuracy and reliability.  Circuit-level AI solutions integrated within control and calibration blocks during circuit design, as proposed in \cite{dosluoglu2022circuit}, can further enhance component predictability and adaptability for predictive maintenance.

\subsubsection{Automated Quality Control}\label{sec7.1.3}
Lightweight AI deployment at production edges enables immediate defect detection through rapid sensor analysis~\cite{raj2021reliable}. Real-time processing reduces reliance on centralized systems while minimizing response latency for corrective actions~\cite{xu2024fusion}.  Moreover, this collaborative paradigm also supports sophisticated quality control mechanisms, such as real-time image analysis for defect identification and visual-disbalance enhancement~\cite{yang2024differentially, wu2025clipae}, by distributing computational tasks across the three layers for efficient processing of large visual datasets.  The edge layer's role in data preprocessing and filtering optimizes bandwidth usage and enhances data security by reducing cloud transmission and keeping sensitive data closer to the source~\cite{wang2024cloudedgeend, chi2024trusted}.  Additionally, federated learning enables training quality control models across multiple sites without sharing raw data, improving model accuracy and robustness while preserving confidentiality~\cite{zhou2024enhancing}.

\subsection{Smart Transportation}\label{sec7.2}
In smart transportation, CETCI is critical for developing intelligent transportation systems that enhance safety and efficiency by processing massive data streams generated by vehicle-to-everything (V2X) communication units, roadside sensors, and in-vehicle IoT devices.

\subsubsection{Autonomous Vehicle Navigation and Control}\label{sec7.2.1}
Safe and efficient vehicular automation requires distributed intelligence across heterogeneous computing layers. Onboard sensors and computers collect real-time environmental data, including images, LiDAR point clouds, and GPS coordinates \cite{tang2018piedge}.  The cloud layer provides high-level functionalities like map updates, traffic prediction, and fleet management \cite{chi2024trusted}.  Specifically, cloud-edge collaboration enables offloading computationally intensive tasks, such as AIGC for trajectory prediction and traffic simulation, to the edge, consequently reducing latency and improving real-time performance \cite{katariya2023vegaedge}.  Similarly, edge-terminal collaboration allows rapid responses to critical events, such as obstacle avoidance, by processing sensor data locally \cite{tang2018piedge}. Moreover, distributed frameworks exemplified by Auto-Split and EdgStr \cite{zhang2024cloudedgeterminal} enhance the robustness of autonomous driving systems by enabling local decision-making and optimizing resource utilization, even with intermittent cloud connectivity \cite{raj2021reliable}.
A concrete CETCI case study is autonomous driving in mixed urban traffic. In this workflow, the terminal layer onboard the vehicle executes safety-critical perception and control loops within milliseconds, the roadside or vehicular edge layer fuses nearby views for cooperative perception and short-horizon trajectory refinement, and the cloud layer aggregates fleet-scale traces for map refresh, long-horizon planning, and model updates \cite{chen2024edgecloud, binh2024reinforcement}. As a result, the division of labor makes the application accessible in practical terms: latency-sensitive braking and lane keeping remain local, cooperative scene understanding is pushed to the edge when line-of-sight or compute asymmetry matters, and cloud coordination is reserved for tasks whose value depends on broader historical context. The case also highlights the main deployment constraint of CETCI in transportation, namely that split placement and offloading policies must remain robust under fast topology changes and intermittent backhaul connectivity \cite{ni2023mscet}.

\subsubsection{Traffic Flow Optimization and Congestion Management}\label{sec7.2.2}
In AIoT-enabled smart cities, cloud-edge-terminal collaboration offers a comprehensive approach to optimizing traffic flow and mitigating congestion. Edge computing's proximity to data sources facilitates real-time responses to changing traffic conditions \cite{vo2022edge}, which is essential for dynamic traffic management.  Furthermore, the collaborative architecture enables coordinated control of traffic signals and other infrastructure, leading to smoother flow and reduced congestion \cite{raj2021reliable}. For example, real-time data from terminal-layer sensors can be processed at the edge to identify congestion, subsequently used to adjust traffic signals or suggest alternative routes through connected vehicle platforms \cite{ni2023mscet}.  Additionally, the cloud layer provides long-term traffic pattern analysis and predictive modeling for proactive congestion management \cite{zhou2024enhancing}.

\subsubsection{Real-time Traffic Monitoring and Incident Detection}\label{sec7.2.3}
Rapid anomaly identification enhances road safety through distributed surveillance and analysis capabilities \cite{katariya2023vegaedge}. Edge processing of traffic camera and GPS data reduces latency while minimizing bandwidth consumption \cite{ghosh2021mobility, becker2020aiops}.  Consequently, AI algorithms deployed at the edge enable real-time analysis of traffic patterns and identify anomalies, such as sudden braking or stopped vehicles \cite{agrawal2022lossless}. Moreover, real-time edge-based detection facilitates immediate responses, including dispatching emergency services or adjusting traffic signals \cite{ni2023mscet}.  Meanwhile, the cloud provides centralized storage and processing for complex analyses like long-term trend identification \cite{raj2021reliable}.  Machine learning models on edge devices further improve the system's adaptability to dynamic traffic conditions and enhance incident detection accuracy \cite{raj2021reliable}.

\subsection{Smart Healthcare}\label{sec7.3}

\begin{figure}[t!]
    \centering
  \includegraphics[width=0.48\textwidth]{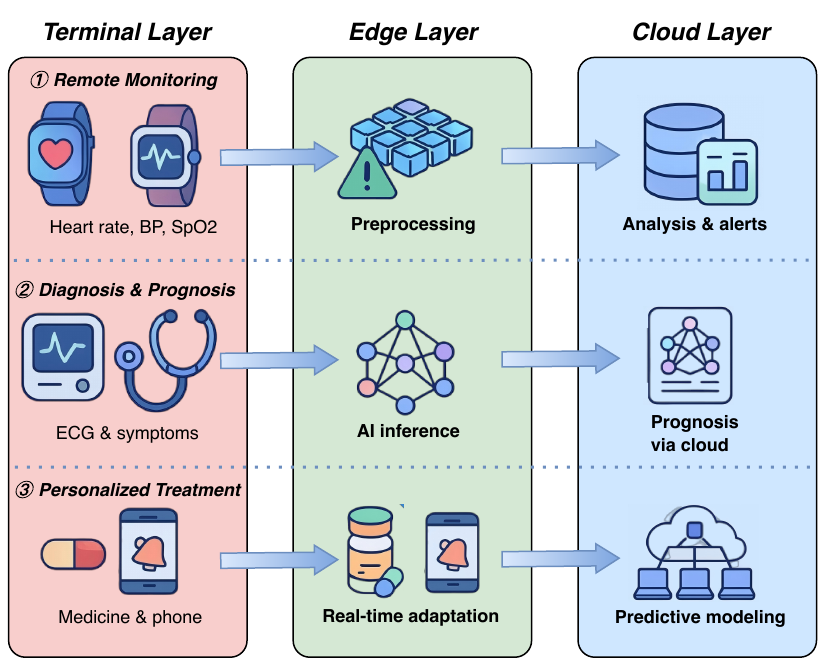}
  \caption{Illustration of CETCI in smart healthcare.}
    \label{fig:smart_healthcare}
    \vspace{-15px}
  \end{figure}

As illustrated in \cref{fig:smart_healthcare}, which shows three core application scenarios: (1) remote patient monitoring, (2) AI-assisted disease diagnosis and prognosis, and (3) personalized treatment. The CETCI paradigm enables three key healthcare services, where each component ensures timely and secure delivery of healthcare interventions.

\subsubsection{Remote Patient Health Monitoring and Vital Signs Tracking}\label{sec7.3.1}
Continuous physiological monitoring transforms healthcare delivery through distributed sensor networks and real-time analytics \cite{shaik2023remote}. Wearable devices and IoT sensors collect vital signs including heart rate, blood pressure, and oxygen saturation \cite{gholipour2023tpto}. Subsequently, edge devices preprocess and aggregate sensor data, reducing transmission volume to the cloud \cite{raj2021reliable} and enabling timely interventions through initial analysis and alerts.  The cloud layer facilitates long-term data storage, advanced analytics, and AI-assisted diagnosis, supporting personalized treatment and medication management \cite{gupta2021hierarchical}.  For example, FL \cite{zhou2024enhancing, gupta2021hierarchical} trains models on decentralized patient data at the edge, preserving privacy while improving vital signs prediction and anomaly detection.  Cloud-edge collaboration also enables continuous monitoring and real-time feedback to patients and providers, promoting proactive healthcare management \cite{buyya2023quality}.  Differentially private federated tensor completion methods \cite{yang2024differentially} and on-device contactless vital sign measurement using MTTS-CAN \cite{liu2020multitask} further enhance privacy and real-time monitoring capabilities, especially as multimodal medical AI systems increasingly combine physiological, imaging, and textual evidence \cite{liu2025multimodala}.
A second practical case study is smart healthcare for chronic-disease monitoring. Here, wearable or bedside terminals capture heart rate, blood pressure, SpO$_2$, and activity signals, an edge gateway at the home or hospital performs denoising, anomaly screening, and alert generation, and the cloud maintains longitudinal records plus more compute-intensive risk stratification models \cite{gupta2021hierarchical, ahmed2024heterogeneous}. In turn, the case clarifies why CETCI is useful beyond abstract architecture diagrams: privacy-sensitive raw biosignals can stay near the patient, urgent anomaly detection remains at the edge for rapid escalation, and only summary statistics, encrypted updates, or model parameters need to participate in wider collaborative learning. The main deployment constraint is therefore not only accuracy but also how privacy budgets, fault tolerance, and clinician-facing response times are jointly managed across the three layers \cite{ni2023mscet}.

\subsubsection{AI-assisted Disease Diagnosis and Prognosis}\label{sec7.3.2}

Intelligent diagnostic systems leverage distributed computing to enhance medical decision-making accuracy \cite{hayyolalam2021edge}. Real-time patient data processing at network edges reduces latency for timely clinical interventions \cite{shaik2023remote}.  For instance, edge-deployed AI algorithms analyze physiological data like ECGs and blood pressure for preliminary diagnoses, rapidly identifying potential issues and triggering alerts \cite{raj2021reliable}.  Meanwhile, the cloud provides access to comprehensive medical databases and sophisticated AI models, enabling more complex analyses and accurate prognoses \cite{yang2024differentially}.  FL trains robust AI models while preserving patient privacy by allowing edge devices to collaboratively train a shared model without exchanging raw data \cite{zhou2024enhancing, maia2024survey}. Notably, the collaborative paradigm, combined with AI and FL, improves diagnostic accuracy and speed, leading to more effective and personalized interventions \cite{yao2023edgecloud, joshi2024integration}.

\subsubsection{Personalized Treatment Plans and Medication Management}\label{sec7.3.3}

Individualized therapeutic approaches emerge through comprehensive patient data integration across distributed healthcare networks~\cite{buyya2023quality}. Multi-source monitoring combines wearable sensors, edge analytics, and cloud platforms for holistic patient assessment~\cite{raj2021reliable}. For example, wearable devices continuously monitor patient vitals, and edge servers perform real-time AI-driven anomaly detection, triggering timely interventions~\cite{zhou2024enhancing}.  Edge computing also facilitates personalized medication reminders and dosage adjustments based on real-time data, improving adherence and minimizing adverse effects. Subsequently, the cloud aggregates data from multiple patients to develop sophisticated predictive models for disease progression and treatment response~\cite{yang2024differentially}. Secure data sharing across this continuum empowers healthcare providers with comprehensive patient information for informed decision-making and collaborative care~\cite{chi2024trusted}.  Distributing computational tasks across different layers addresses privacy concerns and reduces latency, which are crucial for real-time healthcare applications~\cite{luckow2021pilotedge}.

\subsection{Smart Cities}\label{sec7.4}
In smart cities, a dense web of IoT devices, from public surveillance cameras to environmental sensors and smart grid meters, generates heterogeneous data streams that are crucial for urban management. CETCI enables real-time processing of this data for applications like public safety, pollution control, and energy grid optimization. For instance, in smart surveillance, IoT cameras provide video feeds that are analyzed at the edge for real-time threat detection \cite{blasch2019blockchain, nikouei2019decentralized, liu2025networking}. Similarly, data from environmental IoT sensors is processed locally to monitor pollution levels and identify sources quickly \cite{arroba2024sustainable, vo2022edge}. The collaborative framework also supports smart grids by using IoT meter data for predictive maintenance and efficient energy distribution \cite{zhai2024edgecloud, hua2024energyefficient}. For detailed examples, see \cref{sec:supp_4.1} of the \textit{supplementary material}.

\subsection{Smart Agriculture}\label{sec7.5}
Smart agriculture leverages a wide array of IoT devices, including soil moisture sensors, weather stations, agricultural drones, and automated tractors, to optimize farming practices. CETCI facilitates precision agriculture by enabling real-time, data-driven decisions based on information collected from these sensors. For example, edge computing processes data from soil sensors to enable automated, precision irrigation systems, conserving water while maximizing crop health \cite{kumar2021affordable, zhang2024efficient}. For crop monitoring, data from IoT-equipped drones and satellites is integrated and analyzed to predict yields and detect disease outbreaks early \cite{wang2025intelligent, huber2022extreme}. Collaborative learning paradigms like FL allow multiple farms to build robust predictive models without sharing proprietary data, adapting them to local conditions \cite{khuencheng2024integration}. For a detailed analysis, see \cref{sec:supp_4.2} of the \textit{supplementary material}.

Across the five domains, four transferable design patterns recur. First, edge-first closed loops dominate when actuation deadlines are tight, as in manufacturing and transportation~\cite{zhu2023pushing}. Second, privacy-preserving cloud aggregation becomes more important when sensitive data is widely distributed, as in healthcare and smart-city sensing~\cite{liu2026semantic}. Third, event-driven model refresh is preferable when data drift is bursty rather than stationary, which is common in traffic and agricultural monitoring. Fourth, bandwidth-aware filtering, compression, and caching become essential when wide-area links are intermittent or expensive, particularly for complex cross-modal streams~\cite{liu2025scmm}. Consequently, the identified patterns explain why the same CETCI architecture yields different layer responsibilities across applications instead of a single fixed deployment recipe.
A comprehensive comparison of the AIoT application domains, including smart manufacturing, smart transportation, smart healthcare, smart cities, and smart agriculture, is provided in \cref{tab:app_comparison} of the \textit{supplementary material}. The table details the specific IoT device types, key requirements, and primary edge computing roles for each domain, offering a structured overview of how CETCI principles are applied across diverse real-world scenarios.

\section{Discussion}\label{sec8}
\subsection{Technical Challenges}\label{sec8.1}
CETCI still faces several tightly coupled technical challenges. First, \textit{scalability} remains difficult because massive IoT participation increases computation, communication, and model-synchronization pressure across distributed infrastructure \cite{becker2020aiops, zhou2024enhancing}. Second, \textit{heterogeneity and interoperability} complicate cross-layer orchestration because terminals, edge nodes, cloud services, protocols, and AI models differ in resource budgets, data formats, and service requirements \cite{chen2019learning, menetrey2022webassembly}. Third, \textit{security, privacy, trust, and reliability} become more demanding as data, models, and control decisions traverse multiple administrative and physical domains \cite{alwarafy2021survey, chi2024trusted}. Finally, \textit{resource management} must jointly optimize computing, storage, networking, energy, and latency under dynamic conditions rather than treating these objectives separately \cite{wang2024cloudedgeend, li2024energyaware}. Taken together, these four issues are the most immediate blockers to large-scale deployment, because each one directly constrains whether the collaborative strategies reviewed in earlier sections can be executed under practical communication, cost, and governance limits. A detailed discussion of these challenges, together with the challenge overview figure, is provided in \cref{sec:supp_challenges} of the \mbox{\textit{supplementary material}}.

\subsection{Future Research Directions}\label{sec8.2}

\begin{figure}[t!]
\centering
\input{fig8_fr.tex}
\caption{Overview of future research directions.}
\label{fig:future_directions}
\vspace{-10px}
\end{figure}

As illustrated in \cref{fig:future_directions}, the future of CETCI can be framed along two distinct timelines. \textit{Near-term directions} focus on immediate implementation blockers, especially large-scale orchestration under heterogeneity, security and privacy protection under realistic communication budgets, resource-aware support for LLM or agent execution, and timely digital-twin synchronization. In contrast, \textit{long-term directions} explore architectural expansion once those blockers are better controlled, such as the integration of 6G networks for ubiquitous connectivity and the application of quantum computing to solve intractable optimization problems. The proposed ordering matters because emerging technologies influence CETCI differently: some extend current design space, whereas others are only valuable after the present deployment bottlenecks are reduced.

\subsubsection{Advanced Resource Management}\label{sec8.2.1}
Current resource management strategies often rely on static or reactive allocation, which struggles to adapt to the highly dynamic and heterogeneous nature of CETCI environments. The primary limitation lies in the inability to handle extreme scale and fluctuating demands in real-time without latency penalties. A critical open problem is developing adaptive algorithms that can jointly optimize computation, storage, and networking resources across domain boundaries while respecting trust and energy constraints. Potential solutions include leveraging DRL for predictive resource provisioning \cite{xu2024fusion} and integrating DT to simulate and optimize resource flows before execution \cite{bujari2023layered}. Furthermore, trust-based mechanisms \cite{xiao2024domainspecific} and energy-aware strategies \cite{li2024energyaware} must be embedded into these algorithms to ensure sustainable and secure cross-domain collaboration.

\subsubsection{Security and Privacy}\label{sec8.2.2}
{Traditional security perimeters are dissolved in distributed CETCI architectures, exposing data and models to sophisticated attacks across the continuum. A major bottleneck is securing collaborative learning paradigms, such as FL, against model poisoning and inference attacks without compromising model utility or incurring prohibitive communication overhead. The key open research question is how to design lightweight, robust security protocols that scale with the number of heterogeneous devices. Promising approaches include differential privacy (DP)-enabled FL mechanisms that balance privacy budgets with model accuracy \cite{yang2024differentially}, and secure multi-party computation (SMPC) for encrypted collaborative analysis \cite{dhinakaran2024privacypreserving}. Additionally, blockchain-based decentralized trust models \cite{chi2024trusted} and verifiable computing techniques like B-OPRF \cite{yang2024flexible} offer pathways to ensure data integrity and auditability in trustless environments.}

\subsubsection{Standardization}\label{sec8.2.3}
{The fragmentation of communication protocols, data formats, and software interfaces presents a significant barrier to seamless CETCI integration. Current limitations stem from the lack of semantic interoperability between diverse AIoT devices and cloud platforms, leading to data silos and inefficient collaboration. An urgent open problem is establishing unified standards that accommodate the varying resource constraints of terminal devices while supporting complex cloud-edge interactions. Potential technical solutions involve extending lightweight protocols like MQTT and CoAP with semantic descriptions for automated service discovery \cite{sowinski2023autonomous}. Furthermore, developing standardized APIs for cross-layer application deployment and leveraging blockchain for immutable, standardized audit trails of data exchange \cite{chi2024trusted} are critical for fostering a truly interoperable ecosystem.}

\subsubsection{Explainable AI}\label{sec8.2.4}
{The deployment of complex ``black box'' deep learning models in critical AIoT applications faces challenges regarding transparency and user trust. In distributed or split models, the main limitation lies in the difficulty of interpreting decisions when computation is fragmented across layers. The central open problem is developing Explainable AI (XAI) frameworks that can provide coherent explanations for decisions derived from collaborative inference processes. To address this, researchers should explore model-agnostic XAI techniques, such as SHAP or LIME, adapted for distributed environments \cite{sedlak2024equilibrium}. Additionally, federated XAI frameworks that aggregate local explanations into a global interpretation and layer-wise relevance propagation methods \cite{zhou2024enhancing} offer potential pathways to demystify AI decision-making in healthcare and autonomous systems.}

\subsubsection{6G/Next-Generation Networks}\label{sec8.2.5}
As a long-term direction, the integration of 6G will transform CETCI by creating a unified network fabric. A fundamental open question is how to build computing power orchestration protocols for SAGIN that treat computational resources as a native network function. Addressing this challenge requires moving beyond current resource management paradigms to develop a joint communication, sensing, and computation optimization framework. Research should focus on creating intent-based networking protocols that can dynamically allocate and schedule computational tasks across satellites, UAVs, and terrestrial edge nodes based on application-level requirements \cite{masaracchia20256genabled}. Furthermore, integrated sensing and communication will provide the fine-grained context needed for such proactive orchestration \cite{wang2024cooperative}. Within CETCI, 6G is therefore best viewed as an enabler of tighter communication-computation coupling rather than a standalone connectivity upgrade, because its value lies in shrinking the orchestration gap between infrastructure decisions in \cref{sec3} and collaborative-learning requirements in \cref{sec4}.

\subsubsection{Digital Twin}\label{sec8.2.6}
DT technology faces challenges related to synchronization latency and the high computational cost of maintaining high-fidelity virtual replicas. The limitation is the difficulty of achieving real-time synchronization between physical entities and their digital counterparts in bandwidth-constrained AIoT networks. A critical open problem is ensuring the fidelity and timeliness of DTs without overwhelming network resources. Technical solutions include developing edge-native lightweight DTs that prioritize critical state updates \cite{chen2024fast} and employing event-triggered synchronization mechanisms to minimize data transmission. Additionally, leveraging transfer learning for rapid DT evolution and condition-adaptive dynamic evolution methods \cite{li2024conditionadaptive} can enhance the adaptability and efficiency of DT-assisted collaborative computing. For CETCI, DT is most useful when it feeds back into resource prediction, offloading rehearsal, and model-update timing, thereby coupling architectural control with collaborative learning rather than remaining a visualization layer.

\subsubsection{LLMs and Agents}\label{sec8.2.7}
LLM- and agent-enabled CETCI remains an important near-term direction. One open problem is how to perform distributed collaborative inference for LLMs on resource-constrained edge and terminal devices. Dynamic split inference across the cloud-edge-terminal continuum \cite{xu2024when}, quantization \cite{shen2024agilequant}, and parameter-efficient fine-tuning can reduce the memory and latency burden of edge deployment. Beyond compressed LLM execution, recent Agentic AI research suggests a broader opportunity in which cloud-hosted foundation models provide global planning, edge agents perform local context fusion and bounded execution, and terminal agents expose sensing or actuation primitives under safety policies \cite{jiang2026agentic}. The main design implication is that CETCI should coordinate split inference, tool use, verification, memory synchronization, and privacy-aware fallback control rather than deploy monolithic model stacks, because those coordination functions determine whether agentic behavior remains within latency, trust, and energy budgets at scale.

\subsubsection{Quantum Computing}\label{sec8.2.8}
{In the long term, quantum computing offers the potential to revolutionize resource optimization in CETCI. The critical open research question is defining the specific application path for quantum algorithms to solve large-scale, NP-hard resource scheduling problems that are intractable for classical computers. Specifically, this endeavor involves identifying which classes of resource allocation and task offloading problems can be mapped to quantum optimization algorithms, such as the quantum approximate optimization algorithm or quantum annealing. Future work should focus on developing hybrid quantum-edge-cloud computing architectures \cite{hossain2024quantumedge}, where classical edge systems formulate the optimization problem and offload the computationally intensive part to a quantum processor. Furthermore, quantum-enhanced machine learning could provide exponential speedups for training complex resource management policies \cite{paul2025quantumenhanced}.}

\subsection{Lessons Learned}\label{sec8.3}
The central takeaway is that CETCI system design should jointly consider resource volatility, certification needs, communication planning, model partitioning, and privacy budgeting rather than optimizing these concerns independently. The survey evidence across \cref{sec3,sec4,sec5,sec7} also indicates that authoritative technical synthesis in CETCI comes from comparing methods under shared deployment criteria and then tracing how those criteria change from one application domain to another. The \textit{supplementary material} develops the same lesson in greater detail in \cref{sec:supp_lessons}.

\section{Conclusion}\label{sec9}
In this survey, we have provided a comprehensive examination of cloud-edge-terminal collaborative intelligence (CETCI) in AIoT networks, moving beyond traditional architectural reviews to focus on the core mechanisms that enable true end-to-end collaboration. By dissecting CETCI into two synergistic dimensions (intelligent resource management as AI for Collaboration, and collaborative learning as Collaboration for AI), we have presented a holistic framework that highlights the interplay between infrastructure optimization and distributed model evolution. Our analysis systematically reviewed the learning-based strategies for task offloading and resource allocation, which are critical for managing the dynamic and heterogeneous nature of AIoT systems. Furthermore, we delved into collaborative learning paradigms, such as federated and distributed learning, highlighting their role in building intelligent systems while addressing privacy and resource constraints.
By showcasing real-world applications and identifying key challenges and future directions, including the integration of LLMs, agentic workflows, digital twins, and 6G, this survey serves as a guide for researchers aiming to develop the next generation of robust, efficient, and truly intelligent collaborative AIoT systems.

\ifCLASSOPTIONcaptionsoff
  \newpage
\fi

\bibliographystyle{IEEEtranN}
\begin{spacing}{0.86}
  \bibliography{refs.bib}
\end{spacing}

\end{document}


\title{\textit{Supplementary Material}\\A Survey on Cloud-Edge-Terminal\\Collaborative Intelligence in AIoT Networks}

\maketitle

\markboth{IEEE COMMUNICATIONS SURVEYS \& TUTORIALS}%
{Wu \MakeLowercase{\textit{et al}}: Supplement to A Survey on Cloud-Edge-Terminal Collaborative Intelligence in AIoT Networks}

\section{Overview}\label{sec:supp_0}

In this supplementary material, we provide more detailed coverage of key cloud-edge-terminal collaborative intelligence (CETCI) components that are not fully addressed in the main paper due to page limitations. We have moved comprehensive explanations of 1) enabling technologies and performance metrics, 2) optimization techniques for task offloading and resource allocation, 3) data management and communication protocols, 4) security mechanisms and privacy-preserving techniques, 5) specific application domains, and 6) the extended technical challenges summarized in the main manuscript. Specifically, \cref{sec:supp_tech} details enabling technologies and performance metrics. \cref{sec:supp_3} covers mathematical and learning-based optimization techniques including linear programming, convex optimization, and game theory approaches. \cref{sec:supp_2} provides comprehensive analysis of data management and communication protocols in distributed AIoT environments. \cref{sec:supp_1} presents detailed security mechanisms and privacy-preserving techniques for collaborative intelligence systems. \cref{sec:supp_4} showcases specific applications in smart cities and smart agriculture domains, while \cref{sec:supp_challenges} provides the expanded challenge discussion and challenge overview figure.

\section{Enabling Technologies and Performance Metrics}\label{sec:supp_tech}

\subsection{Enabling Technologies}\label{sec:supp_tech.1}
\subsubsection{Network Virtualization}
Resource abstraction revolutionizes network management by creating logical networks independent of physical infrastructure.  Virtual network components enable flexible resource allocation while supporting multiple isolated network slices simultaneously \cite{maia2024survey}.  Decoupling the virtual resources from the underlying physical infrastructure offers flexibility in resource allocation and management.  Specifically, mapping virtual to physical resources introduces a key optimization problem, formally defined as $\min_{\mathbf{M}} f(\mathbf{M}, \mathbf{R}_v, \mathbf{R}_p)$, where $\mathbf{M}$ is the mapping matrix, $\mathbf{R}_v$ represents VN resource requirements, and $\mathbf{R}_p$ denotes available physical resources \cite{wang2025efficient}.  The objective function $f(\cdot)$ aims to minimize metrics such as resource utilization or energy consumption. Consequently, network virtualization facilitates efficient resource sharing among multiple VNs, enhancing resource utilization and reducing costs \cite{babou2024distributed}.  In AIoT systems, network virtualization enables dynamic resource provisioning and scaling, adapting to fluctuating application demands.  Network slicing, a crucial application of network virtualization in AIoT \cite{khalafi2024network}, creates dedicated, isolated virtual networks tailored to specific application or service needs, thereby improving resource allocation and enhancing security and privacy.  For cloud-edge-terminal collaboration, network virtualization enables seamless integration and interoperability between architectural layers \cite{sabbioni2024serverless}, facilitating dynamic resource allocation across the cloud, edge, and terminal devices to optimize performance and reduce latency.  By abstracting the complexities of the underlying physical infrastructure, network virtualization simplifies AIoT application deployment and management \cite{ozyar2022decentralized}, allowing developers to focus on application logic and accelerating solution development.  Additionally, it enhances network reliability and resilience by providing redundancy and fault tolerance \cite{raj2021reliable}, enabling dynamic remapping of virtual networks to available resources in case of physical resource failure.

\subsubsection{Container Orchestration}
Application lifecycle automation across distributed infrastructure relies on container orchestration platforms.  Declarative management approaches simplify deployment complexity while ensuring consistent application behavior across heterogeneous environments \cite{bogo2020componentaware}. Notably, the declarative model simplifies application management and enhances portability across diverse cloud and edge environments.  Central to container orchestration is cluster scheduling, which efficiently places containers onto available nodes within a cluster \cite{rodriguez2020container}.  Given a set of containers $\mathcal{C}$ and nodes $\mathcal{N}$, cluster scheduling seeks a mapping function $f: \mathcal{C} \rightarrow \mathcal{N}$ that optimizes objectives like minimizing resource utilization or maximizing performance, subject to constraints such as resource capacity, application dependencies, and network latency.  Autoscaling dynamically adjusts container instances based on real-time demand by monitoring performance metrics and triggering scaling actions based on predefined thresholds \cite{zhong2022machine}.  Networking solutions within container orchestration systems establish virtual networks, assign IP addresses, and manage traffic routing, often leveraging SDN principles for dynamic configuration and optimization \cite{liu2023enabling}.  Efficient container image distribution, particularly in bandwidth-constrained edge environments, can be achieved through peer-to-peer (P2P) sharing and optimized image layer structures \cite{becker2021edgepier, feng2024break}.  Resource allocation algorithms distribute resources like CPU, memory, and storage among containers, considering application requirements, resource availability, and performance goals, with particular attention to shared resources for real-time applications \cite{monaco2023extensions}.  Integrating container orchestration with lightweight approaches like K3s enhances hybrid deployments across resource-constrained edge nodes and resource-rich cloud environments \cite{wang2021integration, rodriguez2019containerbased, wang2023container}.

\subsubsection{Software-Defined Networking}
Centralized network control emerges through SDN's separation of control and data planes.  Programmable network management enables dynamic traffic optimization while supporting flexible policy enforcement across distributed systems \cite{nunez2023brief}.  An SDN architecture comprises a controller $\mathcal{C}$, a set of switches $\mathcal{S}$, the network topology $\mathcal{N}$, and a set of applications $\mathcal{A}$. The controller interacts with the switches via a southbound interface, typically OpenFlow \cite{ali2020detecting}, using a standardized protocol with a message type and payload for instructions.  Consequently, the controller dynamically manages network traffic by installing flow entries, defined by matching traffic characteristics and corresponding actions \cite{moraes2019pub}. In addition to dynamic flow management, SDN facilitates network virtualization \cite{cardoso2023softwaredefined}, creating virtual networks ($\mathcal{S}_v$, $\mathcal{N}_v$) on the physical infrastructure for efficient resource allocation and isolation \cite{gedia2018netoapp}.  Enhanced security is achieved through centralized security policy enforcement and standardized management of authentication, authorization, and accounting (AAA) infrastructures \cite{ahmad2020improving, lopezgomez2025sdnaaa}.  Furthermore, traffic engineering and load balancing capabilities optimize data flow across multiple network interfaces, improving network resource utilization and AIoT application performance, especially in multi-homed environments \cite{alnajjar2020network}.

\subsection{Performance Metrics}\label{sec:supp_tech.2}
\subsubsection{Response Time}
Real-time responsiveness demands stringent latency control across distributed computing layers.  Response time optimization becomes critical for time-sensitive applications where millisecond delays can significantly impact system performance \cite{abouaomar2021resource}. Latency, the delay between request initiation and response commencement, and response time, the duration from request initiation to complete fulfillment, are key performance indicators.  In a cloud-edge-terminal collaborative system, total latency ($L_{total}$) comprises processing latencies on the terminal ($L_{term}$), edge ($L_{edge}$), and cloud ($L_{cloud}$) servers, along with network transmission latency ($L_{net}$) as: $L_{total} = L_{term} + L_{edge} + L_{cloud} + L_{net}$ \cite{hao2019smartedgecocaco}.

CETCI enables low-latency data processing and real-time responsiveness by distributing computational tasks strategically \cite{raj2021reliable, gholipour2023tpto}. Specifically, processing data closer to the source (at the edge or terminal) reduces transmission delays to the cloud. Response time $R_{total}$ is defined as $R_{total} = L_{total} + P_{proc}$, where $P_{proc}$ is the task processing time.  Within a collaborative framework, response time depends on the task allocation strategy; offloading computationally intensive tasks to the cloud while assigning less demanding tasks to the edge or terminal can significantly reduce $R_{total}$ \cite{xu2024fusion}.  Consequently, efficient task offloading and resource allocation are essential for optimizing response time and ensuring timely service delivery \cite{zhou2024enhancing}, especially given the dynamic nature of AIoT environments, which necessitates adaptive strategies for varying workloads and resource availability \cite{rao2021eco}.

\subsubsection{Energy Consumption}
Power efficiency constraints drive sustainable design practices across resource-limited computing environments.  Energy optimization strategies must balance computational performance with battery life and operational costs while maintaining system reliability \cite{arroba2024sustainable}.  The total energy consumption ($E_{total}$) is the sum of energy consumed by the cloud, edge, and terminal layers, denoted as $E_{cloud}$, $\sum_{i=1}^{N_{edge}} E_{edge,i}$, and $\sum_{j=1}^{N_{term}} E_{term,j}$, respectively, where $N_{edge}$ and $N_{term}$ represent the number of edge nodes and terminal devices. Each term includes energy used for computation, communication, and storage. Because component energy consumption depends on utilization \cite{chen2024fast}, the energy consumed by an edge node $i$ can be modeled as $E_{edge,i} = P_{idle,i} \cdot T_{idle,i} + P_{active,i} \cdot T_{active,i} + E_{comm,i}$, where $P_{idle,i}$ and $P_{active,i}$ represent power consumption in idle and active states, $T_{idle,i}$ and $T_{active,i}$ denote the corresponding time durations, and $E_{comm,i}$ is the communication energy.

Cloud-edge-terminal collaboration is key to balancing computational load and reducing energy consumption \cite{wu2022pecco}. Offloading computationally intensive tasks from terminals to edge servers or the cloud reduces terminal energy consumption, but introduces communication overhead.  Therefore, intelligent task offloading strategies are essential for minimizing overall system energy.  Optimizing energy consumption requires considering task characteristics, resource availability, network conditions, and device energy profiles. Dynamic resource allocation, such as putting underutilized edge servers into sleep mode, enhances energy efficiency \cite{li2024energyaware}.  In addition, using renewable energy sources at the edge and cloud reduces the carbon footprint of AIoT systems \cite{zhai2024edgecloud}.  Federated learning, a distributed learning paradigm, reduces communication overhead and thus energy consumption by training models locally and aggregating updates at the edge or cloud \cite{zhou2024enhancing}.

\subsubsection{Resource Utilization}
Optimal resource allocation maximizes computational efficiency while minimizing operational overhead across heterogeneous infrastructure.  Strategic workload distribution enables cost-effective utilization of available computing, storage, and network resources \cite{tocze2018taxonomy}.  Overall resource utilization $U_{total}$ can be defined as the ratio of the sum of used resources $U_{r}$ to the sum of total available resources $A_{r}$ across $N_{res}$ resources, such as CPU, memory, and bandwidth. Maximizing $U_{total}$ requires strategically distributing workloads and data across the cloud, edge, and terminal layers.  Specifically, allocating $M_{task}$ tasks to $K_{node}$ computing nodes across this continuum involves minimizing the total execution cost, subject to constraints like deadlines, resource availability, and energy limits \cite{wang2024efficient}.  Consequently, the allocation problem, where $c_{task,node}$ represents the cost of executing task $i$ on node $j$, can be effectively addressed using optimization techniques like linear programming and deep reinforcement learning.  In addition, computation reuse, quantified by the reuse rate ($R_{comp}$), optimizes resource utilization by avoiding redundant computations and reducing latency \cite{nour2024networkbased}.  Energy-aware resource allocation and collaborative strategies like federated learning further enhance efficiency by minimizing energy consumption and sharing computational loads \cite{hua2024energyefficient, zhou2024enhancing}.  As a result, mechanisms contribute to both cost-effectiveness and high performance in distributed systems, addressing challenges in sustainable edge computing, including energy efficiency and fault tolerance \cite{arroba2024sustainable}.  Ensuring secure and reliable resource sharing across different domains also necessitates trust-based collaboration mechanisms \cite{li2024energyaware}.

\subsubsection{Quality of Service}
Service reliability guarantees require comprehensive performance monitoring across distributed system components.  QoS metrics encompass latency, throughput, availability, and reliability to ensure consistent application behavior under varying operational conditions \cite{tuli2023ai}. Different AIoT applications have distinct QoS requirements. For instance, autonomous driving systems require low latency for real-time decision-making, while environmental monitoring emphasizes high throughput and data integrity. Cloud-edge-terminal collaboration ensures consistent service delivery by dynamically adapting to application requirements and network conditions \cite{bisicchia2024continuous}. Adaptability stemming from the distributed architecture enables flexible resource allocation and task offloading. Specifically, QoS is a function of resource allocation $R_{alloc}$ and task offloading decisions $T_{off}$, represented as $Q_{total} = f(R_{alloc}, T_{off})$. CETCI optimizes this function by dynamically adjusting $R_{alloc}$ and $T_{off}$ based on real-time conditions and application needs. Service level agreements (SLAs) define specific performance targets, expressed as constraints on the QoS vector: $Q_{total} \in \mathcal{C}_{feasible}$, where $\mathcal{C}_{feasible}$ represents the feasible QoS region \cite{sedlak2024equilibrium}. Consequently, cloud-edge-terminal collaboration strives to maintain $Q_{total}$ within $\mathcal{C}_{feasible}$ by continuously monitoring performance and adapting resource allocation and task offloading strategies \cite{zhou2024enhancing, raj2021reliable}. CETCI collaborative approaches also enhance reliability and availability through redundancy and fault tolerance, distributing tasks and data across multiple nodes to ensure continuous service even with component failures.

\subsubsection{Quality of Experience}
User satisfaction measurement extends beyond technical metrics to encompass subjective interaction quality.  QoE evaluation considers usability, accessibility, and contextual factors that collectively determine overall user perception of system performance \cite{zhang2021empowering}.  QoE can be modeled as a function $Q_{exp} = f(Q_{obj}, U_{pref}, C_{con})$, where $Q_{obj}$ represents objective QoS parameters (e.g., latency, throughput), $U_{pref}$ denotes subjective user preferences, and $C_{con}$ encompasses contextual factors like device capabilities and network conditions \cite{li2024energyaware}. In cloud-edge-terminal collaborative systems, QoE is particularly important due to the system's distributed nature and the diverse application requirements.  Specifically, CETCI collaboration enhances QoE by distributing tasks and resources across different layers, where the overall QoE can be expressed as $Q_{etotal} = \alpha_{cloud} Q_{ecloud} + \alpha_{edge} Q_{eedge} + \alpha_{term} Q_{eterm}$, with weighting factors $\alpha_{cloud}$, $\alpha_{edge}$, and $\alpha_{term}$ reflecting the contribution of each layer \cite{li2024energyaware}.  Processing data closer to the source (at the edge or terminal) minimizes transmission delays and improves real-time performance, which is critical for applications like autonomous driving and interactive gaming \cite{raj2021reliable}.  Additionally, context-aware adaptation of processing and communication strategies, formulated as $A_{adapt}(t) = g(N_{net}(t), D_{device}(t), A_{req}(t))$ where $N_{net}(t)$, $D_{device}(t)$, and $A_{req}(t)$ represent network conditions, device capabilities, and application requirements at time $t$, further enhances QoE by ensuring seamless and efficient interactions \cite{zhou2024enhancing}.  Furthermore, such adaptive mechanisms allow the system to optimize resource utilization while maintaining desired QoE levels, and also address privacy concerns by enabling local data processing, thereby increasing user trust \cite{yang2024differentially, liu2024quasyncfl}.

\section{Optimization Techniques in Intelligent Resource Management}\label{sec:supp_3}

Mathematical and learning-based optimization approaches address the complex challenges of task offloading and resource allocation in distributed computing environments. The following subsections provide detailed coverage of three fundamental optimization paradigms: linear programming, convex optimization, and game theory. Each approach offers unique advantages for solving resource management problems in cloud-edge-terminal collaborative systems. \cref{tab:4} summarizes representative optimization techniques across different algorithm types and scenarios, demonstrating their performance improvements in metrics such as response time, latency, and energy consumption.

\input{t4_opt_3.3.tex}

\subsection{Linear Programming}\label{sec:supp_3.1}
Constrained optimization under linear relationships enables tractable solutions for resource allocation problems. LP formulations transform complex distribution challenges into mathematically solvable frameworks with guaranteed optimal solutions \cite{bandyopadhyay2024delaysensitive}. For example, the objective function might minimize total system cost, encompassing energy consumption at terminal and edge layers~\cite{zeng2024joint}, computation costs at the cloud and edge~\cite{he2024efficient}, and network communication costs~\cite{nguyen2024integrated}. Constraints can represent resource limitations, such as bandwidth~\cite{gan2024optimal}, processing capacity at each layer~\cite{choudhary2024enhancing}, and latency~\cite{laboni2024hyper}. LP also supports various task offloading strategies, including decisions on offloading locations (edge or cloud) and resource allocation~\cite{qiu2024online}. Additionally, incorporating predicted future system states into the LP formulation enables proactive resource management in dynamic AIoT environments~\cite{lu2024dynamic}.

\subsection{Convex Optimization}\label{sec:supp_3.2}
Nonlinear optimization with convergence guarantees addresses complex resource distribution challenges. Convex programming techniques enable globally optimal solutions through gradient-based methods and interior-point algorithms \cite{nguyen2024integrated}. As an example, resource allocation might aim to minimize energy consumption under latency and throughput constraints. If the constraints and the objective function are convex, gradient descent can iteratively update resource allocation based on the objective function's gradient, while interior-point methods traverse the feasible region for optimal solutions \cite{feng2024dependencyaware}. Several studies have explored convex optimization techniques, including \cite{khalafi2024network}, which uses online convex optimization for network slicing in edge-cloud orchestrated networks to minimize system costs. Similarly, \cite{chen2022resource} employs convex optimization for resource and power allocation in cloud-edge collaboration to minimize user overhead. In addition, \cite{fang2024joint} uses successive convex approximation (SCA) for resource optimization in NOMA-aided cloud-edge-terminal cooperation, and \cite{bai2024delayaware} applies convex approximation to a delay minimization problem in multi-UAV enabled edge-cloud cooperation. However, formulating such convex programs often requires simplifications to address the complexity and dynamic nature of cloud-edge-terminal environments.

\subsection{Game Theory}\label{sec:supp_3.3}
Competitive and cooperative interaction modeling provides equilibrium-based optimization frameworks. Strategic decision-making among rational participants enables stable resource allocation through concepts like Nash equilibrium and Stackelberg leadership \cite{yu2024game}. For instance, \cite{wu2024computation} formulates a non-cooperative game for distributed cost minimization in multi-cell networks, demonstrating game theory's applicability to computation offloading in edge-cloud collaboration. Similarly, \cite{zhang2024stackelberggamebased} uses a Stackelberg game to model user, base station, and cloud interactions in multi-user multi-task offloading for mobile edge computing. Beyond resource allocation, game theory also addresses security, as exemplified by \cite{dragotto2024critical}, which introduces the critical node game for assessing cybersecurity risks in cloud networks. Additionally, \cite{duan2024binary} proposes a game-theoretic algorithm for optimizing system completion time and energy consumption in cloud robot task offloading. Cooperative solutions, such as the Nash bargaining solution in \cite{shams2025joint} for joint latency-energy minimization in fog-assisted networks, offer alternative approaches for various collaboration paradigms. For instance, examples, including task offloading \cite{hazra2024distributed}, resource allocation \cite{fang2023large}, and collaborative computation \cite{zhu2024syroc}, highlight game theory's utility in optimizing resource management for cloud-edge-terminal collaboration.

\section{Data Management and Communication}\label{sec:supp_2}
\subsection{Data Storage and Retrieval}\label{sec:supp_2.1}
\subsubsection{Edge Caching}\label{sec:supp_2.1.1}
Proximity-based storage strategies reduce access latency through intelligent content placement near end-users \cite{antonogiorgakis2019view}. Popular data items stored at network edges increase cache hit rates and minimize retrieval delays. Result caching extends this concept by storing computation results, which is particularly advantageous for computationally intensive tasks \cite{wang2020survey}. Similarly, predictive caching anticipates future requests based on historical patterns and preemptively stores data, further minimizing latency \cite{li2023predictive, torabi2024learningbased}. For example, PEC employs deep learning for content popularity prediction and prefetching \cite{li2023predictive}, while a D3QN-based scheme dynamically evaluates content importance for efficient caching and eviction in dynamic networks \cite{zhang2024how}. In vehicular networks, edge collaborative caching, aided by tabu search algorithms, prioritizes high-traffic data and ensures backup within the collaborative domain to reduce latency and prevent single-point failures \cite{zheng2024optimization}.

\subsubsection{Distributed Data Storage}\label{sec:supp_2.1.2}
Multi-tier storage architectures enable efficient data distribution across cloud and edge infrastructures \cite{luckow2021pilotedge}. Proximity-based storage reduces latency for real-time applications while enhancing fault tolerance through geographical distribution \cite{raj2021reliable}. Distributing data across multiple nodes enhances fault tolerance, ensuring system reliability even with individual node failures \cite{chi2024trusted}. Strategic data placement, based on access patterns, caches frequently accessed data at the edge for faster retrieval, while less frequently used data resides in the cloud \cite{becker2020aiops}. Data partitioning techniques divide large datasets into smaller chunks distributed across multiple nodes, facilitating parallel processing and improving system performance \cite{yang2024differentially}. Additionally, essential data synchronization mechanisms maintain consistency and data integrity across distributed locations, preventing conflicts from concurrent updates \cite{zhou2024enhancing}. Security considerations are addressed through encryption techniques for data confidentiality and access control mechanisms to restrict unauthorized access \cite{li2024computation}.

\subsubsection{Data Integrity and Availability}\label{sec:supp_2.1.3}
Comprehensive protection mechanisms ensure information authenticity and continuous accessibility across distributed networks \cite{dui2024maintenance}. Trustworthy decision-making requires robust data integrity, exemplified by blockchain applications for enhanced security \cite{chi2024trusted}. Addressing authentication and preventing model poisoning attacks, as discussed in \cite{javed2018cefiot} and \cite{yang2024differentially}, respectively, are also critical. Data availability, ensuring data accessibility, can be impacted by fluctuating resources and bandwidth in dynamic edge networks, especially for time-sensitive applications \cite{chen2023remr}. Efficient data storage and retrieval, including edge caching techniques \cite{raj2021reliable}, are thus essential for improving data access latency and enhancing availability. Similarly, distributed data storage across multiple nodes, as highlighted in the CEFIoT architecture \cite{javed2018cefiot}, improves fault tolerance and resilience.

\subsection{Communication Protocols and Optimization}\label{sec:supp_2.2}
Efficient data exchange is crucial for distributed collaborative systems. We explore communication protocols and optimization techniques that enable effective data transmission. Established protocols like MQTT and CoAP enable efficient data transmission in complex environments. Furthermore, bandwidth optimization methods, including traffic scheduling, rate adaptation, and link aggregation, maximize data throughput and minimize latency. Additionally, fault tolerance mechanisms, such as error detection and recovery, connection management, and service monitoring, ensure reliable communication, contributing to overall performance and stability.

\subsubsection{Communication Protocols}\label{sec:supp_2.2.1}
Resource-aware messaging protocols optimize data exchange efficiency in constrained CISAIoT environments. MQTT~\cite{paolo2023security} and CoAP~\cite{gundogan2020iot} exemplify lightweight approaches designed for low-overhead communication. MQTT, a publish-subscribe protocol with low overhead and diverse QoS levels, suits resource-constrained environments and various AIoT applications~\cite{giambona2018mqtt}. Open-source MQTT broker implementations, such as Mosquitto and EMQX, are readily available and have been experimentally benchmarked in edge computing~\cite{dizdarevic2023engineering}. CoAP, specifically designed for resource-constrained devices and networks, offers a RESTful interface for interaction with web services and has been explored with security mechanisms like OSCORE~\cite{gundogan2020iot}. In addition, network coding enhances data integrity and reduces redundancy in AIoT communication. For instance, integrating network coding into CoAP's block-wise transfer reduces retransmissions in lossy networks~\cite{phung2020enhancing}, while secure network coding schemes like GS-SNC~\cite{fang2024secure} enhance security and efficiency by protecting against attacks.

\subsubsection{Bandwidth Optimization Techniques}\label{sec:supp_2.2.2}
Network resource management strategies maximize throughput while minimizing latency across distributed systems \cite{li2024cloudedgedevice}. Traffic scheduling prioritizes critical data flows to optimize bandwidth utilization, particularly for real-time applications requiring immediate response \cite{wang2024cloudedgeend}. Rate adaptation, in contrast, dynamically adjusts transmission rates based on network conditions \cite{xu2024fusion}. Specifically, it reduces the rate during congestion and increases it when bandwidth is available. Link aggregation, similarly, combines multiple network links into a single logical link, thereby increasing capacity and resilience against link failures \cite{nezami2021decentralized}. Notably, this approach enhances efficiency for bandwidth-intensive applications.

\subsubsection{Fault Tolerance and Quality Assurance}\label{sec:supp_2.2.3}
Resilient communication infrastructure requires comprehensive error handling and performance monitoring mechanisms \cite{nikolic2021selfhealing}. Fault tolerance employs error detection techniques to identify disruptions, data corruption, or device malfunctions, while recovery mechanisms restore functionality through data replication or task rescheduling \cite{javed2018cefiot}. Redundancy management, often achieved through redundant hardware or communication paths, ensures data availability and service continuity \cite{dui2024maintenance}. Connection management strategies optimize communication performance through link monitoring to identify connectivity issues, failover strategies for seamless transition to backup paths \cite{dui2024maintenance}, and load balancing to distribute traffic and prevent congestion \cite{wang2024cloudedgeend}. In addition, quality assurance practices guarantee reliable service delivery. Specifically, service monitoring tools collect performance data, enabling identification of bottlenecks, while performance optimization techniques, such as bandwidth allocation or data compression, improve efficiency and reduce latency \cite{arroba2024sustainable, zhou2024enhancing}. SLA management ensures that the system meets predefined performance targets. Nonetheless, practices enhance reliability by minimizing errors and reducing latency, the potential trade-off between performance optimization and energy consumption in resource-constrained AIoT environments requires careful consideration \cite{wang2024cloudedgeend}.

\section{Security Mechanisms and Privacy-Preserving Techniques}\label{sec:supp_1}

\subsection{Security Mechanisms and Solutions}\label{sec:supp_1.1}
Protecting collaborative intelligence paradigms against various threats requires robust security mechanisms. We explore three key security solutions: encryption and authentication, intrusion detection and prevention, and blockchain technologies. Encryption and authentication techniques safeguard data transmission and storage across distributed architectures. Additionally, intrusion detection and prevention systems proactively identify and mitigate security breaches. Finally, blockchain technologies offer potential enhancements to security and trust establishment within collaborative environments.

\subsubsection{Encryption and Authentication}\label{sec:supp_1.1.1}
Cryptographic protection mechanisms secure data transmission and storage across distributed computing layers. Strong encryption protocols combined with robust authentication systems prevent unauthorized access while maintaining communication efficiency \cite{wang2018secure}. Lightweight symmetric encryption secures data exchange between resource-constrained devices and edge nodes, while CP-ABE~\cite{xiao2024domainspecific} offers fine-grained access control based on specific criteria, addressing the varying data sensitivity in AIoT environments. Authentication verifies the identity of communicating entities, preventing unauthorized access and manipulation. For example, GS-SNC enhances security through scrambling and encoding, offering advantages in computational complexity and space overhead compared to other secure network coding approaches like DP-SNC and SPOC~\cite{fang2024secure}. Integrating blockchain technology strengthens security by providing immutable records of data transactions and access logs~\cite{chi2024trusted}, and homomorphic signatures enable efficient third-party integrity auditing without accessing raw data~\cite{yao2024efficient}.

\subsubsection{Intrusion Detection and Prevention}\label{sec:supp_1.1.2}
Proactive threat monitoring systems identify and neutralize malicious activities before system compromise. Real-time anomaly detection enables rapid response to security incidents across distributed computing infrastructure \cite{zhou2024enhancing}. For example, unusual data flows between the edge and cloud, or sudden spikes in data transmission at the terminal layer, can signal intrusions \cite{yigit2023digital}. Deploying IDPS at different architectural layers, such as the edge or cloud, provides comprehensive security coverage \cite{saezdecamara2023clustered}. Edge-based IDPS enables faster responses to localized threats by analyzing data near the source, while cloud-based IDPS leverages centralized resources to identify complex attack patterns \cite{chi2024trusted}. In addition to detection, IDPS \cite{raj2021reliable} incorporates prevention mechanisms, including blocking malicious traffic and isolating compromised devices, to actively mitigate threats.

\subsubsection{Blockchain Technologies}\label{sec:supp_1.1.3}
Decentralized trust mechanisms eliminate single points of failure while ensuring immutable transaction records. Distributed consensus protocols enhance system resilience and provide transparent, tamper-proof data management \cite{blasch2019blockchain}. Crzyptographic hashing and consensus mechanisms, integral to blockchain's architecture, ensure data integrity and prevent unauthorized tampering \cite{rahman2023blockchainbased}, which is crucial for secure data aggregation and processing from numerous AIoT devices. Furthermore, blockchain facilitates secure access control and device authentication within the collaborative ecosystem \cite{chi2024trusted}. Through smart contracts, predefined rules and policies are automatically enforced, streamlining device interactions and ensuring compliance for activities such as sensitive data access and resource allocation across the cloud, edge, and terminal layers \cite{gadekallu2022blockchain}. Applications of blockchain integration in AIoT include secure data logging and auditing, decentralized avionics services, and trusted decision architectures \cite{pan2019edgechain}.

\subsection{Privacy-Preserving Techniques}\label{sec:supp_1.2}
Protecting user data within distributed collaborative systems is paramount. Four prominent techniques address privacy concerns: federated learning, differential privacy, homomorphic encryption, and secure multi-party computation. Federated learning enables collaborative model training without sharing raw data, while differential privacy adds noise to further protect individual data points. Homomorphic encryption allows computations on encrypted data without decryption, and secure multi-party computation enables joint computation on private inputs, revealing only the output.

\subsubsection{Federated Learning and Differential Privacy}\label{sec:supp_1.2.1}
Complementary privacy mechanisms combine to strengthen distributed learning frameworks \cite{rodriguezbarroso2020federated}. FL facilitates collaborative model training without direct data sharing \cite{li2023edgecloud}, where edge devices or terminals train local models and share only model updates (e.g., gradients) with a central server \cite{liu2024quasyncfl}. However, potential information leakage from the updates necessitates further privacy enhancements through DP \cite{li2019asynchronous}. Specifically, DP introduces calibrated noise to these shared updates, ensuring that the presence or absence of a single data point does not significantly affect the final model \cite{wang2022privacypreserving}, thereby hindering adversarial attempts to infer individual user data. Techniques like objective perturbation, adding noise to the objective function, achieve DP guarantees in FL \cite{yang2024differentially}. The privacy-utility trade-off is controlled by the privacy budget ($\epsilon$, $\delta$), exemplified by differentially private over-the-air FL where transmit power adjustments influence this trade-off \cite{zhang2024efficient}.

\subsubsection{Homomorphic Encryption}\label{sec:supp_1.2.2}
Computational privacy through encrypted data processing eliminates decryption requirements during analysis \cite{zhang2024efficient, wang2018secure}. Beyond traditional cryptographic approaches, HE enables secure data aggregation, model training, and inference across distributed devices and cloud servers. For example, edge devices can encrypt sensor readings using HE before transmission, ensuring data confidentiality during transit and storage, even if intermediary nodes are compromised \cite{kim2024privacy}. Consequently, cloud servers can perform computations on this encrypted data without decryption, enabling tasks like model training and inference on sensitive data \cite{fahina2022securing}. In collaborative learning, HE facilitates secure model updates by enabling encrypted aggregation of model parameters \cite{aloufi2019collaborative}. Hierarchical federated learning frameworks demonstrate the practical integration of HE with multi-level privacy protection, exemplified by digital twin-based anomaly detection systems in smart healthcare that utilize edge cloudlets for local model training while preserving patient data privacy through encrypted multi-party collaboration \cite{gupta2021hierarchical}. Additionally, verifiable encodings and authenticators built upon HE can ensure the correctness of cloud-based computations on encrypted data \cite{chatel2024verifiable}. Although the benefits are substantial, practical application of HE faces challenges related to computational overhead. However, ongoing research focuses on optimizing HE schemes, including using GPUs for accelerated computations and developing privacy-authority-aware compilers for efficient ciphertext management \cite{zhai2022accelerating}.

\subsubsection{Secure Multi-Party Computation}\label{sec:supp_1.2.3}
Joint computation over private datasets maintains input confidentiality across participating entities \cite{sengupta2022sprite}. Multiple parties (including cloud servers, edge nodes, and devices) can collaboratively compute functions over their combined data while preserving individual input secrecy, particularly crucial for sensitive information such as patient health records \cite{zhou2024enhancing} or manufacturing process data \cite{chi2024trusted}. For example, multiple hospitals could leverage MPC to train a shared machine learning model on patient data without directly sharing the raw data due to privacy regulations. Consequently, model updates can be jointly computed based on local datasets without revealing individual patient records. Similarly, MPC enhances security in task offloading and resource allocation within cooperative networks \cite{fang2024secure}, facilitating collaborative decision-making regarding resource utilization without disclosing individual constraints or priorities. Furthermore, integrating MPC with techniques such as federated learning \cite{yang2024differentially} and differential privacy \cite{duan2023privascissors} provides a more comprehensive privacy solution. For example, this integration enables secure aggregation of model updates or the addition of noise to computation outputs.

\section{Applications}\label{sec:supp_4}

\cref{tab:app_comparison} presents a comprehensive comparison of AIoT application domains in CETCI, highlighting the diversity of IoT devices, key system requirements, and primary roles of edge computing across different sectors. Smart manufacturing demands ultra-low latency and high reliability for real-time defect detection and robotic control, while smart transportation requires extreme reliability for cooperative perception and autonomous driving tasks. Healthcare applications prioritize privacy compliance and secure data aggregation, whereas smart cities balance real-time surveillance with delay-tolerant environmental monitoring. Smart agriculture, by contrast, emphasizes wide-area coverage and precision analytics for sustainable farming. The comparison demonstrates how CETCI adapts to domain-specific constraints while maintaining cross-layer collaboration principles.

\input{t7_app_comparison.tex}

\subsection{Smart Cities}\label{sec:supp_4.1}
Distributed collaboration significantly enhances the development of intelligent urban environments. Consequently, CETCI paradigms improve public safety, environmental monitoring, and urban planning within modern networks. For instance, they enhance smart surveillance and crime prevention, leading to improved security and faster response times. In addition, collaborative frameworks contribute to creating more sustainable and livable urban spaces by enabling advanced environmental monitoring and pollution control. They also optimize smart energy management and grid operations, ultimately improving energy efficiency and resource allocation in smart cities.

\subsubsection{Smart Surveillance and Crime Prevention}\label{sec:supp_4.1.1}
Real-time security monitoring leverages distributed processing to enhance public safety through intelligent video analytics \cite{blasch2019blockchain}. Large-scale surveillance data processing benefits from edge deployment of lightweight frameworks such as LISPS, reducing cloud dependency \cite{nikouei2019decentralized}. Additionally, distributing deep learning models, such as A-YONet, which combines YOLO and MTCNN aspects, across the cloud-edge-terminal continuum enhances multi-target detection and activity tracking in complex environments \cite{zhou2021deeplearningenhanced}. Furthermore, such collaborative approach improves surveillance system efficiency and accuracy while addressing real-time analysis challenges and limited edge computing resources, ultimately facilitating proactive crime prevention through timely suspicious activity identification and efficient resource allocation.

\subsubsection{Environmental Monitoring and Pollution Control}\label{sec:supp_4.1.2}
Sustainable urban management requires comprehensive environmental data collection and analysis across distributed sensor networks \cite{arroba2024sustainable}. City-wide edge devices monitor air quality, water conditions, and noise levels through real-time sensing \cite{raj2021reliable}. Edge computing's localized data processing reduces latency and bandwidth consumption compared to cloud-only solutions \cite{vo2022edge}, enabling faster identification of pollution hotspots and quicker responses to environmental incidents. For complex tasks like predictive modeling and control strategy development, cloud resources are leveraged \cite{zhou2024enhancing}. For example, machine learning models trained on historical and real-time data can predict future pollution levels, enabling proactive mitigation \cite{yang2024differentially}. Cloud platforms also facilitate data sharing and collaboration among stakeholders, including government agencies, research institutions, and citizens \cite{zhou2024enhancing}.

\subsubsection{Smart Energy Management and Grid Optimization}\label{sec:supp_4.1.3}
Grid efficiency improvements emerge through intelligent energy distribution and consumption optimization~\cite{zhai2024edgecloud}. Distributed data collection on energy usage and grid conditions informs real-time decision-making for resource allocation~\cite{hua2024energyefficient}. As a result, collaborative frameworks facilitate more efficient energy allocation, reducing waste and improving overall grid efficiency. In addition, by analyzing data from various sources, distributed collaboration enables predictive maintenance of grid infrastructure, minimizing downtime and improving grid reliability~\cite{wang2024cloudedgeend}. Integrating renewable energy sources is also supported through this framework~\cite{arroba2024sustainable, li2024energyaware}. Specifically, edge devices can manage the intermittent nature of renewable energy by intelligently distributing power based on real-time conditions and predicted demand, thereby enhancing grid stability and reducing reliance on fossil fuels.

\subsection{Smart Agriculture}\label{sec:supp_4.2}
Distributed collaboration offers transformative potential for intelligent and efficient farming practices. Precision agriculture benefits from real-time data processing and decision-making capabilities provided by collaborative frameworks. Additionally, collaborative systems play a crucial role in crop monitoring and yield prediction by leveraging distributed intelligence to analyze data from diverse sources, including sensors, drones, and satellites. Consequently, distributed intelligence analysis provides insights into crop health, growth patterns, and potential yields, enabling timely interventions and optimized resource allocation for increased agricultural productivity and sustainability.

\subsubsection{Precision Agriculture}\label{sec:supp_4.2.1}
Resource optimization in farming benefits from real-time sensor analytics and intelligent decision-making systems~\cite{kumar2021affordable}. Edge processing of IoT sensor data minimizes latency while enabling immediate responses to changing agricultural conditions~\cite{raj2021reliable}. The cloud provides computational power for complex tasks like crop yield prediction and disease detection~\cite{zhang2024efficient}, while federated learning enables distributed model training across farms without sharing sensitive data, enhancing privacy and model robustness~\cite{khuencheng2024integration}. Consequently, farmers gain actionable insights from real-time data analysis, optimizing resource allocation like water and fertilizer usage, and improving efficiency and sustainability~\cite{bassine2023recent}. For example, AI-powered systems analyze sensor data and imagery to detect early signs of crop stress, enabling timely interventions to prevent yield loss~\cite{xu2024fusion}.

\subsubsection{Crop Monitoring and Yield Prediction}\label{sec:supp_4.2.2}
Agricultural intelligence leverages multi-source data integration for enhanced crop management and productivity forecasting. Terminal IoT sensors collect real-time information on soil conditions, weather patterns, and growth stages~\cite{wang2025intelligent}. Edge devices preprocess and aggregate collected sensor data, reducing transmission volume to the cloud and enabling faster localized insights~\cite{chen2024fast}. Accordingly, edge computing facilitates real-time monitoring and control, allowing for prompt responses to changing conditions, such as adjusting irrigation or applying fertilizers based on real-time analysis~\cite{cao2022igrow}. The cloud layer, with its extensive computational resources, performs complex analyses, including ML-based yield prediction using historical data, weather forecasts, and real-time sensor data~\cite{huber2022extreme, cao2022corn}. Furthermore, predictive information supports farmers' decisions regarding harvesting, resource allocation, and market planning. Cloud-edge collaboration also enables ML model deployment and updates at the edge, adapting models to local conditions and improving accuracy~\cite{khuencheng2024integration}. Moreover, the collaborative approach improves agricultural productivity and promotes sustainable practices by optimizing resource utilization and minimizing environmental impact, such as optimizing harvesting logistics and reducing waste~\cite{chowdhury2023artificial, lin2023mmstvit}.

\begin{table*}[ht]
\caption{Comparison of CETCI deployment characteristics across five AIoT application domains.}
\label{tab:app_comparison}
\resizebox{\textwidth}{!}{%
\begin{tabular}{lp{3cm}p{3.5cm}p{3.5cm}p{3cm}p{2.5cm}}
\toprule
\textbf{Domain} & \textbf{Key IoT Devices} & \textbf{Key Requirements} & \textbf{Edge Role} & \textbf{Cloud Role} & \textbf{Primary Pattern} \\
\midrule
Smart Manufacturing & Industrial sensors, robots, PLCs & Ultra-low latency ($<$1\,ms), reliability & Real-time control, anomaly detection & Global scheduling, DT management & Edge-first closed loop \\
Smart Transportation & Cameras, LIDAR, V2X & Latency $<$10\,ms, safety-critical & Local collision avoidance, traffic flow & Route optimization, fleet management & Edge-first closed loop \\
Smart Healthcare & Wearables, medical sensors & Privacy, regulatory compliance & Vital monitoring, local diagnosis & Longitudinal training, research aggregation & Privacy-preserving cloud aggregation \\
Smart Cities & Cameras, env. sensors, meters & Scalability, bandwidth efficiency & Real-time threat detection, local analytics & Smart grid optimization, city-wide models & Event-driven model refresh \\
Smart Agriculture & Soil sensors, drones, weather & Intermittent connectivity, bandwidth & Precision irrigation, local prediction & Yield modeling, FL aggregation & Bandwidth-aware filtering/caching \\
\bottomrule
\end{tabular}}
\end{table*}
\section{Detailed Technical Challenges}\label{sec:supp_challenges}

\begin{figure}[t!]
    \centering
  \includegraphics[width=0.48\textwidth]{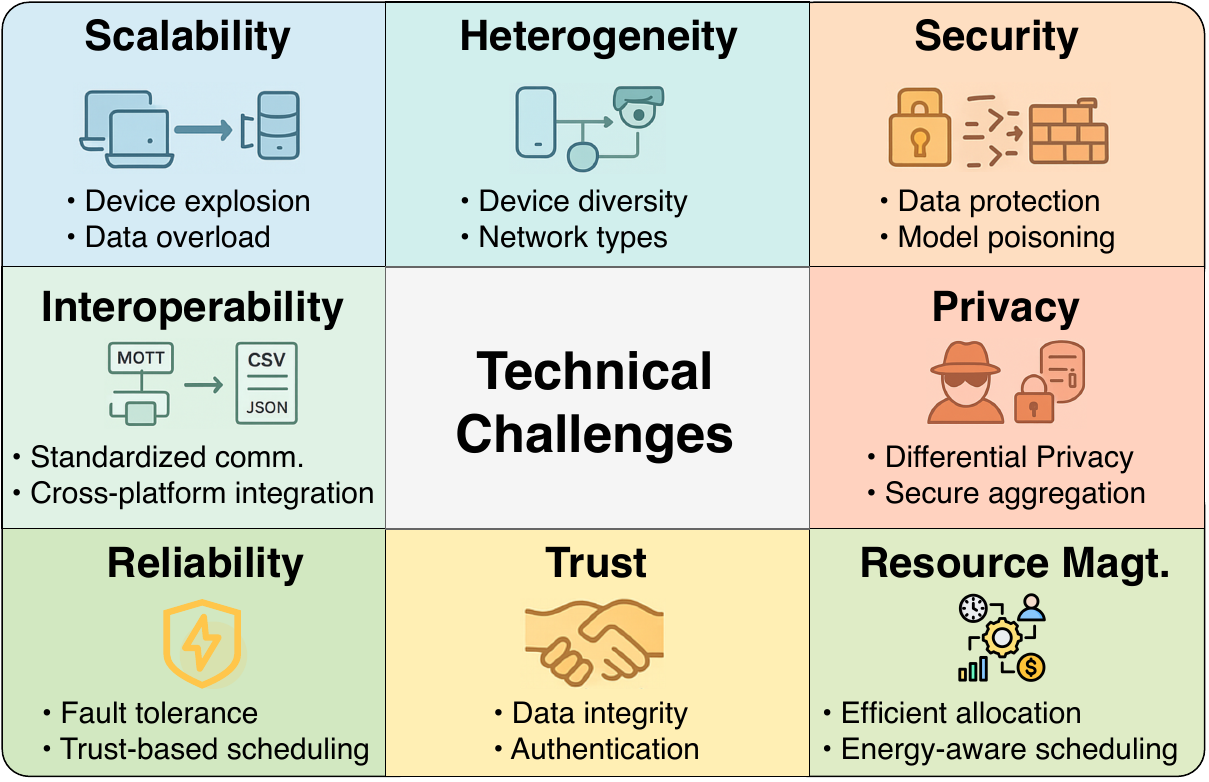}
\caption{Summary of key challenges in CISAIoT networks.}
    \label{fig:supp-challenges}
    \vspace{-10px}
  \end{figure}

This supplementary section provides the extended challenge discussion corresponding to the concise overview in \cref{sec8.1} of the main manuscript. As summarized in \cref{fig:supp-challenges}, the challenges are not isolated obstacles; they interact across scalability, heterogeneity, interoperability, security, privacy, resource management, trust, and reliability.

\subsection{Scalability}\label{sec:supp_challenges_scalability}
Exponential growth in IoT device deployment creates unprecedented computational and communication bottlenecks across distributed infrastructure \cite{becker2020aiops}. Managing massive device populations while maintaining system performance requires innovative distributed computing paradigms \cite{baucas2020using, arroba2024sustainable}. For example, relying solely on cloud or edge resources is inadequate for complex systems like urban sound classification, highlighting the need for optimized resource allocation across the entire continuum. Additionally, the dynamic nature of AIoT networks, characterized by fluctuating device participation, necessitates adaptive and resilient architectures \cite{javed2018cefiot}. Scaling frameworks like the one proposed in \cite{raj2021reliable} for fleet analytics to handle massive device fleets and diverse data streams remains a critical challenge. Efficiently processing data and training models in these distributed environments often leverages techniques like federated learning \cite{zhou2024enhancing, yang2024differentially}, which in turn introduce their own scalability concerns regarding communication overhead and model convergence.

\subsection{Heterogeneity and Interoperability}\label{sec:supp_challenges_heterogeneity}
Device diversity spanning computational capabilities, communication protocols, and data formats introduces fundamental integration challenges \cite{chen2019learning}. Network infrastructure variations from high-bandwidth 5G to resource-constrained environments compound complexity \cite{becker2020aiops}. Additionally, diverse AIoT applications introduce varying performance and security requirements \cite{luckow2021exploring}. Addressing such heterogeneity necessitates flexible and adaptive frameworks, such as Auto-Split and EdgStr, designed for diverse edge and cloud environments \cite{zhou2024enhancing}. Crucially, ensuring interoperability through standardized communication protocols and data formats is essential for efficient data exchange and collaborative processing across the cloud-edge-terminal continuum \cite{yang2024differentially, raj2021reliable}.

Disparate hardware architectures, operating systems, and communication protocols create fundamental barriers to seamless integration \cite{menetrey2022webassembly}. Protocol diversity ranging from lightweight MQTT \cite{javed2018cefiot} to robust enterprise solutions complicates unified system design. Additionally, the diverse nature of AI models, from simple rule-based systems on terminals to complex deep learning models in the cloud, further complicates integration \cite{raj2021reliable}. Without standardized interfaces and data exchange mechanisms, integrating these disparate components hinders efficient collaborative AI tasks and real-time insights. Consequently, deploying and managing AI models across the cloud-edge-terminal continuum becomes challenging, limiting system adaptability and scalability, and creating data silos that restrict AI model training and evolution, ultimately impacting overall performance \cite{zhou2024enhancing}.

\subsection{Security and Privacy}\label{sec:supp_challenges_security}
Multi-layer attack surfaces expose data, models, and system integrity to sophisticated threats \cite{rahman2023blockchainbased}. Sensitive information transmission across distributed layers faces interception and exploitation risks \cite{duan2023privascissors}. Similarly, DoS attacks can disrupt AIoT service availability by overwhelming network resources or computational nodes \cite{saezdecamara2023clustered}. Additionally, model poisoning attacks, particularly relevant in federated learning \cite{yang2024differentially}, compromise AI model integrity and reliability by injecting corrupted data during training \cite{chi2024trusted}. Ensuring data integrity and authentication across the system is therefore crucial for preventing unauthorized modifications and ensuring trustworthy data sources and processing results.

Distributed data processing across multiple infrastructure layers amplifies privacy vulnerabilities and regulatory compliance challenges \cite{alwarafy2021survey}. Information traversing terminals, edge servers, and cloud environments faces escalated breach and eavesdropping risks \cite{yang2024flexible}. Additionally, inferring sensitive information from aggregated model parameters, even without direct data access, poses a significant threat \cite{baseri2024cybersecurity}. Consequently, techniques like DP and secure aggregation become crucial for protecting user data during collaborative learning \cite{yang2024differentially}. Ensuring secure data storage and retrieval across these diverse layers further complicates privacy preservation, demanding robust techniques like those explored in \cite{chi2024trusted} to foster trust and ethical AIoT deployment.

\subsection{Resource Management, Trust, and Reliability}\label{sec:supp_challenges_resource_trust}
Dynamic resource allocation across heterogeneous infrastructure demands sophisticated optimization strategies for computational, storage, and network resources \cite{wang2024cloudedgeend}. Balancing diverse application requirements while minimizing energy consumption and operational costs requires adaptive algorithms \cite{li2024energyaware}. For example, algorithms like ETCRA aim to minimize system latency and enforce trust constraints while optimizing resource allocation \cite{li2024energyaware}. Additionally, resource collaboration and expansion schemes, including real-time resource purchasing based on current demand, are vital for enhancing system efficiency and long-term profit maximization \cite{wang2024efficient}.

System dependability requires robust authentication mechanisms and fault tolerance strategies across distributed environments \cite{venkatraman2022conceptual}. Identity verification prevents unauthorized access while ensuring authenticity and integrity of devices, data, and processes \cite{chi2024trusted}. Data integrity must also be maintained throughout its lifecycle, from terminal acquisition to edge processing and cloud storage \cite{sagar2024can}. Additionally, reliable systems require fault tolerance strategies to address device malfunctions, network disruptions, and security breaches, as well as efficient resource management for reliable performance under varying application demands \cite{javed2018cefiot, li2024energyaware}. For example, trust-based resource allocation can prioritize trustworthy devices for reliable execution of critical tasks.

\section{Additional Lessons Learned}\label{sec:supp_lessons}

The supplementary material retains the extended lessons that complement the main manuscript while keeping the article within the target page budget.

Throughout this comprehensive survey of CETCI, several key lessons have emerged that underscore the persistent challenges and future trajectory of the field. Derived from the extensive body of literature reviewed, these insights provide a high-level perspective on the state of the art beyond the main-text summary.

First, a ``no-free-lunch'' theorem is evident in resource management. As analyzed in \cref{sec3}, a fundamental tension exists between the adaptability of learning-based methods (e.g., DRL) and the theoretical stability of traditional optimization (e.g., game theory). While learning-based approaches excel in dynamic, uncertain environments, they often suffer from high training overhead and a lack of formal guarantees, which are critical for mission-sensitive AIoT applications. The actionable implication is that system designers should map method choice to deployment volatility, certification needs, and retraining budget instead of defaulting to the most accurate benchmarked algorithm.

Second, the interdependence between infrastructure-level and model-level collaboration is profound but often underexplored. The efficiency of sophisticated collaborative learning paradigms like federated or split learning (\cref{sec4}) is fundamentally constrained by the underlying resource allocation and task scheduling mechanisms (\cref{sec3}). An innovative learning algorithm can be rendered ineffective by a suboptimal communication strategy. Therefore, communication planning, model partitioning, and privacy budgeting should be treated as a coupled design problem rather than three independent optimizations.

Finally, a significant gap persists between theoretical models and practical deployment. Many surveyed studies rely on simplified assumptions, such as perfect state information or static network conditions, which do not hold in real-world AIoT systems. The practical lesson is that future research must prioritize robustness, fault tolerance, and adaptability under uncertainty, because trust, reliability, and security are core architectural requirements rather than afterthoughts in large-scale decentralized environments.

\ifCLASSOPTIONcaptionsoff
  \newpage
\fi

\bibliographystyle{IEEEtranN}
\bibliography{refs.bib}

\input{bios.tex}

%% file: preamble.tex
\ifdefined\pdfminorversion\pdfminorversion=4\fi
\ifdefined\pdfobjcompresslevel\pdfobjcompresslevel=0\fi

\makeatletter
\PassOptionsToPackage{table}{xcolor}
\@ifpackageloaded{xcolor}{}{\usepackage{xcolor}}
\@ifpackageloaded{soul}{}{\usepackage{soul}}
\@ifpackageloaded{pifont}{}{\usepackage{pifont}}
\makeatother
\definecolor{revisionHighlightColor}{RGB}{255,255,0}
\providecommand{\blue}[1]{}
\renewcommand{\blue}[1]{{\sethlcolor{revisionHighlightColor}\hl{#1}}}
\providecommand{\RegisterRevisionHighlightCommands}{%
  \ifdefined\cite\soulregister\cite7\fi
  \ifdefined\ref\soulregister\ref7\fi
  \ifdefined\cref\soulregister\cref7\fi
  \ifdefined\Cref\soulregister\Cref7\fi
  \ifdefined\eqref\soulregister\eqref7\fi
  \ifdefined\textbf\soulregister\textbf7\fi
  \ifdefined\textit\soulregister\textit7\fi
  \ifdefined\emph\soulregister\emph7\fi
}
\RegisterRevisionHighlightCommands
\ifdefined\RevisionPreambleLight
  \def\RevisionPreambleMaybeEnd{ }
\else
  \let\RevisionPreambleMaybeEnd\relax
\fi
\RevisionPreambleMaybeEnd

\providecommand{\redcross}{\textcolor{red!70!black}{\ding{55}}}

\usepackage{mathtools,bm,amsmath,amssymb,amsfonts,amsthm,nicefrac,latexsym,dsfont,extarrows,bbm}
\usepackage{cases,subeqnarray}
\usepackage{algorithm,algorithmicx,algpseudocode}
\usepackage{booktabs,multirow,subcaption,graphicx,epsfig,epstopdf,float,stfloats,bigstrut,makecell,array}
\usepackage[most]{tcolorbox}
\usepackage{xcolor,color,colortbl,textcomp,pifont,microtype}
\usepackage{multicol,balance,flushend,setspace,titletoc}
\usepackage[noadjust]{cite}
\usepackage{url,xr-hyper}
\makeatletter
\@ifpackageloaded{wasysym}{}{\usepackage[nointegrals]{wasysym}}
\makeatother

\usepackage[pagebackref=false,breaklinks=true,colorlinks=false,pdfborder={0 0 1},bookmarks=true,linkbordercolor={0 1 0},citebordercolor={0 1 0},urlbordercolor={0 1 0},filebordercolor={0 1 0},menubordercolor={0 1 0},runbordercolor={0 1 0}]{hyperref}
\makeatletter
\@ifundefined{Hy@setpdfversionfalse}{}{\Hy@setpdfversionfalse}
\makeatother
\usepackage[nameinlink]{cleveref}
\RegisterRevisionHighlightCommands

\usepackage{tikz}
\definecolor{lime}{HTML}{A6CE39}
\DeclareRobustCommand{\orcidicon}{\begin{tikzpicture}\draw[lime, fill=lime] (0,0) circle [radius=0.16] node[white] {{\fontfamily{qag}\selectfont \tiny ID}};\draw[white, fill=white] (-0.0625,0.095) circle [radius=0.007];\end{tikzpicture}\hspace{-2mm}}
\foreach \x in {A, ..., Z}{\expandafter\xdef\csname orcid\x\endcsname{\noexpand\href{https://orcid.org/\csname orcidauthor\x\endcsname}{\noexpand\orcidicon}}}

\theoremstyle{plain}

\newtheorem{theorem}{Theorem}
\newtheorem{thm}{Theorem}[section]
\newtheorem{lemma}{Lemma}
\newtheorem{lemm}{Lemma}

\newtheorem{prop}{Proposition}

\newtheorem{corr}{Corollary}

\theoremstyle{remark}

\newtheorem{rem}{Remark}

\crefname{section}{Sec.}{Secs.}
\crefname{subsection}{Sec.}{Secs.}
\crefname{subsubsection}{Sec.}{Secs.}
\crefname{figure}{Fig.}{Figs.}
\crefname{table}{Table}{Tables}
\crefname{equation}{Eq.}{Eqs.}
\crefname{algorithm}{Alg.}{Algs.}
\crefname{assumption}{Assumption}{Assumptions}
\crefname{theorem}{Theorem}{Theorems}
\crefname{lemma}{Lemma}{Lemmas}
\crefname{proposition}{Proposition}{Propositions}
\crefname{corollary}{Corollary}{Corollaries}
\crefname{definition}{Definition}{Definitions}
\crefformat{equation}{#2Eq.~#1#3}
\crefmultiformat{equation}{#2Eqs.~#1#3}{ and~#2#1#3}{, #2#1#3}{ and~#2#1#3}
\Crefformat{equation}{#2Eq.~#1#3}
\Crefmultiformat{equation}{#2Eqs.~#1#3}{ and~#2#1#3}{, #2#1#3}{ and~#2#1#3}

\definecolor{lightkeycolor}{RGB}{255,240,240}
\definecolor{lightgray}{RGB}{234,234,234}

\def\x{{\mathbf x}}

\ifdefined\pdfminorversion\pdfminorversion=4\fi
\ifdefined\pdfinclusioncopyfonts\pdfinclusioncopyfonts=1\fi
\IfFileExists{macros.tex}{\input{macros}}{}

%% file: t1_survey_1.tex
\begin{table*}[tb!]
  \centering
  \caption{%
  \protect Comparison of recent surveys on cloud-edge-terminal collaboration and related AIoT systems, evaluating how each survey covers resource management, model evolution, security aspects, computing paradigms, and application domains.}
  \label{tab:1}
  \vspace{-0.1cm}
  \setlength{\tabcolsep}{0.5mm}{
  \resizebox{0.98\textwidth}{!}{
    \begin{tabular}{@{}cccccccccccccccccc@{}}
      \rowcolor{gray!8}
      \toprule
      \multirow{2}{*}{\textbf{Ref.}} & \multirow{2}{*}{\textbf{Year}} & \multirow{2}{*}{\textbf{Domain}} & \multicolumn{2}{c}{\textbf{Resource Management}} & \multicolumn{2}{c}{\textbf{Model Evolution}} & \multicolumn{2}{c}{\textbf{Security Aspects}} & \multicolumn{5}{c}{\textbf{Computing Paradigms}} & \multicolumn{4}{c}{\textbf{Applications}} \\ \cmidrule(l){4-5} \cmidrule(l){6-7} \cmidrule(l){8-9} \cmidrule(l){10-14} \cmidrule(l){15-18}
      \rowcolor{gray!8}
      &  &  & \textbf{Task Off.} & \textbf{Res. Alloc.} & \textbf{Model Comp.} & \textbf{Know. Trans.} & \textbf{Security} & \textbf{Privacy} & \textbf{Edge} & \textbf{Cloud} & \textbf{ECC} & \textbf{CETC} & \textbf{AIoT} & \textbf{Manuf.} & \textbf{Transport.} & \textbf{Health} & \textbf{Agri.} \\ \midrule
      \rowcolors{3}{blue!4}{white}
      \cite{duc2019machine} & 2019 & Edge-Cloud Systems, Application Mgmt. & \CIRCLE & \CIRCLE & \LEFTcircle & \LEFTcircle & \LEFTcircle & \Circle & \CIRCLE & \CIRCLE & \LEFTcircle & \Circle & \LEFTcircle & \LEFTcircle & \LEFTcircle & \LEFTcircle & \Circle \\
      \cite{hong2019resource} & 2019 & Network Architecture, Resource Allocation & \CIRCLE & \CIRCLE & \Circle & \Circle & \Circle & \Circle & \CIRCLE & \LEFTcircle & \LEFTcircle & \Circle & \LEFTcircle & \LEFTcircle & \LEFTcircle & \LEFTcircle & \LEFTcircle \\
      \cite{ren2019survey} & 2019 & IoT, Distributed Systems & \CIRCLE & \CIRCLE & \Circle & \LEFTcircle & \CIRCLE & \CIRCLE & \CIRCLE & \CIRCLE & \CIRCLE & \LEFTcircle & \CIRCLE & \LEFTcircle & \LEFTcircle & \LEFTcircle & \LEFTcircle \\
      \cite{asim2020review} & 2020 & Resource optimization, AI algorithms & \CIRCLE & \CIRCLE & \Circle & \LEFTcircle & \Circle & \Circle & \CIRCLE & \CIRCLE & \LEFTcircle & \Circle & \LEFTcircle & \LEFTcircle & \LEFTcircle & \Circle & \Circle \\
      \cite{vanhuynh2022edge}     & 2022          & Metaverse, Digital Twin       & \CIRCLE                  & \CIRCLE                      & \Circle                    & \Circle                     & \Circle                 & \LEFTcircle                 & \CIRCLE                 & \LEFTcircle              & \CIRCLE                           & \Circle                                   & \LEFTcircle   & \Circle                & \LEFTcircle             & \Circle             & \Circle              \\
      \cite{ahmad2023deep} & 2023 & Multiple domains, IoT networks & \LEFTcircle & \CIRCLE & \LEFTcircle & \LEFTcircle & \LEFTcircle & \LEFTcircle & \CIRCLE & \CIRCLE & \LEFTcircle & \Circle & \CIRCLE & \LEFTcircle & \LEFTcircle & \LEFTcircle & \LEFTcircle \\
      \cite{duan2023distributed} & 2023 & Distributed AI Systems & \CIRCLE & \CIRCLE & \CIRCLE & \CIRCLE & \CIRCLE & \CIRCLE & \CIRCLE & \CIRCLE & \CIRCLE & \CIRCLE & \CIRCLE & \LEFTcircle & \LEFTcircle & \LEFTcircle & \LEFTcircle \\
      \cite{kumar2023analysis} & 2023 & Industrial IoT, Data Analysis & \LEFTcircle & \CIRCLE & \Circle & \Circle & \LEFTcircle & \LEFTcircle & \CIRCLE & \CIRCLE & \LEFTcircle & \Circle & \CIRCLE & \CIRCLE & \LEFTcircle & \LEFTcircle & \LEFTcircle \\
      \cite{asghari2024server} & 2024 & Mobile Computing, Network Optimization & \LEFTcircle & \CIRCLE & \Circle & \Circle & \Circle & \LEFTcircle & \CIRCLE & \CIRCLE & \Circle & \Circle & \LEFTcircle & \LEFTcircle & \Circle & \Circle & \Circle \\
      \cite{qu2024mobile} & 2024 & Edge AI, LLMs, Mobile Computing & \CIRCLE & \CIRCLE & \CIRCLE & \CIRCLE & \LEFTcircle & \CIRCLE & \CIRCLE & \LEFTcircle & \CIRCLE & \LEFTcircle & \LEFTcircle & \Circle & \LEFTcircle & \LEFTcircle & \Circle \\
      \cite{wang2024endedgecloud} & 2024          & DL, AI Systems Architecture   & \CIRCLE                  & \CIRCLE                      & \CIRCLE                    & \CIRCLE                     & \LEFTcircle             & \LEFTcircle                 & \CIRCLE                 & \CIRCLE                  & \CIRCLE                           & \CIRCLE                                   & \CIRCLE       & \LEFTcircle            & \LEFTcircle             & \LEFTcircle         & \LEFTcircle          \\
      \cite{souza2024maintenance} & 2024 & Cloud Computing, Edge Computing, IoT & \Circle & \CIRCLE & \Circle & \Circle & \CIRCLE & \LEFTcircle & \CIRCLE & \CIRCLE & \LEFTcircle & \Circle & \CIRCLE & \LEFTcircle & \LEFTcircle & \Circle & \LEFTcircle \\
      \cite{xu2024unleashing} & 2024 & Mobile Networks, Generative AI Services & \LEFTcircle & \LEFTcircle & \LEFTcircle & \LEFTcircle & \CIRCLE & \CIRCLE & \CIRCLE & \CIRCLE & \Circle & \Circle & \LEFTcircle & \Circle & \Circle & \Circle & \Circle \\
      \cite{yang2025edge}         & 2025          & Mobile Networks, Edge Comp.   & \CIRCLE                  & \CIRCLE                      & \LEFTcircle                & \LEFTcircle                 & \CIRCLE                 & \LEFTcircle                 & \CIRCLE                 & \LEFTcircle              & \LEFTcircle                       & \Circle                                   & \CIRCLE       & \LEFTcircle            & \CIRCLE                 & \LEFTcircle         & \Circle              \\
      \cite{liu2025edgecloud}     & 2025          & Edge-Cloud Comp., Model Opt. & \CIRCLE                  & \CIRCLE                      & \CIRCLE                    & \CIRCLE                     & \CIRCLE                 & \CIRCLE                     & \CIRCLE                 & \CIRCLE                  & \CIRCLE                           & \Circle                                   & \LEFTcircle   & \LEFTcircle            & \CIRCLE                 & \CIRCLE             & \LEFTcircle          \\
      \cite{jiang2026agentic}     & 2026          & Agentic AI, Future Communications & \LEFTcircle             & \LEFTcircle                  & \CIRCLE                    & \CIRCLE                     & \LEFTcircle             & \LEFTcircle                 & \LEFTcircle             & \CIRCLE                  & \LEFTcircle                       & \Circle                                   & \LEFTcircle   & \Circle                & \LEFTcircle             & \LEFTcircle         & \Circle              \\ \midrule
      \rowcolor{lightkeycolor}
      \textit{Ours} & 2025 & AIoT, CETCI, Res. Mgmt., Model Evo. & \CIRCLE & \CIRCLE & \CIRCLE & \CIRCLE & \CIRCLE & \CIRCLE & \CIRCLE & \CIRCLE & \CIRCLE & \CIRCLE & \CIRCLE & \CIRCLE & \CIRCLE & \CIRCLE & \CIRCLE \\ \bottomrule
           
  \end{tabular}}}

  \fontsize{6.5pt}{2pt}\selectfont {
  \begin{minipage}{0.98\textwidth}
    \vspace{1px}
    \textit{Notes}: $\CIRCLE$: "Fully covered"; $\LEFTcircle$: "Partially covered"; $\Circle$: "Not covered". Column abbreviations: Task Off.: Task Offloading; Res. Alloc.: Resource Allocation; Model Comp.: Model Compression; Know. Trans.: Knowledge Transfer; ECC: Edge-Cloud Collaboration; CETC: Cloud-Edge-Terminal Collaboration; AIoT: Artificial Intelligence of Things; Manuf.: Manufacturing; Transport.: Transportation; Health: Healthcare; Agri.: Agriculture.
  \end{minipage}
  }
\vspace{-15px}
\end{table*}

%% file: t1_acronyms.tex
\begin{table}[t!]
  \centering
  \caption{List of key acronyms used in this survey.}
  \label{tab:acronyms}
  \vspace{-0.1cm}
  \setlength{\tabcolsep}{12mm}{
  \resizebox{0.47\textwidth}{!}{
  \begin{tabular}{@{}cl@{}}
    \toprule
    \textbf{Acronym} & \multicolumn{1}{c}{\textbf{Definition}} \\ \midrule
    AI & Artificial intelligence \\
    AIoT & Artificial intelligence of things \\
    CETCI & Cloud-edge-terminal collaborative intelligence \\
    CISAIoT & Collaborative intelligence systems for AIoT \\
    DDL & Distributed deep learning \\
    DP & Differential privacy \\
    DRL & Deep reinforcement learning \\
    DT & Digital twin \\
    FL & Federated learning \\
    HE & Homomorphic encryption \\
    IDPS & Intrusion detection and prevention systems \\
    IIoT & Industrial Internet of Things \\
    IoT & Internet of Things \\
    LLMs & Large language models \\
    MARL & Multi-agent reinforcement learning \\
    MEC & Mobile edge computing \\
    MPC & Multi-party computation \\
    PEFT & Parameter-efficient fine-tuning \\
    PQC & Post-quantum cryptography \\
    QKD & Quantum key distribution \\
    QoE & Quality of experience \\
    QoS & Quality of service \\
    RAG & Retrieval-augmented generation \\
    RL & Reinforcement learning \\
    SDN & Software-defined networking \\
    SLA & Service level agreement \\
    VN & Virtual network \\
    XAI & Explainable AI \\
    \bottomrule
  \end{tabular}}}
\vspace{-20px}
\end{table}

%% file: t_metrics.tex
\begin{table*}[t!]
\centering
\caption{Key Performance Metrics in CETCI Systems.}
\label{tab:metrics}
  \setlength{\tabcolsep}{8mm}{
  \resizebox{0.96\textwidth}{!}{
\begin{tabular}{lll}
\toprule
\multicolumn{1}{c}{\textbf{Metric}} & \multicolumn{1}{c}{\textbf{Focus in AIoT}} & \multicolumn{1}{c}{\textbf{Key Optimization Goal}} \\
\midrule
Response Time & Inference latency, end-to-end delay & minimize total latency ($L_{total}$) via optimal offloading \\
Energy Consumption & Battery life, green computing & minimize total energy ($E_{total}$) across layers \\
Resource Utilization & CPU/GPU/Bandwidth efficiency & maximize utilization ($U_{total}$) without congestion \\
Quality of Service (QoS) & Reliability, throughput, availability & maintain QoS within feasible region ($\mathcal{C}_{feasible}$) \\
Quality of Experience & User satisfaction, perceived quality & maximize subjective satisfaction ($Q_{exp}$) \\
\bottomrule
\end{tabular}%
}}
\end{table*}

%% file: fig3_sec3.tex
\tikzstyle{my-box}=[
    rectangle,
    draw=hidden-draw,
    rounded corners,
    text opacity=1,
    minimum height=3.5em,
    minimum width=5em,
    align=center,
    fill opacity=1,
    line width=0.8pt,
    font=\fontsize{13.5}{16}\selectfont,
    text=black,
]

\tikzstyle{tklevel0}=[my-box,
    fill=level0,
    font=\fontsize{16}{16}\selectfont,
]

\tikzstyle{tklevel1}=[my-box,
    fill=level1,
]

\tikzstyle{tklevel2}=[my-box,
    fill=level2,
]
\tikzstyle{tklevel3}=[my-box,
    fill=level3,
]

\tikzstyle{leaf}=[my-box,
    fill=level4,
]

\begin{figure}[t]
    \centering
    \resizebox{0.49\textwidth}{!}
    {
        \begin{forest}
            forked edges,
            for tree={
                fill=level0,
                grow=east,
                reversed=true,
                anchor=base west,
                parent anchor=east,
                child anchor=west,
                base=center,
                font=\large,
                rectangle,
                draw=hidden-draw,
                rounded corners,
                align=center,
                minimum width=4em,
                edge+={darkgray, line width=1pt},
                s sep=3pt,
                line width=0.8pt,
                ver/.style={rotate=90, child anchor=north, parent anchor=south, anchor=center},
            },
            where level=0{ver,text width=35em, tklevel0}{},
            where level=1{ver, text width=12em, tklevel1}{},
            where level=2{text width=10em,tklevel2,}{},
            where level=3{text width=42.8em,tklevel3,minimum height=5em}{},
            [Intelligent Resource Management Paradigms (§\ref{sec3})
                [Task\\Offloading (§\ref{sec3.1})
                    [Learning-Based\\Offloading,align=center
                        [\citet{ji2024task}{,} MADDPG~\cite{she2024efficient}{,} \citet{sharma2024deep}{,} \citet{chen2024dynamica}, leaf]
                    ]
                    [Game Theory\\\& Optimization,align=center
                        [\citet{zhang2024stackelberggamebased}{,} \citet{duan2024binary}{,} DARC-DE~\cite{bandyopadhyay2024delaysensitive}{,} SA-BPSO~\cite{chen2022resource}, leaf]
                    ]
                    [Context-Aware\\Strategies,align=center
                        [\citet{feng2024dependencyaware}{,} \citet{li2024blockchainbased}{,} \citet{bai2024delayaware}{,} \citet{li2024twotimescale}, leaf]
                    ]
                ]
                [Resource\\Allocation (§\ref{sec3.2})
                    [Learning-Based\\Allocation
                        [\citet{fang2023large}{,} OTFAC~\cite{gan2024optimal}{,} \citet{huang2024joint}{,} \citet{zhang2024securityaware}, leaf]
                    ]
                    [Energy-Aware\\Allocation
                        [ETCRA~\cite{li2024energyaware}{,} Alopex-DE~\cite{chen2024fast}{,} \citet{hua2024energyefficient}{,} \citet{nguyen2024integrated}, leaf]
                    ]
                    [QoS-Driven\\Allocation
                        [\citet{alieldin2021hidden}{,} \citet{li2024cloudedgedevice}{,} \citet{zhai2024edgecloud}{,} \citet{asghar2022survey}, leaf]
                    ]
                ]
                [Optimization\\Techniques (§\ref{sec3.3})
                    [Linear\\Programming
                        [\citet{bandyopadhyay2024delaysensitive}{,} \citet{zeng2024joint}{,} \citet{he2024efficient}{,} \citet{lu2024dynamic}, leaf]
                    ]
                    [Convex\\Optimization
                        [\citet{nguyen2024integrated}{,} \citet{feng2024dependencyaware}{,} \citet{khalafi2024network}{,} SCA~\cite{fang2024joint}, leaf]
                    ]
                    [Game Theory\\Approaches
                        [\citet{wu2024computation}{,} \citet{zhang2024stackelberggamebased}{,} \citet{dragotto2024critical}{,} \citet{shams2025joint}, leaf]
                    ]
                    [Deep Reinfor.\\Learning
                        [\citet{yu2024game}{,} \citet{she2024efficient}{,} \citet{huang2024joint}{,} \citet{zhang2024securityaware}, leaf]
                    ]
                ]
            ]
        \end{forest}
    }    
\vspace{-10px}
\caption{\protect Taxonomy of intelligent resource management.}
\label{fig:sec3}
\vspace{-15pt}
\end{figure}

%% file: t2_task_3.1.tex
\begin{table}[t!]
  \centering
  \caption{Summary of learning-based and optimization-based task offloading strategies in CETCI AIoT systems.}
  \label{tab:2}
  \vspace{-0.1cm}
  \setlength{\tabcolsep}{0.6mm}{
  \resizebox{0.49\textwidth}{!}{
    \begin{tabular}{@{}l>{\raggedright\arraybackslash}p{2.2cm}>{\raggedright\arraybackslash}p{3.2cm}>{\raggedright\arraybackslash}p{4cm}cc@{}}
      \toprule
      {\textbf{Method}} & {\textbf{Key Tech.}} & {\textbf{Metrics}} & {\textbf{Limitations}} & {\textbf{CETC}} & {\textbf{TO}} \\ \midrule
      {Binh et al. \cite{binh2024reinforcement}} & {Reinforcement Learning (RL)} & {Tolerance Time} & {Convergence speed dependent on state space size.} & {\redcross} & {$\checkmark$} \\
      {ADPRL \cite{chen2024dynamica}} & {Deep RL (DRL)} & {Task Completion Time, Energy Consumption} & {High training overhead and requires extensive data.} & {\redcross} & {$\checkmark$} \\
      {ECIVA \cite{ji2024task}} & {Heuristic Algorithm} & {Accuracy, Throughput} & {Provides sub-optimal solutions; lacks adaptability.} & {\redcross} & {$\checkmark$} \\
      {Sharma et al. \cite{sharma2024deep}} & {Meta-RL} & {Time and Energy Consumption} & {Complex training process for meta-learning.} & {\redcross} & {$\checkmark$} \\
      {MADDPG \cite{she2024efficient}} & {Multi-Agent RL (MARL)} & {Offloading Delay, Energy Consumption} & {Assumes rational agents; complex coordination.} & {$\checkmark$} & {$\checkmark$} \\
      {DARC-DE \cite{bandyopadhyay2024delaysensitive}} & {Differential Evolution} & {Total Delay, Bandwidth Utilization} & {Prone to converging to local optima.} & {\redcross} & {$\checkmark$} \\
      {SA-BPSO \cite{chen2022resource}} & {Hybrid Swarm Optimization} & {User Overhead (Latency, Energy, Cost)} & {Performance is highly sensitive to parameter tuning.} & {\redcross} & {$\checkmark$} \\
      {Duan et al. \cite{duan2024binary}} & {Game Theory} & {Completion Time, Energy Consumption} & {High computational complexity to find equilibrium.} & {\redcross} & {$\checkmark$} \\
      {SRA-E-ABCO \cite{jiao2024sraeabco}} & {Swarm Intelligence (ABC)} & {Task Offloading Delay, System Overhead} & {Can exhibit slow convergence in complex problems.} & {$\checkmark$} & {$\checkmark$} \\
      {Zhang et al. \cite{zhang2024stackelberggamebased}} & {Stackelberg Game} & {Cost Minimization, Task Completion Delay} & {Restricted to hierarchical leader-follower models.} & {\redcross} & {$\checkmark$} \\
      {Bai et al. \cite{bai2024delayaware}} & {Convex Optimization} & {Completion Delay} & {Requires the problem to be formulated as convex.} & {\redcross} & {$\checkmark$} \\
      {PRAISE \cite{feng2024dependencyaware}} & {Auction-based Mechanism} & {System Benefits, Resource Costs} & {Incurs significant communication overhead.} & {\redcross} & {$\checkmark$} \\
      {BRTO \cite{li2024blockchainbased}} & {Blockchain + Optimization} & {Latency, Energy Efficiency, Security} & {High latency and overhead from blockchain operations.} & {$\checkmark$} & {$\checkmark$} \\
      {Li et al. \cite{li2024twotimescale}} & {Lyapunov Optimization} & {Long-term Network Performance} & {Assumes time-scale separation, which may not hold.} & {\redcross} & {$\checkmark$} \\ \bottomrule
    \end{tabular}}}

  \fontsize{7.5pt}{2pt}\selectfont {
  \begin{minipage}{0.49\textwidth}
   \vspace{1px}
    \textit{Notes}: TO = Task Offloading. Learning-based methods usually offer stronger adaptability but higher training and convergence cost; optimization- and game-theoretic methods offer clearer feasibility conditions but depend more heavily on accurate system models; heuristic and swarm methods are attractive when fast reaction is prioritized over strict optimality.
  \end{minipage}
  }
\vspace{-15px}
\end{table}

%% file: t3_res_alloc_3.2.tex
\begin{table}[t!]
  \centering
  \caption{Summary of resource allocation techniques.}
  \label{tab:3}
  \vspace{-0.1cm}
  \setlength{\tabcolsep}{0.2mm}{
  \resizebox{0.49\textwidth}{!}{
    \begin{tabular}{@{}l>{\raggedright\arraybackslash}p{3cm}>{\raggedright\arraybackslash}p{4cm}>{\raggedright\arraybackslash}p{5cm}cc@{}}
      \toprule
      {\textbf{Method}} & {\textbf{Key Tech.}} & {\textbf{Metrics}} & {\textbf{Pros \& Cons}} & {\textbf{CETC}} & {\textbf{RA}} \\ \midrule
      {OTFAC \cite{gan2024optimal}} & {Federated Learning + Actor-Critic} & {Delay, Energy Consumption} & {Adaptive and communication-efficient; complex to implement.} & {\redcross} & {$\checkmark$} \\
      {Fang et al. \cite{fang2023large}} & {Deep Reinforcement Learning} & {Latency, Resource Utilization} & {Handles dynamic environments well; can have slow convergence.} & {\redcross} & {$\checkmark$} \\
      {Huang et al. \cite{huang2024joint}} & {DRL with Hybrid Action Space} & {Energy Consumption, Latency} & {Enables joint optimization; high training cost.} & {\redcross} & {$\checkmark$} \\
      {DRLS-VNE \cite{zhang2024securityaware}} & {DRL} & {Security, Resource Utilization} & {Security-aware approach; state space can be very large.} & {$\checkmark$} & {$\checkmark$} \\
      {SA-BPSO \cite{chen2022resource}} & {Hybrid Swarm Optimization} & {Total User Overhead} & {Effective heuristic; performance is sensitive to parameters.} & {\redcross} & {$\checkmark$} \\
      {Hua et al. \cite{hua2024energyefficient}} & {Joint Optimization (Power, Scheduling)} & {Energy Consumption} & {Energy-efficient design; assumes accurate state information.} & {\redcross} & {$\checkmark$} \\
      {ETCRA \cite{li2024energyaware}} & {Heuristic Algorithm} & {Execution Time, Energy, Reliability} & {Multi-objective; trade-offs can be complex to manage.} & {\redcross} & {$\checkmark$} \\
      {Alopex-DE \cite{chen2024fast}} & {Differential Evolution} & {Latency, Energy Consumption} & {Fast convergence; risk of being trapped in local optima.} & {$\checkmark$} & {$\checkmark$} \\
      {Nguyen et al. \cite{nguyen2024integrated}} & {Convex Optimization} & {Energy Consumption, Delay} & {Guarantees optimality; requires problem to be convex.} & {\redcross} & {$\checkmark$} \\
      {DT-MADDPG \cite{wang2024cooperative}} & {MARL + Digital Twin} & {Success Rate, Latency, Energy} & {Proactive via simulation; high overhead to maintain DT.} & {$\checkmark$} & {$\checkmark$} \\
      {Zhai et al. \cite{zhai2024edgecloud}} & {Optimization} & {Delay, Carbon Emissions, Profit} & {Considers multiple practical objectives; complex modeling.} & {\redcross} & {$\checkmark$} \\
      {Ali-Eldin et al. \cite{alieldin2021hidden}} & {Queuing Theory} & {End-to-End Latency} & {Provides analytical insights; based on simplified models.} & {\redcross} & {\redcross} \\
      {Li et al. \cite{li2024cloudedgedevice}} & {Heuristic Algorithm} & {Load Balancing Degree, Queuing Delay} & {Simple and effective for load balancing; may be sub-optimal.} & {$\checkmark$} & {$\checkmark$} \\
      {Zhou et al. \cite{zhou2024enhancing}} & {Heuristic Algorithm} & {Quality of Service, Privacy} & {Privacy-aware by design; potential performance overhead.} & {\redcross} & {\redcross} \\ \bottomrule
  \end{tabular}}}

  \fontsize{7.5pt}{2pt}\selectfont {
  \begin{minipage}{0.49\textwidth}
   \vspace{1px}
    \textit{Notes}: RA = Resource Allocation.
  \end{minipage}
  }
\vspace{-15px}
\end{table}

%% file: fig4_sec4.tex
\tikzstyle{my-box}=[
    rectangle,
    draw=hidden-draw,
    rounded corners,
    text opacity=1,
    minimum height=3.5em,
    minimum width=5em,
    align=center,
    fill opacity=1,
    line width=0.8pt,
    font=\fontsize{13.5}{16}\selectfont,
    text=black,
]

\tikzstyle{tklevel0}=[my-box,
    fill=level0,
    font=\fontsize{16}{16}\selectfont,
]

\tikzstyle{tklevel1}=[my-box,
    fill=level1,
]

\tikzstyle{tklevel2}=[my-box,
    fill=level2,
]
\tikzstyle{tklevel3}=[my-box,
    fill=level3,
]

\tikzstyle{leaf}=[my-box,
    fill=level4,
]

\begin{figure}[t]
    \centering
    \resizebox{0.49\textwidth}{!}
    {
        \begin{forest}
            forked edges,
            for tree={
                fill=level0,
                grow=east,
                reversed=true,
                anchor=base west,
                parent anchor=east,
                child anchor=west,
                base=center,
                font=\large,
                rectangle,
                draw=hidden-draw,
                rounded corners,
                align=center,
                minimum width=3em,
                edge+={darkgray, line width=1pt},
                s sep=3pt,
                line width=0.8pt,
                ver/.style={rotate=90, child anchor=north, parent anchor=south, anchor=center},
            },
            where level=0{ver,text width=35em, tklevel0}{},
            where level=1{ver,text width=8em, tklevel1}{},
            where level=2{text width=10em,tklevel2,}{},
            where level=3{text width=42.8em,tklevel3,minimum height=5em}{},
            [Intelligent Collaboration Learning Paradigms (§\ref{sec4})
                [FL (§\ref{sec4.1})
                    [Privacy-Preserving
                        [zCDP~\cite{hu2020concentrated}{,} MAPA~\cite{li2019asynchronous}{,} \citet{stevens2022efficient}{,} \citet{yang2024differentially}, leaf]
                    ]
                    [Robustness \&\\Security
                        [SparseFed~\cite{panda2022sparsefed}{,} QuAsyncFL~\cite{liu2024quasyncfl}{,} FlexibleFL~\cite{zhao2024flexiblefl}{,} \citet{ma2022privacypreserving}, leaf]
                    ]
                ]
                [DDL (§\ref{sec4.2})
                    [Model\\Parallelism
                        [AlpaServe~\cite{li2023alpaserve}{,} AutoDiCE~\cite{guo2022autodice}{,} \citet{joshi2023enabling}{,} Auto-Split~\cite{banitalebidehkordi2021autosplit}, leaf]
                    ]
                    [Data\\Parallelism
                        [HierTrain~\cite{liu2020hiertrain}{,} QuAsyncFL~\cite{liu2024quasyncfl}{,} ARENA~\cite{qi2023arena}{,} \citet{yang2024differentially}, leaf]
                    ]
                ]
                [ECME (§\ref{sec4.3})
                    [Model Compr.\\\& Distillation
                        [AKD~\cite{wang2024clouddevice}{,} \citet{yang2025growthadaptive}{,} \citet{kim2024dnn}{,} \citet{tang2024anomaly}, leaf]
                    ]
                    [Incr. Learning\\\& Model Upd.
                        [CADCL-DTM~\cite{li2024conditionadaptive}{,} MEDIA~\cite{zhao2024media}{,} EdgeC3~\cite{lin2024online}{,} ADMM-FedMeta~\cite{yue2021inexactadmm}, leaf]
                    ]
                ]
                [RLO (§\ref{sec4.4})
                    [Multi-Agent\\RL
                        [\citet{li2024load}{,} \citet{lu2024dynamic}{,} MASAC~\cite{gao2024cloudedge}{,} Coop-MADRL~\cite{suzuki2023multiagent}, leaf]
                    ]
                    [Reinforcement\\Learning
                        [\citet{luo2019resource}{,} \citet{sharma2024deep}{,} MADRLOM~\cite{guo2024madrlom}{,} \citet{zhang2024lsia3cs}, leaf]
                    ]
                ]
            ]
        \end{forest}
    }    
\vspace{-10px}
\caption{%
\protect Taxonomy of intelligent collaboration learning paradigms, including federated learning (FL), distributed deep learning (DDL), edge-cloud model evolution (ECME), and reinforcement learning optimization (RLO).}
\label{fig:4}
\vspace{-15pt}
\end{figure}

%% file: t5_ec_mod_evo_4.3.tex
\begin{table}[t!]
  \centering
  \caption{Summary of model compression and knowledge distillation techniques for CETCI AIoT systems, with emphasis on update cost, edge deployability, and sensitivity to teacher, partition, or synchronization choices.}
  \label{tab:5}
  \vspace{-0.1cm}
  \setlength{\tabcolsep}{0.6mm}{
  \resizebox{0.49\textwidth}{!}{
    \begin{tabular}{@{}l>{\raggedright\arraybackslash}p{3.5cm}>{\raggedright\arraybackslash}p{4.5cm}>{\raggedright\arraybackslash}p{4cm}cc@{}}
      \toprule
      {\textbf{Method}} & {\textbf{Key Tech.}} & {\textbf{Optimization Objectives}} & {\textbf{Pros \& Cons}} & {\textbf{ECC}} & {\textbf{CETC}} \\ \midrule
      {PECo \cite{di2024peco}} & {Point Cloud Compression} & {Geometric Accuracy, Structural Compactness} & {Efficient for 3D data; specialized use case.} & {\redcross} & {\redcross} \\
      {CDC \cite{dong2020cdc}} & {Collaborative Compression} & {Bandwidth Efficiency, Accuracy} & {Reduces data transmission; requires coordination.} & {$\checkmark$} & {\redcross} \\
      {Hawlader et al. \cite{hawlader2024leveraging}} & {Knowledge Distillation} & {Detection Quality, Latency} & {Improves edge model accuracy; teacher dependent.} & {$\checkmark$} & {\redcross} \\
      {BPS \cite{hou2024bps}} & {Model Partitioning} & {Throughput, Latency} & {Balances load; sensitive to partition point.} & {$\checkmark$} & {$\checkmark$} \\
      {Joshi et al. \cite{joshi2024integration}} & {Incremental Learning + DT} & {Energy, Latency, Accuracy} & {Adapts to changes; high DT maintenance cost.} & {$\checkmark$} & {$\checkmark$} \\
      {CO-PILOT \cite{kim2024dnn}} & {Split Computing} & {End-to-end Latency, Accuracy} & {Lowers device computation; network sensitive.} & {$\checkmark$} & {\redcross} \\
      {EdgeActNet \cite{luo2024edgeactnet}} & {Lightweight CNN Design} & {Accuracy, Memory, Inference Time} & {Optimized for edge; may have lower accuracy.} & {\redcross} & {\redcross} \\
      {DTS-KD \cite{tang2024anomaly}} & {Knowledge Distillation} & {F1 Score, Accuracy, Precision, Recall} & {High accuracy on edge; requires powerful teacher.} & {$\checkmark$} & {$\checkmark$} \\
      {Wang et al. \cite{wang2024clouddevice}} & {Knowledge Distillation for LLMs} & {Generalization Capabilities} & {Adapts LLMs to device; complex distillation.} & {$\checkmark$} & {$\checkmark$} \\
      {Wu et al. \cite{wu2024systematic}} & {CNN Pruning} & {Edge Detection Accuracy, Model Size} & {Reduces model size; can degrade accuracy.} & {\redcross} & {\redcross} \\
      {VLMCE \cite{yang2023approximately}} & {Graph Neural Network} & {VNR Cost, Blocking Ratio} & {Efficient mapping; high computational complexity.} & {$\checkmark$} & {\redcross} \\
      {Yang et al. \cite{yang2025growthadaptive}} & {Growth-Adaptive Distillation} & {Identification Accuracy, Model Size} & {Adapts student model size; complex training.} & {$\checkmark$} & {\redcross} \\
      {CEMA \cite{chen2023robust}} & {Model Ensemble} & {Adaptation Performance, Robustness} & {Robust to shifts; high inference cost.} & {$\checkmark$} & {\redcross} \\
      {MEDIA \cite{zhao2024media}} & {Incremental Learning} & {Offloading Accuracy, Training Cost} & {Adapts to new data; risk of catastrophic forgetting.} & {$\checkmark$} & {\redcross} \\
      {FedITA \cite{he2024fedita}} & {Federated Transfer Learning} & {F1-score, Privacy} & {Leverages knowledge; risk of negative transfer.} & {$\checkmark$} & {\redcross} \\
      {FlexibleFL \cite{zhao2024flexiblefl}} & {Federated Learning} & {Attack Defense Success Rate} & {Robust to attacks; adds computation overhead.} & {$\checkmark$} & {\redcross} \\
      {CADCL-DTM \cite{li2024conditionadaptive}} & {Continual Learning + DT} & {Forgetting Performance, Accuracy} & {Avoids forgetting; high DT synchronization cost.} & {\redcross} & {$\checkmark$} \\
      {EdgeC3 \cite{lin2024online}} & {Collaborative Continuous Learning} & {Model Accuracy, Cost} & {Continual adaptation; complex aggregation.} & {$\checkmark$} & {\redcross} \\
      {ADMM-FedMeta \cite{yue2021inexactadmm}} & {Federated Meta-Learning} & {Computational Cost, Convergence} & {Fast adaptation; sensitive to meta-parameters.} & {\redcross} & {\redcross} \\
      {Cross-FCL \cite{zhang2024crossfcl}} & {Federated Continual Learning} & {Accuracy, Forgetting} & {Handles data drift; high communication cost.} & {$\checkmark$} & {\redcross} \\ \bottomrule
  \end{tabular}}}
\vspace{-15px}
\fontsize{7.5pt}{2pt}\selectfont {
\begin{minipage}{0.49\textwidth}
\textit{Notes}: Methods in this table should be read through four recurring trade-offs: memory or latency relief, communication cost of model refresh, robustness to drift or attack, and implementation maturity on heterogeneous edge hardware.
\end{minipage}
}
\end{table}

%% file: t6_rl_4.4.tex
\begin{table}[t!]
  \centering
  \caption{Summary of RL and MARL approaches, emphasizing their convergence sensitivity, coordination overhead, and suitability for decentralized CETCI control.}
  \label{tab:6}
  \vspace{-0.1cm}
  \setlength{\tabcolsep}{0.6mm}{
  \resizebox{0.49\textwidth}{!}{
    \begin{tabular}{@{}l>{\raggedright\arraybackslash}p{2.5cm}>{\raggedright\arraybackslash}p{4cm}>{\raggedright\arraybackslash}p{4.5cm}@{}}
      \toprule
      {\textbf{Method}} & {\textbf{Learning Paradigm}} & {\textbf{Metrics}} & {\textbf{Pros \& Cons}} \\ \midrule
      {Zhu et al. \cite{zhu2022federated}} & {MARL} & {Average Age of Information} & {Decentralized and privacy-preserving; high communication overhead.} \\
      {Park et al. \cite{park2022coordinated}} & {MADRL} & {Quality of Service} & {Scalable for large agent teams; requires centralized controller.} \\
      {DRL-CCMCO \cite{chen2022deepa}} & {Deep RL} & {Execution Delay, Energy Consumption} & {Fast convergence and stable; sensitive to hyperparameters.} \\
      {Gao et al. \cite{gao2024cloudedge}} & {MADRL} & {Optimization Performance, Latency} & {Handles non-stationary environments; complex actor-critic design.} \\
      {Li et al. \cite{li2024load}} & {MARL} & {Task Completion Latency, Energy Consumption} & {Simple implementation; struggles with large state spaces.} \\
      {Coop-MADRL \cite{suzuki2023multiagent}} & {Cooperative MADRL} & {Network Utilization, Task Latency} & {Promotes cooperation; credit assignment is challenging.} \\
      {Xu et al. \cite{xu2021multiagent}} & {MARL} & {Network Delay, QoS Satisfaction Rate} & {Effective for continuous action spaces; high variance.} \\
      {Kim et al. \cite{kim2024collaborative}} & {Federated RL} & {Learning Speed, Adaptability} & {Improves sample efficiency; vulnerable to non-IID data.} \\
      {BiAAC \cite{afachao2024efficient}} & {Policy-Gradient RL} & {Execution Time, Network Usage, Migration Delay} & {Balances exploration-exploitation; can be sample inefficient.} \\
      {A2C-DRL \cite{lu2024a2cdrl}} & {Deep RL} & {Reward, Task Rejection, Load-balancing} & {More stable than policy gradient; can be slow to converge.} \\
      {Nieto et al. \cite{nieto2024deep}} & {DRL} & {Quality-of-Experience, Energy} & {Effective for discrete actions; overestimates Q-values.} \\
      {Sharma et al. \cite{sharma2024deep}} & {Meta-RL} & {Time and energy consumption} & {Fast adaptation to new tasks; computationally expensive.} \\
      {MADRLOM \cite{guo2024madrlom}} & {MADRL} & {Total system overhead} & {Good for coordination; may not scale to many agents.} \\
      {Zhang et al. \cite{zhang2024hybrid}} & {Hybrid DRL} & {Task completion time, Energy consumption} & {Handles hybrid action spaces; complex architecture.} \\ \bottomrule
  \end{tabular}}}
\vspace{-15px}
\fontsize{7.5pt}{2pt}\selectfont {
\begin{minipage}{0.49\textwidth}
\textit{Notes}: Single-agent RL is better aligned with localized adaptation, whereas MARL improves coordination under coupled decisions at the expense of non-stationarity, credit-assignment difficulty, and greater communication overhead.
\end{minipage}
}
\end{table}

%% file: fig5_sec5.tex
\tikzstyle{my-box}=[
    rectangle,
    draw=hidden-draw,
    rounded corners,
    text opacity=1,
    minimum height=3.5em,
    minimum width=5em,
    align=center,
    fill opacity=1,
    line width=0.8pt,
    font=\fontsize{13.5}{16}\selectfont,
    text=black,
]

\tikzstyle{tklevel0}=[my-box,
    fill=level0,
    font=\fontsize{16}{16}\selectfont,
]

\tikzstyle{tklevel1}=[my-box,
    fill=level1,
]

\tikzstyle{tklevel2}=[my-box,
    fill=level2,
]
\tikzstyle{tklevel3}=[my-box,
    fill=level3,
]

\tikzstyle{leaf}=[my-box,
    fill=level4,
]

\begin{figure}[t]
    \centering
    \resizebox{0.49\textwidth}{!}
    {
        \begin{forest}
            forked edges,
            for tree={
                fill=level0,
                grow=east,
                reversed=true,
                anchor=base west,
                parent anchor=east,
                child anchor=west,
                base=center,
                font=\large,
                rectangle,
                draw=hidden-draw,
                rounded corners,
                align=center,
                minimum width=4em,
                edge+={darkgray, line width=1pt},
                s sep=3pt,
                line width=0.8pt,
                ver/.style={rotate=90, child anchor=north, parent anchor=south, anchor=center},
            },
            where level=0{ver,text width=35em, tklevel0}{},
            where level=1{ver, text width=12em, tklevel1}{},
            where level=2{text width=11em,tklevel2,}{},
            where level=3{text width=42.8em,tklevel3,minimum height=5em}{},
            [Security \& Privacy Protection (§\ref{sec5})
                [Security Threats \&\\Vulnerabilities (§\ref{sec5.1})
                    [Data Breaches \&\\Eavesdropping
                        [\citet{rahman2023blockchainbased}{,} \citet{fang2024secure}{,} \citet{chi2024trusted}{,} \citet{yang2024differentially}, leaf]
                    ]
                    [Denial-of-Service\\Attacks
                        [FLEAM~\cite{li2022fleam}{,} ThingPot~\cite{wang2018thingpot}{,} \citet{chen2024intelligenta}{,} \citet{zhou2024enhancing}, leaf]
                    ]
                    [Data Integrity \&\\Authentication
                        [\citet{dui2024maintenance}{,} \citet{chi2024trusted}{,} \citet{xiao2024domainspecific}{,} \citet{yang2024differentially}, leaf]
                    ]
                ]
                [Security Mechanisms\\Solutions (§\ref{sec5.2})
                    [Encryption \&\\Authentication
                        [CP-ABE~\cite{xiao2024domainspecific}{,} GS-SNC~\cite{fang2024secure}{,} \citet{yao2024efficient}{,} \citet{chi2024trusted}, leaf]
                    ]
                    [Intrusion Detection\\\& Prevention
                        [\citet{zhou2024enhancing}{,} \citet{yigit2023digital}{,} \citet{saezdecamara2023clustered}{,} \citet{raj2021reliable}, leaf]
                    ]
                    [Blockchain\\Technologies
                        [\citet{blasch2019blockchain}{,} EdgeChain~\cite{pan2019edgechain}{,} \citet{gadekallu2022blockchain}{,} \citet{chi2024trusted}, leaf]
                    ]
                ]
                [Privacy-Preserving\\Techniques (§\ref{sec5.3})
                    [FL \& Differential\\Privacy
                        [\citet{li2023edgecloud}{,} \citet{liu2024differentially}{,} \citet{yang2024differentially}{,} QuAsyncFL~\cite{liu2024quasyncfl}, leaf]
                    ]
                    [Homomorphic\\Encryption
                        [\citet{bujari2023layered}{,} \citet{zhang2024efficient}{,} \citet{kim2024privacy}{,} \citet{chatel2024verifiable}, leaf]
                    ]
                    [Secure Multi-Party\\Computation
                        [SPRITE~\cite{sengupta2022sprite}{,} \citet{zhou2024enhancing}{,} \citet{fang2024secure}{,} PrivaScissors~\cite{duan2023privascissors}, leaf]
                    ]
                ]
            ]
        \end{forest}
    }    
\vspace{-10px}
\caption{\protect Taxonomy of security \& privacy protection.}
\label{fig:taxonomy_sec5}
\vspace{-10pt}
\end{figure}

%% file: t7_security_5.1.tex
\begin{table}[t!]
  \centering
  \caption{Summary of methods addressing security threats and vulnerabilities in CISAIoT networks.}
  \label{tab:7}
  \vspace{-0.1cm}
  \setlength{\tabcolsep}{3mm}{
  \resizebox{0.49\textwidth}{!}{
    \begin{tabular}{@{}c>{\centering\arraybackslash}p{2.5cm}>{\centering\arraybackslash}p{3cm}cc@{}}
      \toprule
      \textbf{Method} & \textbf{Scenarios} & \textbf{Performance Metrics} & \textbf{ECC} & \textbf{CETC} \\ \midrule
            Yang et al. \cite{yang2024differentially} & AIoT Data Prediction & Privacy and Accuracy & \checkmark & \checkmark \\
      Rahman et al. \cite{rahman2023blockchainbased} & Industry 4.0 CPS & Security, Detection Performance & \checkmark & \redcross \\
      Fang et al. \cite{fang2024secure} & CETC-AIoT & Computational Complexity & \checkmark & \checkmark \\
      Chen et al. \cite{chen2024intelligenta} & Workflow Scheduling & Execution Cost, Deadline Constraints & \checkmark & \redcross \\
      Wang et al. \cite{wang2022privacypreserving} & Security Inference & Privacy-Utility Tradeoff & \checkmark & \redcross \\
      Zhou et al. \cite{zhou2024enhancing} & AIoT Services & QoS, Effectiveness, Efficiency & \checkmark & \redcross \\
      Wang et al. \cite{wang2018thingpot} & IoT Security & Attack Detection & \redcross & \redcross \\
      Li et al. \cite{li2022fleam} & Industrial IoT & Attack Cost, Detection Accuracy & \checkmark & \redcross \\
      Chi et al. \cite{chi2024trusted} & AIoT Edge Applications & Security, Model Efficiency & \checkmark & \checkmark \\
      Raj et al. \cite{raj2021reliable} & AIoT Applications & Efficiency, Robustness & \checkmark & \redcross \\
      Dui et al. \cite{dui2024maintenance} & IIoT Systems & Reliability, Gateway Cost & \checkmark & \redcross \\
      Xiao et al. \cite{xiao2024domainspecific} & Cloud-Edge IoT & Security, Efficiency & \checkmark & \checkmark \\ \bottomrule  
      
  \end{tabular}}}
\vspace{-15px}
\fontsize{7.5pt}{2pt}\selectfont {
\begin{minipage}{0.49\textwidth}
\textit{Notes}: Security mechanisms should be interpreted jointly with resource management and collaborative learning, because stronger cryptographic or trust guarantees often trade against latency, energy efficiency, and synchronization cost.
\end{minipage}
}
\end{table}

%% file: fig6_sec6.tex
\tikzstyle{my-box}=[
    rectangle,
    draw=hidden-draw,
    rounded corners,
    text opacity=1,
    minimum height=3.5em,
    minimum width=5em,
    align=center,
    fill opacity=1,
    line width=0.8pt,
    font=\fontsize{13.5}{16}\selectfont,
    text=black,
]

\tikzstyle{tklevel0}=[my-box,
    fill=level0,
    font=\fontsize{16}{16}\selectfont,
]

\tikzstyle{tklevel1}=[my-box,
    fill=level1,
]

\tikzstyle{tklevel2}=[my-box,
    fill=level2,
]
\tikzstyle{tklevel3}=[my-box,
    fill=level3,
]

\tikzstyle{leaf}=[my-box,
    fill=level4,
]

\begin{figure}[t]
    \centering
    \resizebox{0.49\textwidth}{!}
    {
        \begin{forest}
            forked edges,
            for tree={
                fill=level0,
                grow=east,
                reversed=true,
                anchor=base west,
                parent anchor=east,
                child anchor=west,
                base=center,
                font=\large,
                rectangle,
                draw=hidden-draw,
                rounded corners,
                align=center,
                minimum width=4em,
                edge+={darkgray, line width=1pt},
                s sep=3pt,
                line width=0.8pt,
                ver/.style={rotate=90, child anchor=north, parent anchor=south, anchor=center},
            },
            where level=0{ver,text width=30em, tklevel0}{},
            where level=1{ver, text width=11.5em, tklevel1}{},
            where level=2{text width=10em,tklevel2,}{},
            where level=3{text width=42.8em,tklevel3,minimum height=5em}{},
            [Data Management \& Communication (§\ref{sec6})
                [DAP (§\ref{sec6.1}),text width=8em,
                    [Data Filtering \&\\Aggregation
                        [\citet{zhu2023pushing}{,} \citet{raj2021reliable}{,} \citet{becker2020aiops}{,} \citet{liu2024quasyncfl}, leaf]
                    ]
                    [Data Compression\\\& Encoding
                        [\citet{underwood2024understanding}{,} \citet{noura2023deep}{,} IDEALEM~\cite{lee2019idealem}, leaf]
                    ]
                ]
                [DSR (§\ref{sec6.2})
                    [Edge\\Caching
                        [PEC~\cite{li2023predictive}{,} D3QN~\cite{zhang2024how}{,} \citet{torabi2024learningbased}{,} \citet{zheng2024optimization}, leaf]
                    ]
                    [Distributed Data\\Storage
                        [PilotEdge~\cite{luckow2021pilotedge}{,} \citet{raj2021reliable}{,} CEFIoT~\cite{javed2018cefiot}{,} \citet{chen2023remr}, leaf]
                    ]
                    [Data Integrity \&\\Availability
                        [\citet{dui2024maintenance}{,} \citet{chi2024trusted}{,} \citet{javed2018cefiot}{,} \citet{chen2023remr}, leaf]
                    ]
                ]
                [CPO (§\ref{sec6.3})
                    [Communication\\Protocols
                        [\citet{paolo2023security}{,} \citet{gundogan2020iot}{,} GS-SNC~\cite{fang2024secure}{,} \citet{phung2020enhancing}, leaf]
                    ]
                    [Bandwidth\\Optimization
                        [\citet{li2024cloudedgedevice}{,} \citet{wang2024cloudedgeend}{,} \citet{xu2024fusion}{,} \citet{nezami2021decentralized}, leaf]
                    ]
                    [Fault Tolerance \&\\Quality Assurance
                        [\citet{nikolic2021selfhealing}{,} \citet{javed2018cefiot}{,} \citet{dui2024maintenance}{,} \citet{arroba2024sustainable}, leaf]
                    ]
                ]
            ]
        \end{forest}
    }    
\vspace{-10px}
\caption{\protect Taxonomy of data management and communication techniques, including data acquisition \& preprocessing (DAP), data storage \& retrieval (DSR), and communication protocols \& optimization (CPO).}
\label{fig:taxonomy_sec6}
\vspace{-10pt}
\end{figure}

%% file: t8_data_6.1.tex
\begin{table}[t!]
  \centering
  \caption{Summary of data preprocessing and management for transmission in resource-constrained environments.}
  \label{tab:8}
  \vspace{-0.1cm}
  \setlength{\tabcolsep}{0.4mm}{
  \resizebox{0.49\textwidth}{!}{
    \begin{tabular}{@{}c>{\centering\arraybackslash}p{2.5cm}>{\centering\arraybackslash}p{2.5cm}>{\centering\arraybackslash}p{3.5cm}cc@{}}
      \toprule
      \textbf{Method} & \textbf{Type} & \textbf{Scenarios} & \textbf{Performance Metrics} & \textbf{ECC} & \textbf{CETC} \\ \midrule
      Raj et al. \cite{raj2021reliable} & Edge MLOps & AIoT Fleet Analytics & Forecasting Accuracy & \checkmark & \redcross \\
      Becker et al. \cite{becker2020aiops} & AIOps & Edge Computing & Resource Utilization, Detection Performance & \redcross & \redcross \\
      QuAsyncFL \cite{liu2024quasyncfl} & Federated Learning & Cloud-Edge-Terminal AIoT & Communication Efficiency, Convergence & \checkmark & \checkmark \\
      Underwood et al. \cite{underwood2024understanding} & Data Compression & ML/AI Training & Compression Ratio, Model Quality & \redcross & \redcross \\
      Noura et al. \cite{noura2023deep} & DL-based Compression & Multimedia IoT & Visual Quality, Compression Ratio & \checkmark & \redcross \\
      Edge Source Coding \cite{lu2020user} & Lossless Compression & Edge Networks & Number of Bits Needed & \redcross & \redcross \\
      Taurone et al. \cite{taurone2023change} & Floating-point Compression & IoT Time-Series Data & Compression Ratio, Recovery Error & \redcross & \redcross \\
      Shadura et al. \cite{shadura2020root} & Lossless Compression & HEP Analysis & I/O Bandwidth, Storage Space & \redcross & \redcross \\
      Melissaris et al. \cite{melissaris2020optimizing} & Selective Compression & IoT and Web Traffic & Network Transfer Speed, Data Savings & \redcross & \redcross \\
      Banerjee et al. \cite{banerjee2022scalable} & Hybrid Compression & Scientific Data & Compression Factor, QoI Preservation & \checkmark & \redcross \\
      LTSDC \cite{agrawal2022lossless} & Statistical Compression & IoT Time-Series Data & Compression Ratio & \redcross & \redcross \\
      Zhou et al. \cite{zhou2024enhancing} & Federated Learning & Edge-Cloud AIoT & QoS, Latency & \checkmark & \redcross \\
      Arroba et al. \cite{arroba2024sustainable} & Sustainable Computing & Edge Infrastructure & Energy Efficiency & \redcross & \redcross \\
      QOMVN \cite{zheng2024optimization} & Edge Collaborative Caching & Vehicular Networks & Data Query Latency & \checkmark & \redcross \\
      Rac et al. \cite{rac2021edge} & Containerization & Edge Computing & Application Execution Efficiency & \redcross & \redcross \\ \bottomrule
  \end{tabular}}}
\vspace{-15px}
\fontsize{7.5pt}{2pt}\selectfont {
\begin{minipage}{0.49\textwidth}
\textit{Notes}: Compression, filtering, and federated preprocessing mainly differ in how they trade communication reduction against information loss, convergence speed, and implementation complexity under heterogeneous links.
\end{minipage}
}
\end{table}

%% file: fig7_sec7.tex
\tikzstyle{my-box}=[
    rectangle,
    draw=hidden-draw,
    rounded corners,
    text opacity=1,
    minimum height=3.5em,
    minimum width=5em,
    align=center,
    fill opacity=1,
    line width=0.8pt,
    font=\fontsize{13.5}{16}\selectfont,
    text=black,
]

\tikzstyle{tklevel0}=[my-box,
    fill=level0,
    font=\fontsize{16}{16}\selectfont,
]

\tikzstyle{tklevel1}=[my-box,
    fill=level1,
]

\tikzstyle{tklevel2}=[my-box,
    fill=level2,
    draw=none,
]
\tikzstyle{tklevel3}=[
    rectangle,
    text opacity=1,
    minimum width=5em,
    align=center,
    font=\fontsize{13.5}{16}\selectfont,
    text=black,
    draw=hidden-draw,
    fill=none,
]

\tikzstyle{leaf}=[my-box,
    fill=level4,
]

\begin{figure}[t]
    \centering
    \resizebox{0.49\textwidth}{!}
    {
        \begin{forest}
            forked edges,
            for tree={
                fill=level0,
                grow=east,
                reversed=true,
                anchor=base west,
                parent anchor=east,
                child anchor=west,
                base=center,
                font=\large,
                rectangle,
                draw=hidden-draw,
                rounded corners,
                align=center,
                minimum width=4em,
                edge+={darkgray, line width=1pt},
                s sep=3pt,
                line width=0.8pt,
                ver/.style={rotate=90, child anchor=north, parent anchor=south, anchor=center},
            },
            where level=0{ver,text width=35em, tklevel0}{},
            where level=1{ver, text width=12em, tklevel1}{},
            where level=2{text width=14em,tklevel2,}{},
            where level=3{text width=42.8em,tklevel3}{},
            [Applications (§\ref{sec7})
                [Manufacturing\\(§\ref{sec7.1})
                    [Real-Time Defect\\Detection
                        [{Terminal sensors and IoT devices capture real-time production data, edge\\processes information for rapid identification of critical flaws \cite{bujari2023layered,becker2020aiops}.}]
                    ]
                    [Predictive\\Maintenance
                        [{Data analysis and machine learning algorithms predict equipment failures\\enabling proactive maintenance and minimizing downtime \cite{dosluoglu2022circuit,xu2024fusion}.}]
                    ]
                    [Automated Quality\\Control
                        [{Lightweight AI models on edge devices enable real-time defect\\detection through rapid sensor analysis and assessment \cite{raj2021reliable,wang2024cloudedgeend}.}]
                    ]
                ]
                [Transportation\\(§\ref{sec7.2})
                    [Autonomous Vehicle\\Nav. \& Control
                        [{Onboard sensors and computers collect data, edge processes information\\for object detection while cloud provides traffic management \cite{tang2018piedge,zhang2024cloudedgeterminal}.}]
                    ]
                    [Traffic Flow Optim.\\\& Congestion Mgmt.
                        [{Edge computing proximity to data sources enables real-time responses\\to changing traffic conditions for optimization strategies \cite{vo2022edge,ni2023mscet}.}]
                    ]
                    [Real-time Traffic\\Monitor. \& Detection
                        [{Edge devices process traffic camera and sensor data, AI algorithms\\identify anomalies for immediate emergency responses \cite{katariya2023vegaedge,ghosh2021mobility}.}]
                    ]
                ]
                [Healthcare (§\ref{sec7.3})
                    [Remote Patient\\Monitoring
                        [{Wearable sensors and IoT devices collect physiological data, edge\\preprocessing enables timely interventions and real-time alerts \cite{shaik2023remote,gupta2021hierarchical}.}]
                    ]
                    [AI-assisted Diagnosis\\\& Prognosis
                        [{Edge comput. processes patient data in real-time for preliminary analysis,\\cloud provides sophisticated AI models for diagnosis \cite{hayyolalam2021edge,joshi2024integration}.}]
                    ]
                    [Personalized Treatment\\\& Medication Mgmt.
                        [{Integrating wearable data with edge and cloud platforms enables\\ patient monitoring and analysis for treatment plans \cite{buyya2023quality,luckow2021pilotedge}.}]
                    ]
                ]
                [Cities (§\ref{sec7.4})\&\\Agriculture (§\ref{sec7.5}), text width=14em
                    [Smart Surveillance\\\& Crime Prevention
                        [{Distributed approach processes surveil. data from multiple sources, edge\\deployment enables real-time processing for prevention \cite{nikouei2019decentralized,zhou2021deeplearningenhanced}.}]
                    ]
                    [Environmental Monitor.\\\& Pollution Control
                        [{Edge devices collect environmental data city-wide from various sensors,\\localized processing reduces latency for responses to incidents \cite{arroba2024sustainable,vo2022edge}.}]
                    ]
                    [Smart Energy Mgmt.\\\& Grid Optimization
                        [{Edge devices collect energy usage and grid condition data, collaborative\\approach facilitates efficient energy allocation and optimization \cite{zhai2024edgecloud,hua2024energyefficient}.}]
                    ]
                    [Precision Agriculture\\\& Crop Monitoring
                        [{Edge computing processes IoT sensor data in real-time for immediate\\responses, cloud performs complex predictions and analytics \cite{kumar2021affordable,cao2022igrow}.}]
                    ]
                ]
            ]
        \end{forest}
    }    
\vspace{-10px}
\caption{\protect Overview of applications across different domains.}
\label{fig:taxonomy_sec7}
\vspace{-15pt}
\end{figure}

%% file: fig8_fr.tex
\newcommand{\researchicon}{\textcolor{red!70}{\faCompass}}
\newcommand{\resourceicon}{\textcolor{orange!70}{\faCogs}}
\newcommand{\securityicon}{\textcolor{red!70}{\faLock}}
\newcommand{\standardicon}{\textcolor{green!70}{\faClipboardList}}
\newcommand{\aiicon}{\textcolor{purple!70}{\faBrain}}
\newcommand{\networkicon}{\textcolor{cyan!70}{\faBroadcastTower}}
\newcommand{\twinicon}{\textcolor{teal!70}{\faUsers}}
\newcommand{\llmicon}{\textcolor{magenta!70}{\faRobot}}
\newcommand{\quantumicon}{\textcolor{violet!70}{\faAtom}}
\resizebox{0.49\textwidth}{!}{%
\begin{tikzpicture}[
    level1/.style={rectangle, rounded corners=10pt, fill=level0!80, text=black, 
                   text width=3.5cm, align=center, font=\Large\bfseries, 
                   minimum height=1.5cm, draw=black, line width=1.5pt},
    level2/.style={rectangle, rounded corners=8pt, fill=level3!80, text=black, 
                   text width=3.2cm, align=center, font=\fontsize{10}{12}\selectfont, 
                   minimum height=1.2cm, draw=black, line width=1pt},
    arrow/.style={-stealth, very thick, color=black!70}
]
\coordinate (pos1) at (-4.5, 2.1);   
\coordinate (pos2) at (0, 2.1);    
\coordinate (pos3) at (4.5, 2.1);    
\coordinate (pos4) at (-4.5, 0);   
\coordinate (pos5) at (0, 0);    
\coordinate (pos6) at (4.5, 0);    
\coordinate (pos7) at (-4.5, -2.1);  
\coordinate (pos8) at (0, -2.1);   
\coordinate (pos9) at (4.5, -2.1);   
\node[level1] (center) at (pos5) {\researchicon~Future Directions};
\node[level2] (node1) at (pos1) {\resourceicon\quad Resource Magt.\\[0.3em]\cite{levin2019aiops}, \cite{chen2022resource}, \cite{xiao2024domainspecific}};
\node[level2] (node2) at (pos2) {\securityicon~Security \& Privacy\\[0.3em]\cite{yang2024differentially}, \cite{zhou2024enhancing}, \cite{dhinakaran2024privacypreserving}};
\node[level2] (node3) at (pos3) {\standardicon~ Standardization\\[0.3em]\cite{maia2024survey}, \cite{sowinski2023autonomous}, \cite{chi2024trusted}};
\node[level2] (node4) at (pos4) {\aiicon~ Explainable AI\\[0.3em]\cite{zolanvari2023trust}, \cite{zhou2024enhancing}, \cite{zhang2021empowering}};
\node[level2] (node6) at (pos6) {\networkicon~ 6G+ Networks\\[0.3em]\cite{christophorou2023adroit6g}, \cite{lin2024satellitemec}, \cite{wang2024cooperative}};
\node[level2] (node7) at (pos7) {\twinicon~ Digital Twin\\[0.3em]\cite{chen2024fast}, \cite{vanhuynh2022edge}, \cite{wang2024cooperative}};
\node[level2] (node8) at (pos8) {\llmicon~ LLMs \& Agents\\[0.3em]\cite{qu2024mobile}, \cite{he2024large}, \cite{xu2024cached}};
\node[level2] (node9) at (pos9) {\quantumicon~ Quantum Comput.\\[0.3em]\cite{baseri2024cybersecurity}, \cite{hossain2024quantumedge}, \cite{lu2024quantum}};
\draw[arrow] (center) -- (node1);
\draw[arrow] (center) -- (node2);
\draw[arrow] (center) -- (node3);
\draw[arrow] (center) -- (node4);
\draw[arrow] (center) -- (node6);
\draw[arrow] (center) -- (node7);
\draw[arrow] (center) -- (node8);
\draw[arrow] (center) -- (node9);

\end{tikzpicture}
}%

%% file: t4_opt_3.3.tex
\begin{table}[t!]
  \centering
  \caption{Summary of optimization techniques.}
  \label{tab:4}
  \vspace{-0.1cm}
  \setlength{\tabcolsep}{0.5mm}{
  \resizebox{0.49\textwidth}{!}{
    \begin{tabular}{@{}ccc>{\centering\arraybackslash}p{2.5cm}>{\centering\arraybackslash}p{2.5cm}>{\centering\arraybackslash}p{3cm}>{\centering\arraybackslash}p{3.5cm}@{}}
      \midrule
      {\textbf{Method}} & {\textbf{Year}} & {\textbf{Venue}} & {\textbf{Algorithm Type}} & {\textbf{Scenarios}} & {\textbf{Metrics}} & {\textbf{Performance Results}} \\ \midrule
      {TIGO \cite{he2024efficient}} & {2024} & {IEEE TSC} & {Two-stage Iterated Greedy Optimization} & {Cloud-Edge Collaborative Computing} & {Average Response Time} & {Better response times in less time compared to existing methods} \\
      {DSCF \cite{ko2024dynamic}} & {2024} & {IEEE IoTJ} & {CMDP-based Framework} & {Distributed Serverless Edge Clouds} & {Inference Latency} & {Less than half the inference latency compared to local computing} \\
      {AWSH \cite{laboni2024hyper}} & {2024} & {IEEE TMC} & {Hyper-Heuristic Algorithm} & {5G Mobile Edge Clouds} & {Latency, Load Balancing} & {Superior performance compared to SOTA solutions} \\
      {Lu et al. \cite{lu2024dynamic}} & {2024} & {IEEE TVT} & {Multi-Agent Reinforcement Learning} & {IoV Edge Cloud} & {Quality of Service} & {Effective performance in complex network situations} \\
      {Nguyen et al. \cite{nguyen2024integrated}} & {2024} & {IEEE TCC} & {Alternating Optimization} & {Space-Air-Ground Networks} & {Energy Consumption, Delay} & {Superior effectiveness compared to baselines} \\
      {Qiu et al. \cite{qiu2024online}} & {2024} & {IEEE TMC} & {Online Algorithm} & {Edge Intelligence Cloud} & {Security, Reliability, Throughput} & {Promising performance compared to baseline algorithms} \\
      {Wei et al. \cite{wei2024joint}} & {2024} & {IEEE TPDS} & {Two-stage Algorithm} & {Federated Learning in Edge Clouds} & {Learning Costs} & {33.5\%-39.6\% lower average total costs vs. previous methods} \\
      {SemCom-SEC \cite{zheng2024semantic}} & {2024} & {IEEE JSAC} & {Federated Learning, Game Theory} & {Satellite-Borne Edge Cloud Network} & {Delay, Energy Consumption, Privacy} & {Reduced delay and energy consumption while enhancing privacy} \\
      {Asgarian et al. \cite{asgarian2024efficient}} & {2024} & {IEEE IoTJ} & {Approximation Algorithm} & {Edge-Cloud Computing Industrial IoT} & {Delay, Resource Consumption} & {Near-optimal solution with polynomial time complexity} \\
      {Bai et al. \cite{bai2024delayaware}} & {2024} & {IEEE TMC} & {Convex Approximation, Lyapunov Optimization} & {UAV-Enabled Edge-Cloud Computing} & {Task Completion Delay} & {Significant reduction in task completion delay} \\
      {JTOCC \cite{fang2024joint}} & {2024} & {IEEE TWC} & {DRL, LSTM} & {NOMA-Aided CETCI} & {Network Latency} & {Lower-delay content distribution and fast convergence} \\
      {PRAISE \cite{feng2024dependencyaware}} & {2024} & {IEEE TMC} & {Task Reconfiguration} & {Multi-Access Edge Cloud Networks} & {Total Offloading Benefit} & {Higher system benefits and lower resource costs than baselines} \\
      {CORA \cite{fu2024datadriven}} & {2024} & {IEEE TWC} & {Data-driven Optimization} & {Mobile Edge Clouds} & {Quality of Experience} & {18\% reduction in user complaint rate on YouTube dataset} \\
      {Co-NOMA \cite{reifert2024extended}} & {2024} & {IEEE IoTJ} & {Cooperative NOMA} & {Reality in Cloud/Edge Computing} & {Logarithmic Rate, Delay} & {Outperforms traditional multiple access and NOMA methods} \\
      {Xiao et al. \cite{xiao2024transcodingenabled}} & {2024} & {IEEE IoTJ} & {Online Learning} & {Video Caching in IoT Networks} & {Delay, Resource Utilization} & {Performance advantages compared to SOTA algorithms} \\
      {ADMM-FedMeta \cite{yue2021inexactadmm}} & {2021} & {MobiHoc} & {Federated Meta-Learning} & {Edge Learning in IoT} & {Learning Performance, Computational Cost} & {Substantially outperforms existing baselines} \\
      {RL-ROIS \cite{zhang2024reinforcementlearningbased}} & {2024} & {IEEE IoTJ} & {Reinforcement Learning} & {RIS-Aided IoT Networks} & {Capacity, Latency, Energy Consumption} & {Superior efficiency and robustness} \\
      {Zhao et al. \cite{zhao2024taming}} & {2024} & {IEEE TMC} & {Online Algorithm} & {Serverless Cloud Inference} & {Inference Latency, Cost, Accuracy} & {Less than half inference latency compared to local computing} \\
      {MESP \cite{hazra2024distributed}} & {2024} & {IEEE IoTJ} & {Game Theory} & {Industry 5.0} & {End-to-End Execution Delay} & {20\%-30\% reduction in computational overhead and delay} \\
      {COMEC \cite{wu2024computation}} & {2024} & {IEEE TMC} & {Algorithm} & {Multi-Cell Networks} & {Total cost (delay and energy)} & {Superior system cost compared to baselines} \\
      {Zhou et al. \cite{zhou2024enhancing}} & {2024} & {ADHOC} & {Framework} & {AIoT} & {Quality of Service} & {Higher effectiveness and efficiency than centralized approach} \\
      {Dragotto et al. \cite{dragotto2024critical}} & {2024} & {JCC} & {Game-theoretic model} & {Cloud Networks} & {Network security posture} & {Efficacy demonstrated through computational tests} \\
      {Raj et al. \cite{raj2021reliable}} & {2021} & {ArXiv} & {Framework} & {Edge IoT} & {Efficiency, Robustness} & {Validated through air quality forecasting experiment} \\
      {PGRA \cite{gao2022virtual}} & {2020} & {ArXiv} & {Algorithm} & {Satellite Edge Computing} & {Network payoff, Deployment cost} & {Effectively addresses VNF placement problem} \\
      {MADRLOM \cite{guo2024madrlom}} & {2024} & {COMNET} & {Mechanism} & {Software-Defined Cloud-Edge Networks} & {System overhead} & {Significant reduction in total system overhead} \\
      {Zhang et al. \cite{zhang2024stackelberggamebased}} & {2024} & {IEEE TCC} & {Game-theoretic} & {Mobile Edge Computing} & {Task completion delay, User cost} & {Superior performance compared to several approaches} \\
      {Li et al. \cite{li2024computation}} & {2024} & {ArXiv} & {Framework} & {Wireless Powered Edge Computing} & {Weighted sum computation rate} & {Comparable to exhaustive search with reduced execution time} \\ \bottomrule
           
  \end{tabular}}}
\vspace{-15px}
\end{table}

%% file: t7_app_comparison.tex
\begin{table*}[t!]
  \centering
  \caption{Comparison of AIoT application domains in cloud-edge-terminal collaborative intelligence.}
  \label{tab:app_comparison}
  \vspace{-0.1cm}

  \setlength{\tabcolsep}{5pt}{
  \resizebox{0.98\textwidth}{!}{
    \begin{tabular}{@{}p{2.2cm}>{\raggedright\arraybackslash}p{4.2cm}>{\raggedright\arraybackslash}p{4.5cm}>{\raggedright\arraybackslash}p{5cm}@{}}
      \toprule
      \textbf{Application} & \textbf{IoT Device} & \textbf{Key Requirements} & \textbf{Edge Role} \\ \midrule
      {Smart\par Manufacturing} & High-speed cameras, acoustic sensors, vibration sensors, robotic arms & Ultra-low latency ($<$10ms), high bandwidth, high reliability, data security & Real-time inference for defect detection, predictive maintenance analytics, data filtering and aggregation, robotic control \\
      {Smart\par Transportation} & LiDAR, radar, HD cameras, V2X communication units, roadside units (RSUs) & Ultra-low latency ($<$5ms), extreme reliability (99.999\%), high mobility support, security & Cooperative perception, sensor fusion, real-time trajectory prediction, offloading autonomous driving tasks, local traffic management \\
      {Smart\par Healthcare} & Wearable sensors (ECG, SpO2), smart beds, remote monitoring devices, medical imaging equipment & Low latency, high privacy and security (HIPAA compliance), high reliability, data integrity & Real-time anomaly detection in vital signs, preprocessing of medical data, secure data aggregation, executing local alerts, FL for diagnostics \\
      {Smart Cities} & Surveillance cameras, environmental sensors (air/water quality), smart meters, traffic sensors & Varies by application: real-time for public safety, delay-tolerant for environmental monitoring, high bandwidth for video & Real-time video analytics for surveillance, event detection (e.g., traffic incidents), data aggregation from sensors, smart grid management \\
      {Smart\par Agriculture} & Soil moisture sensors, drones with multispectral cameras, weather stations, automated farm machinery & Generally delay-tolerant, low-to-medium bandwidth, long battery life for sensors, wide-area coverage & Data preprocessing from sensors, drone control and navigation, localized analytics for precision irrigation, initial pest/disease detection \\ \bottomrule
  \end{tabular}}}
\vspace{-15px}
\end{table*}

%% file: bios.tex
\begin{IEEEbiography}[{\includegraphics[width=1in,height=1.25in,clip,keepaspectratio]{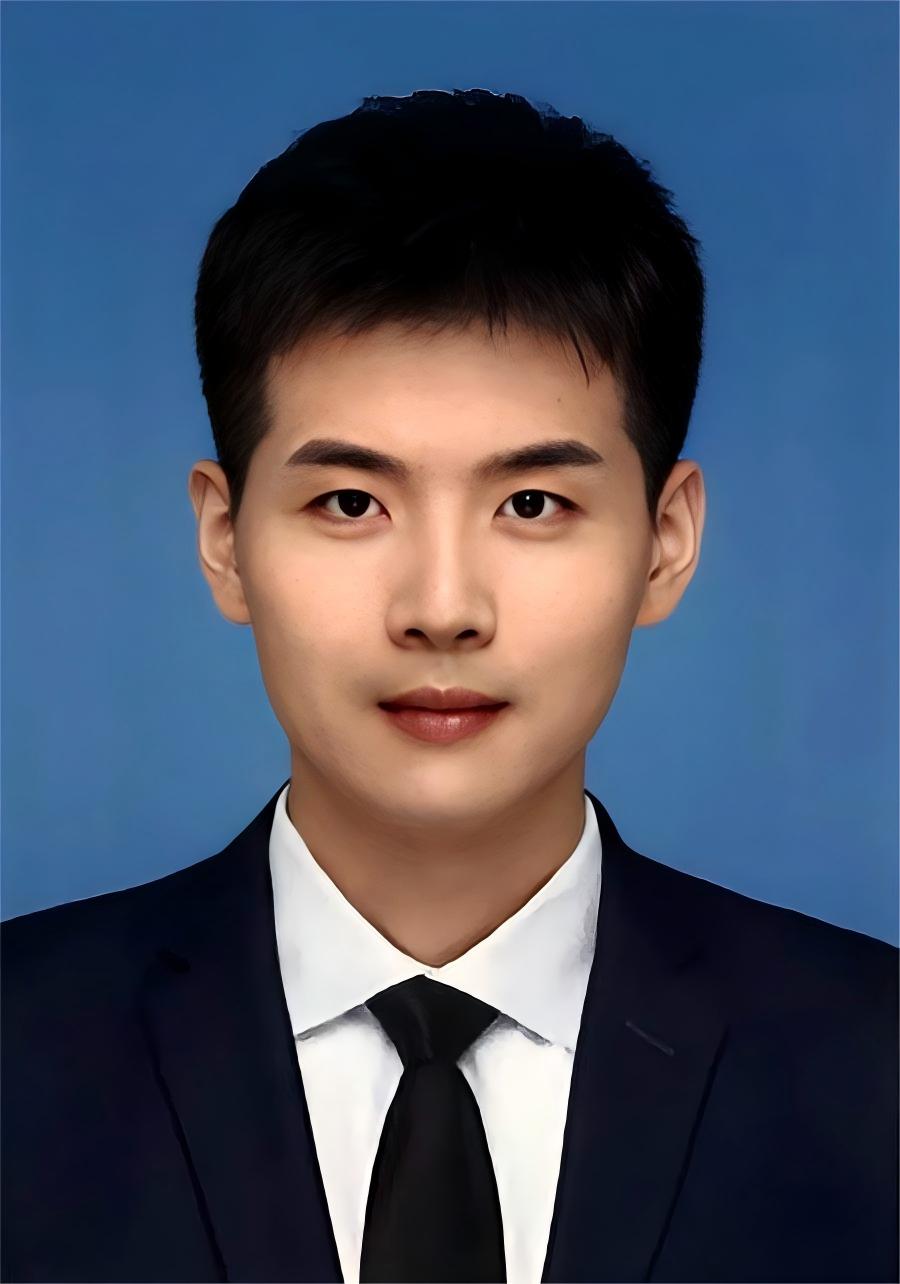}}]{Jiaqi Wu}
received his Ph.D. degree in 2025 from China University of Mining and Technology (Beijing). His research interests include computer vision, computer graphics, the Internet of Things, and blockchain technology. He was a visiting joint doctoral student at the University of British Columbia supported by the China Scholarship Council in 2023. Currently, he is a postdoctoral researcher with the Department of Automation, Tsinghua University.
\end{IEEEbiography}

\begin{IEEEbiography}[{\includegraphics[width=1in,height=1.25in,clip,keepaspectratio]{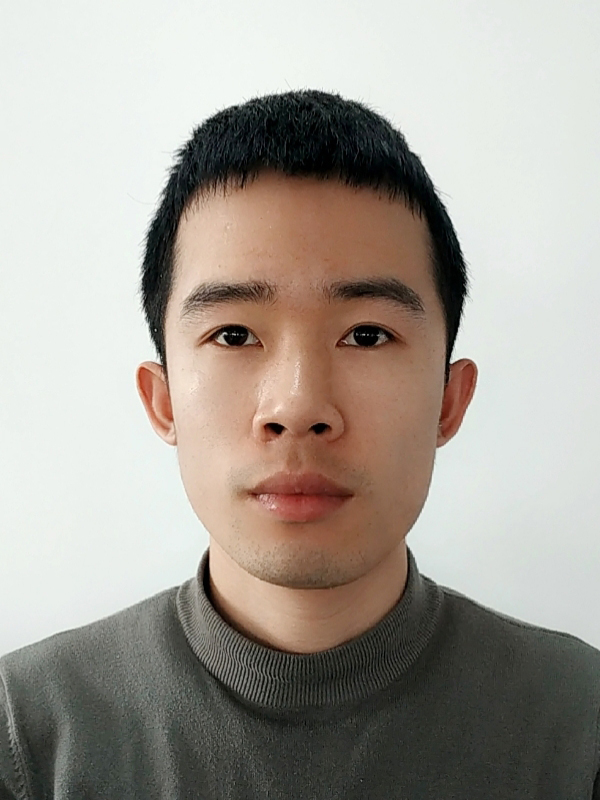}}]{Jing Liu}
(Member, IEEE) received the B.S. and M.S. degrees in Computer Science and Technology and Software Engineering from the University of Shanghai for Science and Technology, Shanghai, China, in 2014 and 2017, respectively, and the Ph.D. degree in Electronics and Information from Fudan University, Shanghai, China, in 2023. 

He is currently a Postdoctoral Research Fellow at the Department of Electrical and Computer Engineering, the University of British Columbia, Vancouver, Canada. His research interests include distributed learning, edge-cloud collaboration, anomaly detection, multimodal learning, and knowledge transfer.
\end{IEEEbiography}

\begin{IEEEbiography}[{\includegraphics[width=1in,height=1.25in,clip,keepaspectratio]{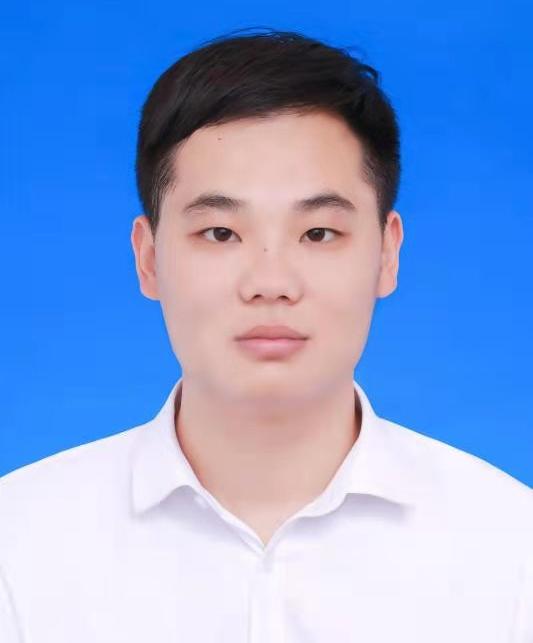}}]{Yang Liu}
received the B.S. degree (with honors) in automation from Nanjing University, China, in 2020, and the Ph.D. degree in computer science from Fudan University, China, in 2025. From 2023 to 2025, he was a visiting scholar at the University of Toronto, Canada, under a joint Ph.D. program. His research focuses on anomaly detection with applications in industry and intelligent transportation systems.

Dr. Liu has published multiple first-author papers in prestigious journals including ACM CSUR, IEEE TIP, IEEE TII, and IEEE TIE. He has served as Area Chair for BMVC 2025 and IEEE ICIP 2025, Workshop/Special Session Co-Chair for IEEE ICIP, IEEE DSP, and IEEE WF-IoT, and TPC member for top-tier conferences including CVPR, ICCV, ICLR, NeurIPS, AAAI, and ACM MM. He is an active reviewer for ACM CSUR, PR, INFU, and multiple IEEE Transactions including TIP, TII, TNNLS, TCSVT, TMM, and TCYB. His work has received over 1,300 citations with an h-index of 21.
\end{IEEEbiography}

\begin{IEEEbiography}[{\includegraphics[width=1in,height=1.25in,clip,keepaspectratio]{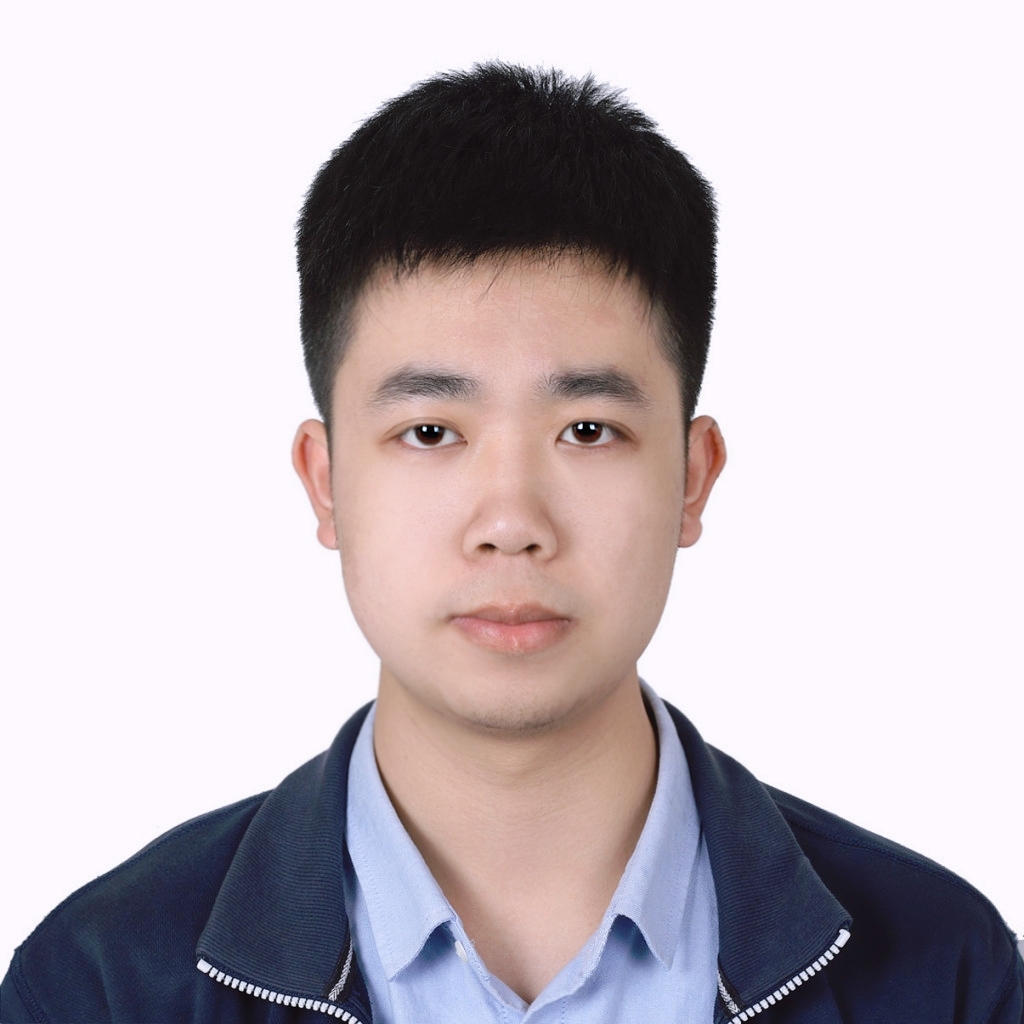}}]{Lixu Wang}
received the Ph.D. degree at Northwestern University, Evanston, IL, USA, in 2024, and the B.E. degree from Zhejiang University, Hangzhou, China, in 2020. He is currently working as a research fellow at Nanyang Technological University, Singapore. He has published over 20 papers at top-tier conferences including NeurIPS, ICLR, CVPR, ICCV, AAAI, Usenix Security. His current research focuses on achieving Data-centric Robust AI by boosting utility, guarding privacy, and securing copyright.
\end{IEEEbiography}

\begin{IEEEbiography}[{\includegraphics[width=1in,height=1.25in,clip,keepaspectratio]{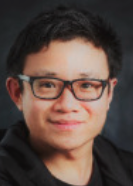}}]{Zehua Wang}
received Ph.D. degree from The University of British Columbia (UBC), Vancouver in 2016. He is an Adjunct Professor in the Department of Electrical and Computer Engineering at UBC, Vancouver and the CTO at Intellium Technology Inc. He implemented the payment system for commercialized mobile ad-hoc networks with the payment/state channel technology on blockchain. His is interested in protocol and mechanism design with optimization and game theories to improve the efficiency and robustness for communication networks, distributed systems, and social networks.
\end{IEEEbiography}

\begin{IEEEbiography}[{\includegraphics[width=1in,height=1.25in,clip,keepaspectratio]{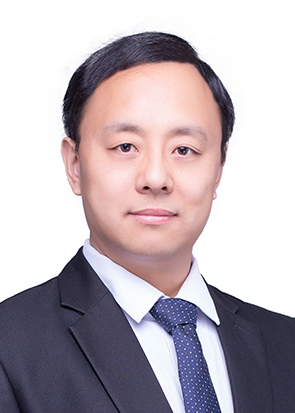}}]{Wei Chen}
[M'18] received the Ph.D. degree in communications and information systems from China University of Mining and Technology, Beijing, China, in 2008. In 2008, he joined the School of Computer Science and Technology, China University of Mining and Technology, where he is currently a professor. His research interests include machine learning, image processing, and computer networks.
\end{IEEEbiography}

\begin{IEEEbiography}[{\includegraphics[width=1in,height=1.25in,clip,keepaspectratio]{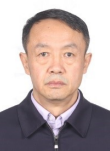}}]{Zijian Tian}
received the Ph.D. degree in communications and information systems from China University of Mining and Technology, Beijing, China, in 2003. In 2003, he joined the School of Mechanical Electronic and Information Engineering, China University of Mining and Technology(Beijing), where he is currently a professor. He is a member of National Expert Group of Safety Production, and is a member and convener of Expert Committee of Information and Automation. His research interests include image processing, analysis of coal mine surveillance video.
\end{IEEEbiography}

\begin{IEEEbiography}[{\includegraphics[width=1in,height=1.25in,clip,keepaspectratio]{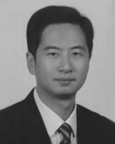}}]{F. Richard Yu}
received the PhD degree in electrical engineering from the University of British Columbia (UBC) in 2003. His research interests include connected/autonomous vehicles, artificial intelligence, blockchain, and wireless systems. He has been named in the Clarivate's list of ``Highly Cited Researchers" in computer science since 2019, Standford's Top 2$\%$ Most Highly Cited Scientist since 2020. He received several Best Paper Awards from some first-tier conferences, Carleton Research Achievement Awards in 2012 and 2021, and the Ontario Early Researcher Award (formerly Premiers Research Excellence Award) in 2011. He is a Board Member the IEEE VTS and the Editor-in-Chief for IEEE VTS Mobile World newsletter. He is a Fellow of the IEEE, Canadian Academy of Engineering (CAE), Engineering Institute of Canada (EIC), and IET. He is a Distinguished Lecturer of IEEE in both VTS and ComSoc.
\end{IEEEbiography}

\begin{IEEEbiography}[{\includegraphics[width=1in,height=1.25in,clip,keepaspectratio]{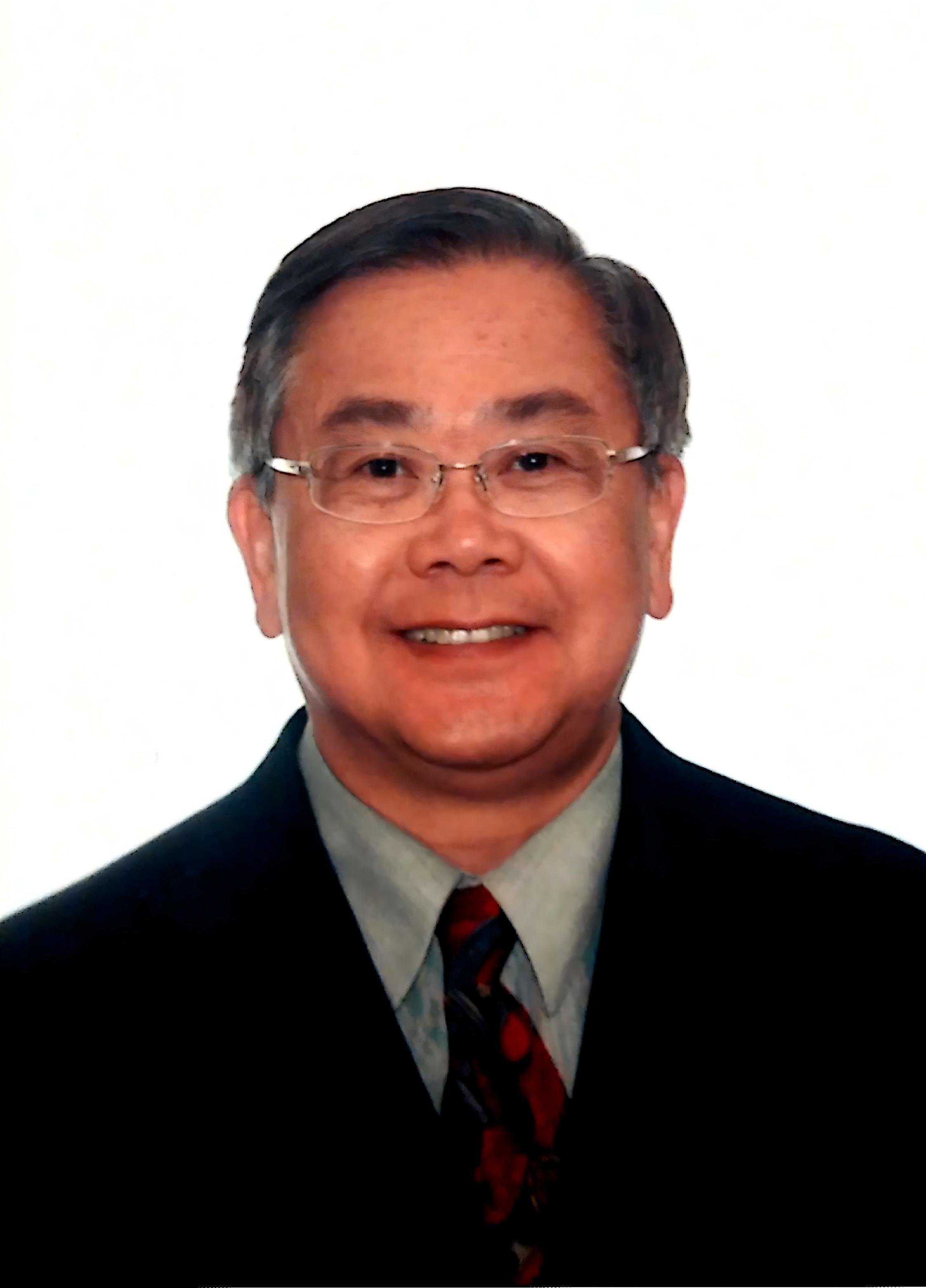}}]{Victor C. M. Leung}
(Life Fellow, IEEE) is the Dean of the Artificial Intelligence Research Institute at SMBU (SMBU), China, a Distinguished Professor of Computer Science and Software Engineering at Shenzhen University, China, and also an Emeritus Professor of Electrical and Computer Engineering and Director of the Laboratory for Wireless Networks and Mobile Systems at the University of British Columbia (UBC), Canada.  His research is in the broad areas of wireless networks and mobile systems, and he has published widely in these areas. His published works have together attracted more than 60,000 citations. He is named in the current Clarivate Analytics list of "Highly Cited Researchers". Dr. Leung is serving on the editorial boards of the IEEE Transactions on Green Communications and Networking, IEEE Transactions on Computational Social Systems, and several other journals. He co-authored papers that were selected for the 2017 IEEE ComSoc Fred W. Ellersick Prize, 2017 IEEE Systems Journal Best Paper Award, 2018 IEEE CSIM Best Journal Paper Award, and 2019 IEEE TCGCC Best Journal Paper Award.  He is a Life Fellow of IEEE, and a Fellow of the Royal Society of Canada (Academy of Science), Canadian Academy of Engineering, and Engineering Institute of Canada.
\end{IEEEbiography}